\documentclass[prd,preprint,nofootinbib,superscriptaddress]{revtex4-2}
\usepackage{amsmath,amssymb,graphicx,subfigure}
\usepackage{dcolumn,multirow}
\usepackage{bm}
\usepackage{soul}
\usepackage{xcolor}
\usepackage{lipsum}

\definecolor{DarkBlue}{rgb}{0.15,0.15,0.85}
\usepackage[linktocpage=true]{hyperref}
\hypersetup{colorlinks=true, citecolor=DarkBlue, linkcolor=magenta, urlcolor=magenta}
\graphicspath{{Figures/}}

\begin{document}
\title{Generic no-scale inflation inspired from string theory compactifications}
\author{Lina Wu }
\email{wulina@xatu.edu.cn}
\affiliation{School of Sciences, Xi'an Technological University, Xi'an 710021, China}
\author{Tianjun Li}
\email{tli@itp.ac.cn}
\affiliation{School of Sciences, Xi'an Technological University, Xi'an 710021, China}
\affiliation{CAS Key Laboratory of Theoretical Physics, Institute of Theoretical Physics, \\Chinese Academy of Sciences, Beijing 100190, China}
\affiliation{School of Physical Sciences, University of Chinese Academy of Sciences, No.~19A Yuquan Road, Beijing 100049, China}

\begin{abstract}

We propose the generic no-scale inflation inspired from string theory compactifications.
We consider the K\"ahler potentials with an inflaton field $\varphi$, as well as one, two, and three K\"ahler moduli.
Also, we consider the renormalizable superpotential of $\varphi$ in general.
We study the spectral index and tensor-to-scalar ratio in details,  and find the viable parameter spaces which are consistent with the Planck and BICEP/Keck experimental data on the cosmic microwave background (CMB). The spectral index is  $n_s\simeq 1-2/N \sim 0.965$ for all models,
and the tensor-to-scalar ratio $r$ is $r\simeq12/N^2$, $ 83/N^4$ and $  4/N^2$ for the one, two and three moduli models, respectively. The particular $r$ for two moduli model  comes from the contributions of the non-negligible higher order term in potential.
In the three moduli model, the scalar potential is similar to the global supersymmetry, but the K\"ahler potential is different. The E-model with $\alpha=1$ and T-model with $\alpha=1/3$ can be realized in the one modulus model and the three moduli model, respectively.  
Interestingly, the models with quadratic and quartic potentials still satisfy the current tight bound on $r$ 
after embedding into no-scale supergravity.

\end{abstract}
\maketitle
\newpage
\section{Introduction}\label{sec:intro}

It is well known that inflation solves several problems in the standard cosmology theory 
such as the horizon problem, flatness problem, and large structure of the Universe, 
etc~\cite{Starobinsky:1980te, Guth:1980zm, Linde:1981mu, Albrecht:1982wi, Lyth:1998xn}.  The almost scale-invariant density perturbation spectrum predicted by inflation is 
qualitatively consistent with the cosmological observations, in particular, 
the cosmic microwave background radiation (CMB). 
With the advent of the era of precise cosmology, more and more observations have given or will give strong
 constraints on the inflationary models. From the Planck 2018 results on the CMB measurements~\cite{Akrami:2018odb},
the scalar spectral index $n_s$,  tensor-to-scalar ratio $r$, and scalar amplitude $A_s$ 
for the power spectrum of the curvature perturbations are constrained to be
 $n_s=0.9649\pm0.0042$, $r_{0.002}\leq0.056$, and $A_s=2.10\times10^{-9}$, respectively. 
Combining with BICEP/Keck data, the tensor-to-scalar ratio is further 
limited to $r_{0.05}\leq 0.036$ at 95\% confidence level (C.L.)~\cite{BICEP:2021xfz}. 
Such a tight bound on $r$ is a big challenge to a lot of previously popular inflationary models.
Interestingly, the inflationary models inspired from the string low-energy effective actions
can have small tensor-to-scalar ratios 
$r<0.01$~\cite{Burgess:2013sla,Bhattacharya:2017pws,Cicoli:2018asa,Kallosh:2021mnu,Ellis:2021kad}.

Supersymmetry provides a natural solution to the gauge hierarchy problem in the 
particle physics Standard Model (SM), and is the promising
new physics beyond the SM. Especially, the scalar masses can be stabilized, and the
superpotential is nonrenormalized. Thus, to stabilize the inflaton potential, 
we need to consider supersymmetry. 
Moreover, gravity plays an important role in the early Universe during inflation,
so it seems to us that supergravity theory inspired from string theory 
is a natural framework for inflationary model building.
However, supersymmetry breaking scalar masses are at the order of the gravitino mass 
in general, and then at the order of  
the Hubble parameter due to the large vacuum energy density during inflation.
Thus, the slow-roll parameter $\eta$ is at the order one ${\cal O}(1)$ during inflation, which conflicts 
with the slow-roll conditions. This gives rise to 
the so-called $\eta$ problem~\cite{Copeland:1994vg,Stewart:1994ts,Dine:1995uk}.
As we know, there are a few elegant solutions: 
no-scale supergravity~\cite{Cremmer:1983bf,Ellis:1983sf, Lahanas:1986uc, Witten:1985xb, Li:1997sk, Ellis:2013xoa, Ellis:2013nxa,Ellis:2015xna,Ellis:2020lnc},  
shift symmetry in the K\"ahler potential~\cite{Kawasaki:2000yn,Yamaguchi:2000vm,Yamaguchi:2001pw,Kawasaki:2001as,Kallosh:2010ug,Kallosh:2010xz,Li:2013nfa,Nakayama:2013jka,Nakayama:2013txa,Takahashi:2013cxa,Li:2014xna,Pallis:2014xva,Li:2015mwa}, and helical phase inflation~\cite{Li:2014vpa,Li:2014unh,Li:2015taa,Sabir:2019xwk}, etc.

No-scale supergravity has vanishing cosmological constant naturally, and then
evades the Anti-de Sitter (AdS) vacua in the generic supergravity theory~\cite{Cremmer:1983bf, Ellis:1983sf,Lahanas:1986uc}.
Interestingly, no-scale supergravity can be realized by the Calabi-Yau compactification 
with standard embedding of the weakly coupled heterotic $E_8\times E_8$ theory~\cite{Witten:1985xb},
as well as by the similar compactification of M-theory on $S^1/Z_2$ \cite{Li:1997sk}.
The K\"ahler potential of no-scale supergravity inspired by 
the above string theory compactifications~\cite{Witten:1985xb, Li:1997sk} is
\begin{align}
	K&=-3\log({T+\overline{T}}-2|\varphi_i|^2)~,~
\label{One-Moduli-Kahler-Potential}
\end{align}
where $T$ is the K\"ahler moduli, and $\varphi_i$ denote the matter, Higgs and inflaton fields. 
In this paper, for simplicity, we shall neglect dilaton field and complex structure moduli,
and only consider inflaton field $\varphi$ in the following.
Because K\"ahler potential is a logarithmic real function, the $\eta$ problem is solved,
and the inflaton potential can have flat directions as well.
In particular, considering the no-scale supergravity and a Wess-Zumino superpotential, one can obtain
the $R+R^2$ Starobinsky model elegantly~\cite{Ellis:2013xoa, Ellis:2013nxa}. 
In addition to $\varphi$, $T$ can be inflaton field as well ~\cite{Ellis:2013nxa,Ellis:2015xna}.
Of course, the predicted $n_s$ and $r$ are compatible with the CMB observations. For the relevant studies
on string inflations, please see Refs.~\cite{Broy:2015zba,Ellis:2014cma,Cicoli:2011ct,Gao:2013hn,Cicoli:2013oba,Kobayashi:2017jeb,Bhattacharya:2020gnk,Cicoli:2020bao,Cicoli:2022sih}.

On the other hand, a generalized model named $\alpha$-attractor inflation is build by introducing a parameter $\alpha$ 
related to the curvature of the inflaton  K\"ahler manifold \cite{Kallosh:2013yoa}. 
The inflationary attractors \cite{Kallosh:2015lwa,Linde:2015uga,Linde:2016uec,Ellis:2019bmm,Tang:2019olx,Akrami:2020zxw,Rodrigues:2020fle,Kallosh:2021mnu,Ellis:2021kad} such as T-models and E-models 
predict the tensor-to-scalar ratio $r$ by a factor $\alpha$, which are consistent 
with the current observations and can be tested in the future. 
In particular, to generalize the above no-scale inflation,
one can introduce an $\alpha$ factor in the K\"ahler moduli
\begin{align}
	K&=-3 \alpha \log({T+\overline{T}}-2|\varphi|^2)~,~
\end{align}
and then study the unified no-scale attractors~\cite{Ellis:2019bmm}. 

Because the above simple no-scale inflation is very interesting, 
we have strong motivation to study the generic no-scale inflation inspired by the string theory compactifications.
The first question is what are the generic no-scale supergravity theories inspired by the string theory compactifications.
Previously, one of us (T. L.) has already studied various orbifold compactifications of M-theory on 
$T^6/Z_3$, $T^6/Z_6$, $T^6 /Z_{12}$, as well as the compactification by keeping
singlets under $SU (2)\times U (1)$ symmetry, and then the compactification on $S_1/Z_2$~\cite{Li:1998sq}.
Thus, the generic no-scale inflation can be inspired by these compactifications.
Inspired by the compactification by keeping   singlets under $SU(2)\times U(1)$
symmetry and then the compactification on $S^1/Z_2$~\cite{Li:1998sq}, 
we can consider the K\"ahler potential with two K\"ahler moduli $T_{1,2}$ and one chiral field $\varphi$ as follows
\begin{equation}
	K=-2\log({T_1+\overline{T}_1 }-2|\varphi|^2)-\log({T_2+\overline{T}_2})~.~
\label{Two-Moduli-Kahler-Potential}
\end{equation}
Also, inspired by the orbifold compactifications of M-theory on  $T^6/Z_{12}$ and $S^1/Z_2$~\cite{Li:1998sq}, 
we can consider the K\"ahler potential with three K\"ahler moduli $T_{1,2,3}$ and one chiral field $\varphi$ as follows
\begin{equation}
	K=-\log({T_1+\overline{T}_1}-2|\varphi|^2)-\log({T_2+\overline{T}_2})-\log({T_3+\overline{T}_3})~.~
\label{Three-Moduli-Kahler-Potential}
\end{equation}
Previously, a few relevant studies have been done.  
A chaotic inflation in no-scale supergravity with string inspired moduli stabilization,
which has two moduli, is obtained from Type IIB string compactification with an anomalous $U(1)_X$ gauged symmetry~\cite{Li:2014owa}. 
The tensor-to-scalar ratio is consistent with BICEP/Keck experimental data.
In a recent paper \cite{Wu:2021zta} with Gong, we have studied the primordial black holes and secondary gravitational waves
for the generic no-scale inflation with 
K\"ahler potentials in Eq.~\eqref{Two-Moduli-Kahler-Potential}. However, the systematical studies on the generic no-scale inflation 
have not been done yet.

In this paper, we shall perform the systematical studies on the generic no-scale inflation 
with K\"ahler potentials in Eqs.~\eqref{One-Moduli-Kahler-Potential}, \eqref{Two-Moduli-Kahler-Potential}, 
and \eqref{Three-Moduli-Kahler-Potential}. We consider $\varphi$ as an inflaton field, and
the renormalizable superpotential of $\varphi$ in general.
We study the spectral index and tensor-to-scalar ratio in detail,
 and find the viable parameter spaces which are consistent with 
the Planck and BICEP/Keck experimental data on the cosmic microwave background (CMB). The spectral index is  $n_s\simeq 1-2/N \sim 0.965$ for all models, 
and the tensor-to-scalar ratio is $r\simeq12/N^2$, $ 83/N^4$ and $  4/N^2$ for the one, two and three moduli models, respectively. The predicted $r$ for two moduli models is clearly different from that for other two models due to the non-negligible  higher order term, which will be explained in detail in Sec. \ref{sec:M2}.
In the three moduli model, the scalar potential is similar to the global supersymmetry, but the K\"ahler potential
is different. The E-model with $\alpha=1$ and T-model with $\alpha=1/3$ can be realized in the one modulus model 
and the three moduli model, respectively. 
Interestingly, the models with quadratic and quartic potentials still satisfy the current tight bound on $r$ 
after embedding into no-scale supergravity.

Based on the string compactifications,  the coefficients of the logarithmic term  are integers in K\"ahler potential. Thus, it is interesting to build the generic no-scale $\alpha$-attractor inflation. We propose the following K\"ahler potential
\begin{equation}
	K=-3\alpha \log({T+\overline{T} }-2|\varphi|^2)- 3(1-\alpha)\log({T'+\overline{T}'})~,~
\end{equation}
where $0 < \alpha \leq 1$.  From the phenomenological point of view, the generic attractor will relate the above three kinds of models with each other, and the predicted $n_s-r$ curves are consistent with CMB experimental results. We shall preform the detailed study, which will be given elsewhere in the future.

This paper is organized as follows. We briefly discuss the generic no-scale supergravity theories inspired by the
string theory compactifications in Sec. \ref{sec:sugra}.  We show the generic slow-roll inflation model and field transformation  in Sec. \ref{sec:inf}. Then we study the cosmological predictions $n_s$ and $r$ for the inflationary models with one, two and three moduli in Secs. \ref{sec:M1}, \ref{sec:M2}, and \ref{sec:M3}.  Finally in Sec. \ref{sec:conclusion}, we make conclusions with a brief discussion.
In the following, we set the reduced Planck mass $M_{\rm{Pl}}^2=8\pi G=1$ for simplicity.


\section{The Generic No-scale Supergravity Theories Inspired by the String Theory Compactifications} \label{sec:sugra}

The Lagrangian of the $\mathcal{N}=1$ supergravity can be written in the form
\begin{equation}\label{eq:lag}
	\mathcal{L}=-\frac{1}{2}R+K_i^{\overline{j}} \partial_{\mu}\varphi^i\partial^{\mu}\overline{\varphi}_{\overline{j}}-V,
\end{equation}
where the K\"ahler metric is $K_i^{\overline{j}}\equiv \partial^2K/(\partial\varphi^i\partial\overline{\varphi}_{\overline{j}})$.  The effective scalar potential is
\begin{equation}
	V=e^G\left[\frac{\partial G}{\partial\varphi^i}\left(K^{-1}\right)^i_{\overline{j}}\frac{\partial G}{\partial \overline{\varphi}_{\overline{j}}}-3\right],
\end{equation}
where the K\"ahler function is $G\equiv K+\ln{|W|^2}$, and $\left(K^{-1}\right)^i_{\overline{j}}$ is the inverse of the K\"ahler metric. Introducing K\"ahler covariant derivative
\begin{equation}
	D_i W\equiv  W_i+K_i W,
\end{equation}
we obtain the scalar potential 
\begin{equation}
	V=e^K\left[D_iW\left(K^{-1}\right)^i_{\overline{j}}D^{\overline{j}}\overline{W}-3|W|^2\right]~.~\,\label{eq:pot-sugra}
\end{equation}

To study the inflation in the generic no-scale supergravity theories inspired by the
string theory compactifications, 
we parametrize the generic K\"ahler potential and superpotential as follows:
\begin{align}
	K&=-N_1\log({T_1+\overline{T}_1}-2|\varphi|^2)-\sum_{i=2,3}N_i\log({T_i+\overline{T}_i}),\label{eq:genK}\\
	W&=\sum_{i=0}^{3} a_i(\sqrt{2}\varphi)^i,\label{eq:genW}
\end{align}
where $N_1+N_2+N_3=3$, $T_i$ are K\"ahler moduli, and $\varphi$ is inflaton field. 
According to the number of K\"ahler moduli $T_i$,
we shall classify the inflationary models as one moduli model, two moduli model,
and three moduli model~\cite{Wu:2021zta}.
We also consider $W_T=0$, and the $\eta$ problem is avoided 
since no large mass term is generated \cite{Gaillard:1995az,Diamandis:1986zg}. 
For simplicity, we assume inflation along the ${\rm Re}(\varphi)$ direction as well.
Moreover, the K\"ahler metric and covariant derivative are
\begin{equation}
	K_i^{\bar{j}}=\left(
	\begin{array}{cccc}
		\frac{N_1}{X^2} & 0 & 0 & -\frac{2N_1 \varphi }{X^2} \\
		0 & \frac{N_2}{Y_2^2} & 0 & 0 \\
		0 & 0 & \frac{N_3}{Y_3^2} & 0 \\
		-\frac{2N_1 \overline{\varphi}}{X^2} & 0 & 0 & \frac{2N_1 Y_1}{X^2} \\
	\end{array}
	\right)
\end{equation}
\begin{equation}
	D_iW=W\begin{pmatrix}
		 -\frac{N_1}{X}~&~-\frac{N_2}{Y_2} ~&~-\frac{N_3}{Y_3} ~&~ \frac{2N_1\bar{\varphi}}{X}+\frac{W_{\varphi}}{W}
	\end{pmatrix}
\end{equation}
where $X\equiv T_1+\overline{T}_1-2|\varphi|^2$, $Y_i\equiv T_i+\overline{T}_i$. Thus, the general scalar potential can be written as
\begin{equation}
	V=\frac{|W_{\varphi}|^2}{2N_1 X^{N_1-1} Y_2^{N_2}Y_3^{N_3}}.\label{Eq:genV}
\end{equation}

In the following section, we will study inflationary models with typical K\"ahler potential in the general no-scale 
supergravity model. Then we will compare the predictions of these models with CMB observations in order to find the observational constraints on the parameter  space of the models. 

\section{Inflationary models}\label{sec:inf}

We assume that all the real components of the complex fields which do not drive inflation have been stabilized, 
whereas the inflaton field remains dynamical.  Following Refs.~\cite{Ellis:2013nxa,Ellis:2013xoa,Garg:2015mra,Garg:2017tds,Khalil:2018iip}, the real and imaginary parts of modulus $T_i$ can be stabilized by adding $(T_i+\overline{T}_i-1)^4$ terms and $(T_i-\overline{T}_i)^4$ terms into the log terms of the K\"ahler potential. In this paper, we fix the moduli $T_i$ with 
 vacuum expectation values (VEVs) $2\langle {\rm{Re}} (T_i)\rangle = c_i$ and $\langle{\rm{Im}}(T_i)\rangle=0$. 
Also, without loss of generality, we assume the inflation trajectory along the real part of the $\varphi$ direction.  
The scalar potential in the Jordan frame is 
\begin{equation}
\begin{split}
		V&= V_0\frac{|W_{\varphi}|^2}{\left(c_1-2|\varphi| ^2\right)^{N_1-1}},\\
\end{split}
\end{equation}
where $V_0=1/(2N_1c_2^{N_2}c_3^{N_3})$.
The inflationary model with $N_1=1$ is similar to that with the global supersymmetry.

The kinetic term in terms of the field $\varphi$ in Eq. \eqref{eq:lag} is noncanonical, 
so we need to define a new canonical field $\chi$, which satisfies
\begin{equation}
	\frac{1}{2}\partial_{\mu}\chi \partial^{\mu}\chi=K_{\varphi\overline{\varphi}}\partial_{\mu}\varphi\partial^{\mu}\overline{\varphi}
\end{equation}
with
\begin{equation}\label{eq:fieldtrans}
	K_{\varphi\overline{\varphi}}=\frac{2N_1 (T_1+\overline{T}_1)}{(T_1+\overline{T}_1-2|\varphi|^2  )^2}.
\end{equation}
By integrating the above equation, we get the field transformation 
\begin{equation}
\varphi=\sqrt{\frac{c_1}{2}}\tanh{\left(\frac{\chi}{\sqrt{2N_1}}\right)}.
\end{equation}
Then the scalar potential in the Einsten frame is 
\begin{equation}
\begin{split}
		V=V_1  \text{sech}^{2m}\left(b\chi \right)
		 \left(a_1+2 a_2 \sqrt{c_1} \tanh \left(b\chi \right)+3 a_3 c_1 \tanh ^2\left(b\chi \right)\right)^2
\end{split}
\end{equation}
where $V_1=N_1^{-1}c_1^{m}c_2^{-N_2}c_3^{-N_3}$ and $b=1/\sqrt{2N_1}$, $m=1-N_1$.  Furthermore, the T-model and E-model \cite{Kallosh:2013yoa} will be realized by setting $N_1=1$ with $a_1=a_3=0$ and $N_1=3$ with $a_1=0$, respectively. Defining $\varphi=(\phi+i \xi)/\sqrt{2}$ and taking T-models for example, the squared masses of $\xi$ are $m_{\xi}^2=8a_2^2/(c_2 c_3)$ for the $\varphi^2$ model and $m_{\xi}^2=36a_3^2 \phi^2/(c_2 c_3)$ for the $\varphi^4$ model.

In the following calculations, the new canonical field $\chi$ is regarded as inflaton. The slow-roll parameters are 
\begin{equation}
	\varepsilon(\chi)=\frac{M_{\rm{Pl}}^2}{2}\left(\frac{V_{\chi}}{V}\right)^2~,~~\eta(\chi)=M_{\rm{Pl}}^2\frac{V_{\chi\chi}}{V}
\end{equation}
The CMB observations in terms of slow-roll parameters are $n_s=1-2\varepsilon+6\eta$ and $r=16\varepsilon$. Under the slow-roll approximation, the power spectrum can be expressed by
\begin{equation}
	A_s= \frac{1}{24\pi^2M_{\rm{Pl}}^2}\frac{V}{\varepsilon},
\end{equation}
and is fixed to $2.10\times10^{-9}$ by choosing the proper parameter $V_0$ in our calculations.

\section{One Modulus Inflationary Model}\label{sec:M1}

The simple no-scale supergravity model with one modulus \cite{Cremmer:1983bf, Lahanas:1986uc}
can be realized via the Calabi-Yau compactification with standard embedding of
the weakly coupled heterotic $E_8\times E_8$ theory~\cite{Witten:1985xb} and M-theory on $S^1/Z_2$~\cite{Li:1997sk}.
Assuming  K\"ahler potential of the form
\begin{equation}
	K=-3 \ln{\left(T+\overline{T}-2|\varphi|^2\right)}
\end{equation}
and substituting $N_1=3$ and $N_2=N_3=0$ into Eq. \eqref{Eq:genV}, the scalar potential in the Jordan frame and the Einstein frame become
\begin{equation}
\begin{split}
	V&= \frac{|W_{\varphi}|^2}{6 \left(c_1-2|\varphi| ^2\right)^2} \\
	&=V_0 e^{-2\sqrt{\frac{2}{3}}\chi } \left(1+A_1 e^{ \sqrt{\frac{2}{3}} \chi }+B_1 e^{2\sqrt{\frac{2}{3}} \chi }\right)^2~,~\label{Eq:V_simple}
\end{split}
\end{equation}
where  $V_0=\frac{\left(a_1-2 a_2 \sqrt{c}+3 a_3 c_1\right)^2}{48 c_1^2}$, $A_1=\frac{2 (a_1-3 a_3 c_1)}{a_1-2 a_2 \sqrt{c_1}+3 a_3 c_1}$, and $B_1=\frac{a_1+2 a_2 \sqrt{c_1}+3 a_3 c_1}{a_1-2 a_2 \sqrt{c_1}+3 a_3 c_1}$.   

\subsection{Zero parameter models}
First, we consider the inflationary case only with one term in the numerator of the potential \eqref{Eq:V_simple}, which can be rewritten as
\begin{equation}
	\begin{split}
		a_1=a_2=0; & ~V
		=3 a_3^2 \sinh ^4\left(\frac{\chi }{\sqrt{6}}\right),\\
		a_1=a_3=0; & ~V
		=\frac{a_2^2 }{3 c}\sinh ^2\left(\sqrt{\frac{2}{3}} \chi \right),\\
		a_2=a_3=0; & ~V
		=\frac{a_1^2 }{3 c^2}\cosh ^4\left(\frac{\chi }{\sqrt{6}}\right).
	\end{split}
\end{equation}
The minimum of potential is at $\chi_m=0$ and inflation will occur on the left or right branches. However, in these models, the second order slow-roll parameters are 
\begin{equation}
	\begin{split}
		a_1=a_2=0;~\eta(\chi)&=2 \coth ^2\left(\frac{\chi }{\sqrt{6}}\right)+\frac{2}{3},\\
		a_1=a_3=0;~\eta(\chi)&=\frac{4}{3} \left(\coth ^2\left(\sqrt{\frac{2}{3}} \chi \right)+1\right),\\
		a_2=a_3=0;~\eta(\chi)&=2 \tanh ^2\left(\frac{\chi }{\sqrt{6}}\right)+\frac{2}{3}.
	\end{split}
\end{equation}
Therefore, inflation is unbearable since slow-roll conditions $\varepsilon,|\eta|\ll 1$ are unsatisfied and $\eta\to8/3$ in the limit $\chi\to \infty$.

\subsection{$a_1=0$ case: $A_1+B_1=-1$} 
\begin{figure}[t]
	\subfigure[]{\includegraphics[width=0.45\linewidth]{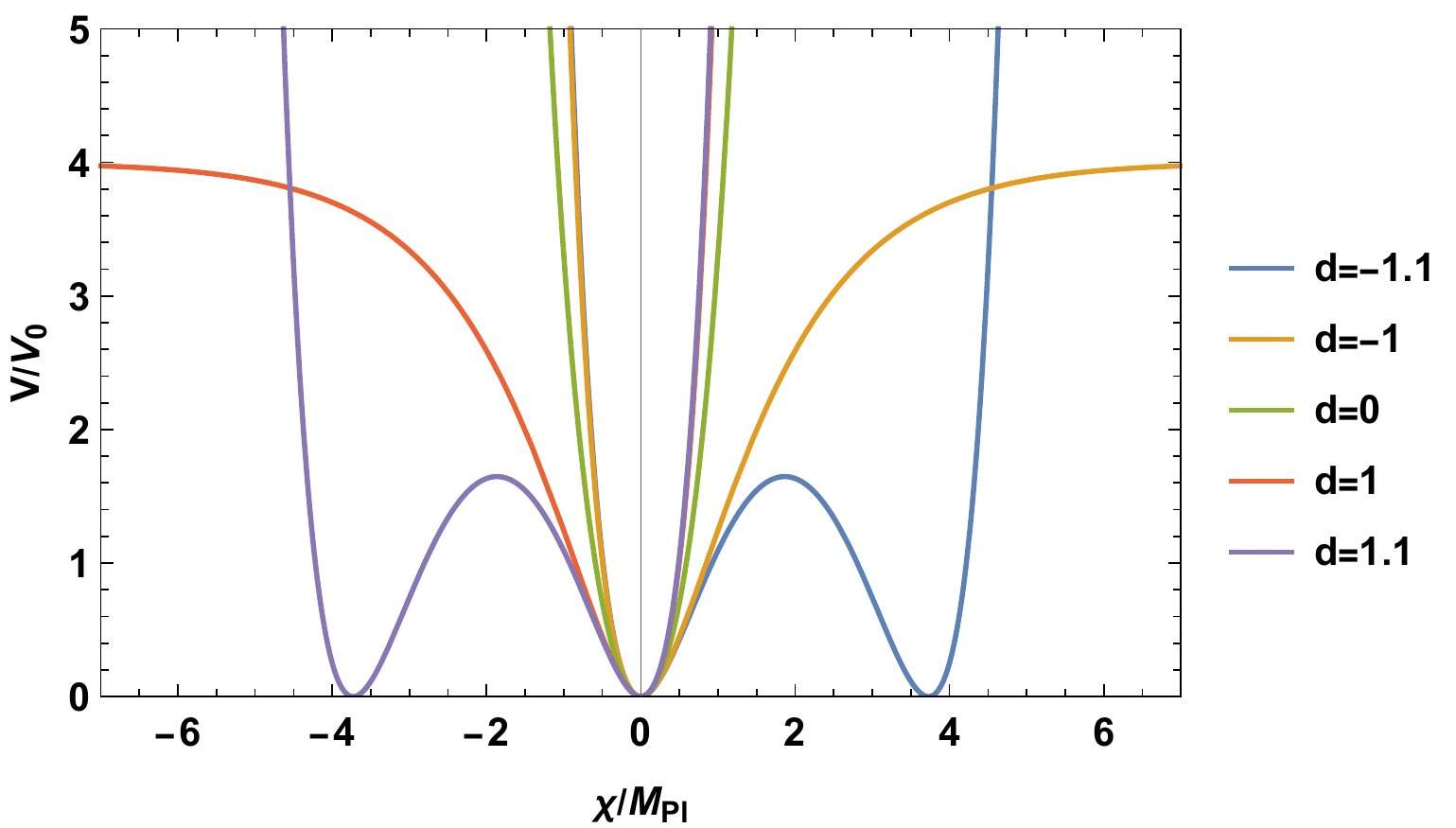}}
	\subfigure[]{\includegraphics[width=0.4\linewidth]{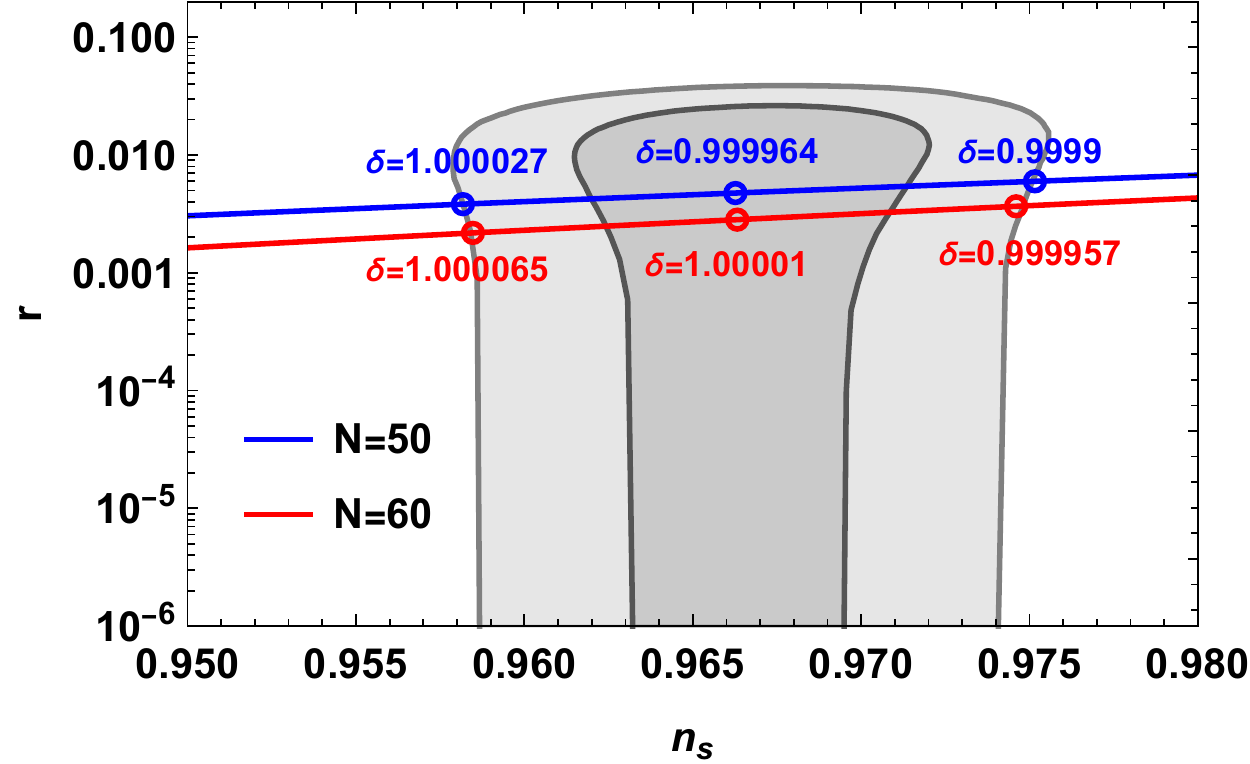}}
	\caption{Inflationary potential (a)  and the $r$ versus $n_s$ predictions (b) for the inflaton potential
in Eq.~\eqref{Eq:V_oneD} with parameter $d=\pm \delta$.}\label{fig:pot-oneD}
\end{figure}

For $a_1=0$, the two parameters $A_1$ and $B_1$ relate with each other as $A_1+B_1=-1$, and defining a new parameter $d=3a_3\sqrt{c_1}/2a_2$, the potential \eqref{Eq:V_simple} becomes 
\begin{equation}
	V=V_0 \left(1-e^{-\sqrt{\frac{2}{3}} \chi }\right)^2 \left( \left(e^{\sqrt{\frac{2}{3}} \chi }+1\right)+d \left(e^{\sqrt{\frac{2}{3}} \chi }-1\right)\right)^2,\label{Eq:V_oneD}
\end{equation}
with $V_0=a_2^2/12c_1$. The similar inflation in the simple Wess-Zumino model has been previously discussed by Ellis \textit{et al.}~\cite{Ellis:2013xoa}. Here for completeness, we will go through it as well.   The potential with different parameter $d$ are shown in Fig. \ref{fig:pot-oneD}(a). We note that the potential remains the same after both parameter $d$ and field $\chi$ becoming $-d$ and $-\chi$. Therefore, we will only discuss the negative $d$ in the following calculation. 

\begin{figure}[t]
	\subfigure[]{\includegraphics[width=0.4\linewidth]{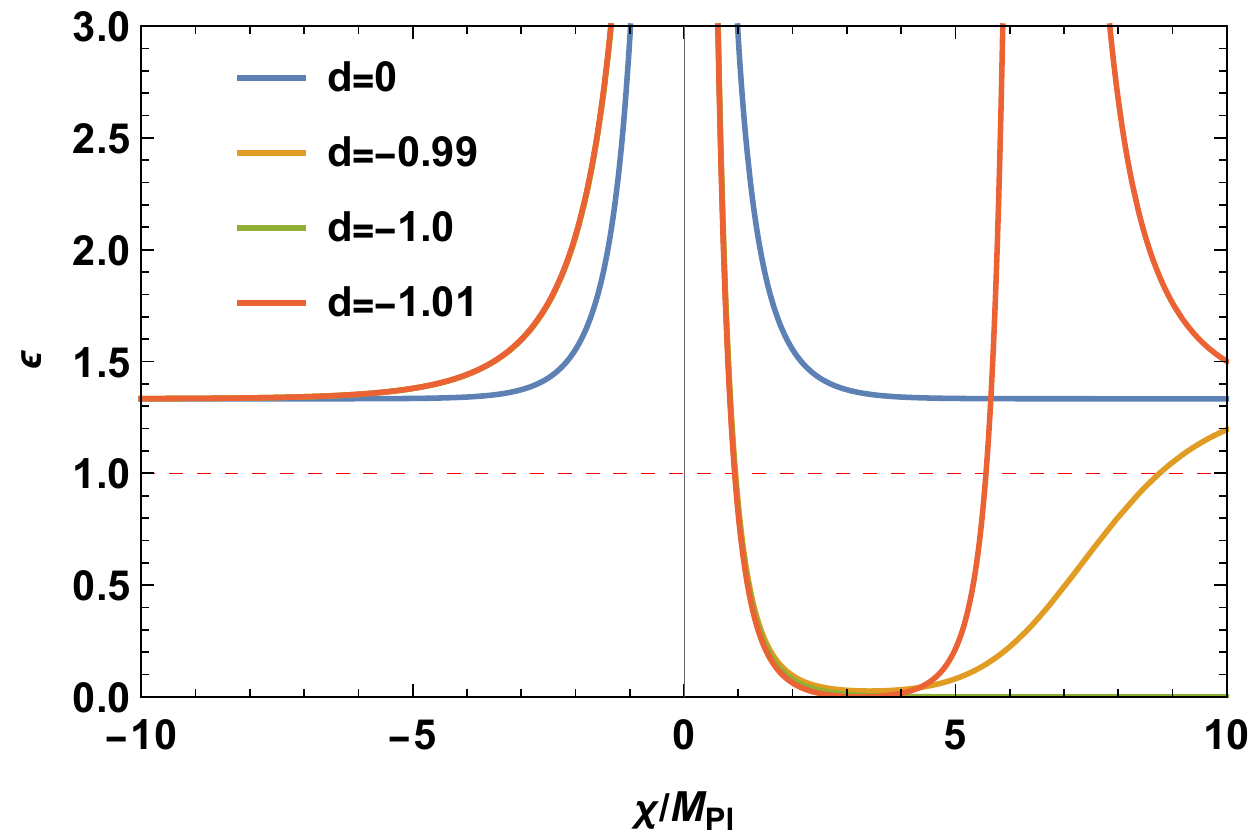}}
	\subfigure[]{\includegraphics[width=0.4\linewidth]{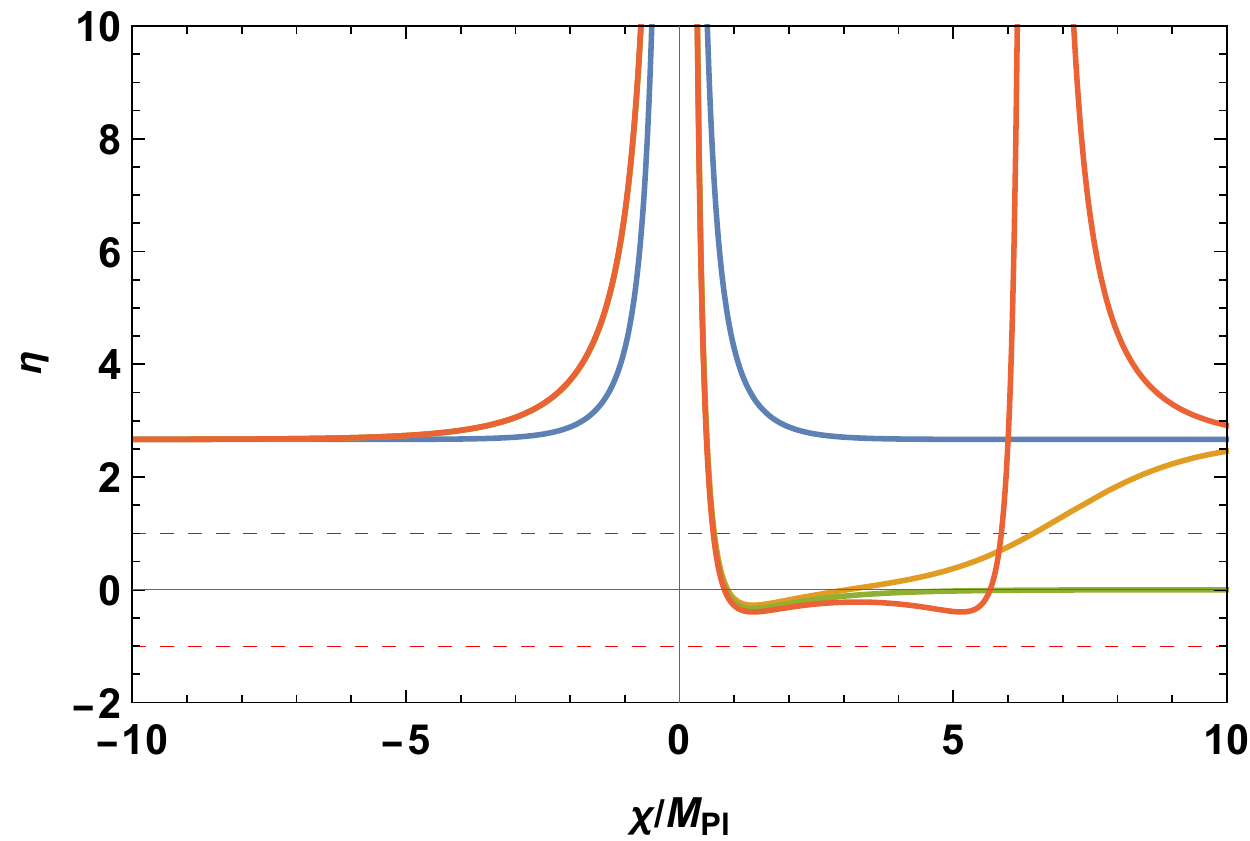}}
	\caption{The evolution of the slow-roll parameters (a) $\varepsilon$ and (b) $\eta$ for the inflaton potential in Eq. \eqref{Eq:V_oneD}.}\label{fig:slowroll-oneD}
\end{figure}

When $d$ is larger than $-1$, there is a vacuum at $\chi=0$. When $d$ is smaller than $-1$, there are two minima at $\chi_{m_1}=0$ and $\chi_{m_2}=\frac{\sqrt{6}}{2} \log \left(\frac{d+1}{d-1}\right)$, and a maximum at $\chi_M=\frac{\sqrt{6}}{4} \log \left(\frac{d+1}{d-1}\right)$. From the evolution of the slow-roll parameters in Fig. \ref{fig:slowroll-oneD}, the possible inflationary trajectories locate in the region $[0,\infty]$ for  $d\gtrsim -1$ and the region $[\chi_{m_1},\chi_{m_2}]$ for $d\lesssim -1$. Others are ruled out due to the steep potential, and the slow-roll conditions $\varepsilon,|\eta| <1$ are not satisfied. The cosmological predictions $n_s$ and $r$ for the model in Eq. \eqref{Eq:V_oneD} are numerically shown in Fig.~\ref{fig:pot-oneD}(b) and the inflationary trajectory is in the large field region $\chi\sim 5M_{Pl}$. The parameter $d$ locates in a tiny range around $\pm1$.

\subsubsection{E-model realization }
Particularly, when $d=-1$ or $2a_2=-3a_3\sqrt{c_1}$,  the E-model for $\varphi^2$ \cite{Kallosh:2013yoa,Kallosh:2021mnu} or Starobinsky inflation model \cite{Ellis:2013xoa,Ellis:2020lnc} with potential $V=V_0(1-e^{-\sqrt{2/3}\chi})^2$  is realized, shown in Table~\ref{tab:T-Emodel}. 
The potential can be expanded  as
\begin{equation}
	V=V_0\left(1-2e^{-\sqrt{2/3}\chi}+e^{-2\sqrt{2/3}\chi}\right) .
\end{equation}   
The spectrum index, tensor-to-scalar ratio, and \textit{e}-folding number are expressed in the form \cite{Ellis:2013xoa,Ellis:2020lnc}
\begin{equation}
\begin{split}
	n_s &\simeq 1-\frac{8}{3}e^{-\sqrt{2/3}\chi}+\mathcal{O}(e^{-2\sqrt{2/3}\chi}) \\
	r&\simeq \frac{64}{3}e^{-2\sqrt{2/3}\chi}+\mathcal{O}(e^{-4\sqrt{2/3}\chi}) \\
	N&\simeq\frac{3}{4}e^{\sqrt{2/3}\chi}+\mathcal{O}(e^{-2\sqrt{2/3}\chi}) 
\end{split}
\end{equation}
and 
\begin{equation}
	n_s\simeq 1-\frac{2}{N},~~r\simeq \frac{12}{N^2}.
\end{equation}
Thus, the predictions $n_s=[0.960,0.967]$ and $r=[0.0033,0.0048]$ for $N=[50,60]$ are consistent with Planck and BICEP/Keck experiments \cite{Akrami:2018odb,BICEP:2021xfz}. Similarly with the generalization in Ref. \cite{Kallosh:2013yoa}, replacing $N_1$ with $\alpha N_1$, a factor $\alpha$ will be added to the leading term of $r$ and the E-models of $\alpha$-attractors are achieved.
\begin{table*}
	\caption{The realization of T- and E-models in one modulus and three moduli models.}\label{tab:T-Emodel}
	\begin{tabular}{p{120pt}p{90pt}p{110pt}p{80pt}}
		\hline
		Cases&\multicolumn{3}{c}{$V\propto  |W_{\varphi}|^2/(c_1-2\varphi^2)^{2}$}\\
		\hline
		$a_{1}=0, 2a_2=-3a_3\sqrt{c_1}$&$V_J\propto\frac{|\varphi(1-\sqrt{2}\varphi)|^2}{(c_1-2|\varphi|^2)^2}$ & $V_E\propto (1-e^{-\sqrt{2/3}\chi})^2$& E-model for $\varphi^2$\\
		\hline\hline
		Cases&\multicolumn{3}{c}{$V \propto |W_{\varphi}|^2$}\\
		\hline
		$a_{1,3}=0$ &$V_J\propto |\varphi|^2$ & $V_E\propto \tanh^2\left(\chi/\sqrt{2}\right)$& T-model for $\varphi^2$\\
		$a_{1,2}=0$ &$V_J\propto |\varphi^2|^2$& $V_E\propto \tanh^4\left(\chi/\sqrt{2}\right)$&T-model for $\varphi^4$\\
		\hline
	\end{tabular}
\end{table*}

\subsection{$a_2=0$ case}
\begin{figure}[t]
	\subfigure[]{\includegraphics[width=0.45\linewidth]{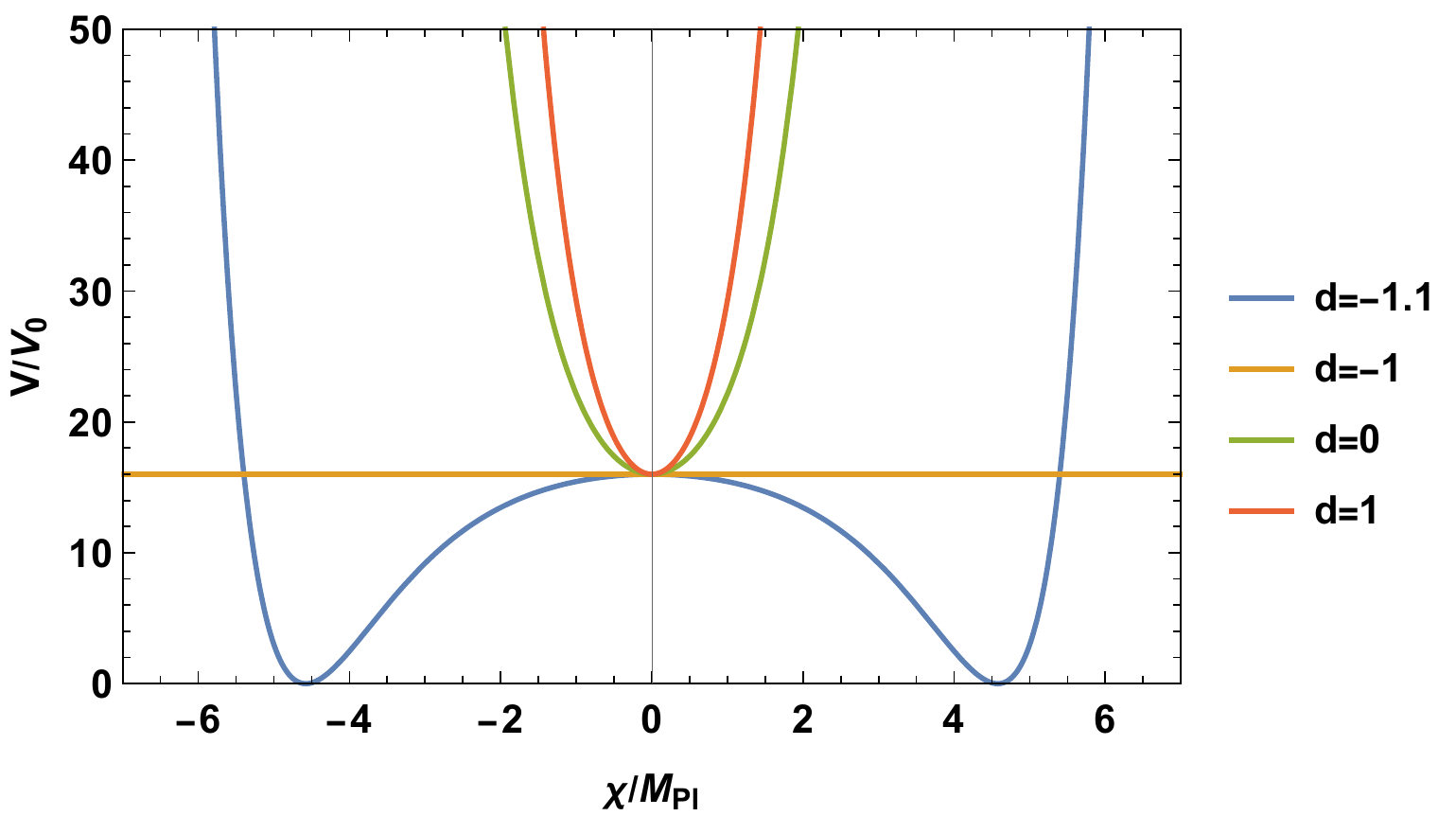}}
	\subfigure[]{\includegraphics[width=0.4\linewidth]{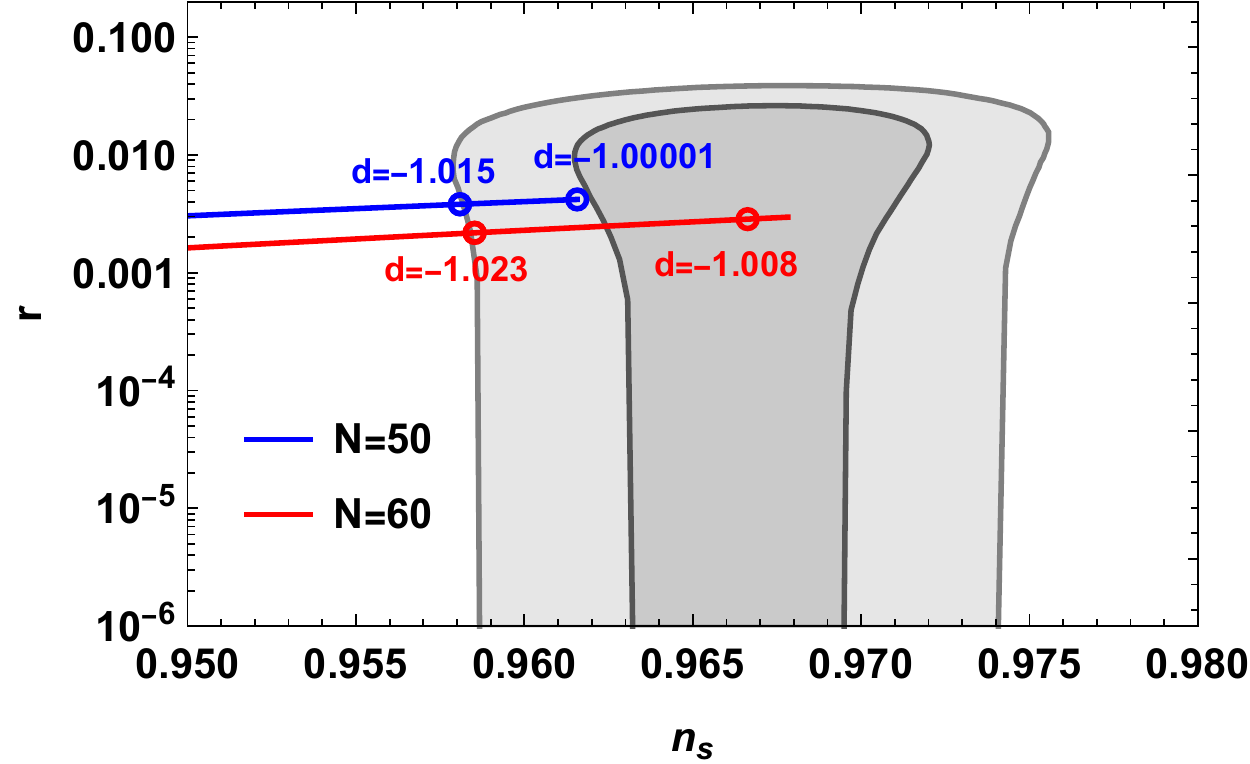}}
	\caption{Inflationary potential (a) and the $r$ versus $n_s$ predictions (b) for the inflaton potential 
in Eq. \eqref{Eq:V_oneE} with parameter $d<-1$.}\label{fig:pot-oneE}
\end{figure}
\begin{figure}[!h]
	\subfigure[]{\includegraphics[width=0.4\linewidth]{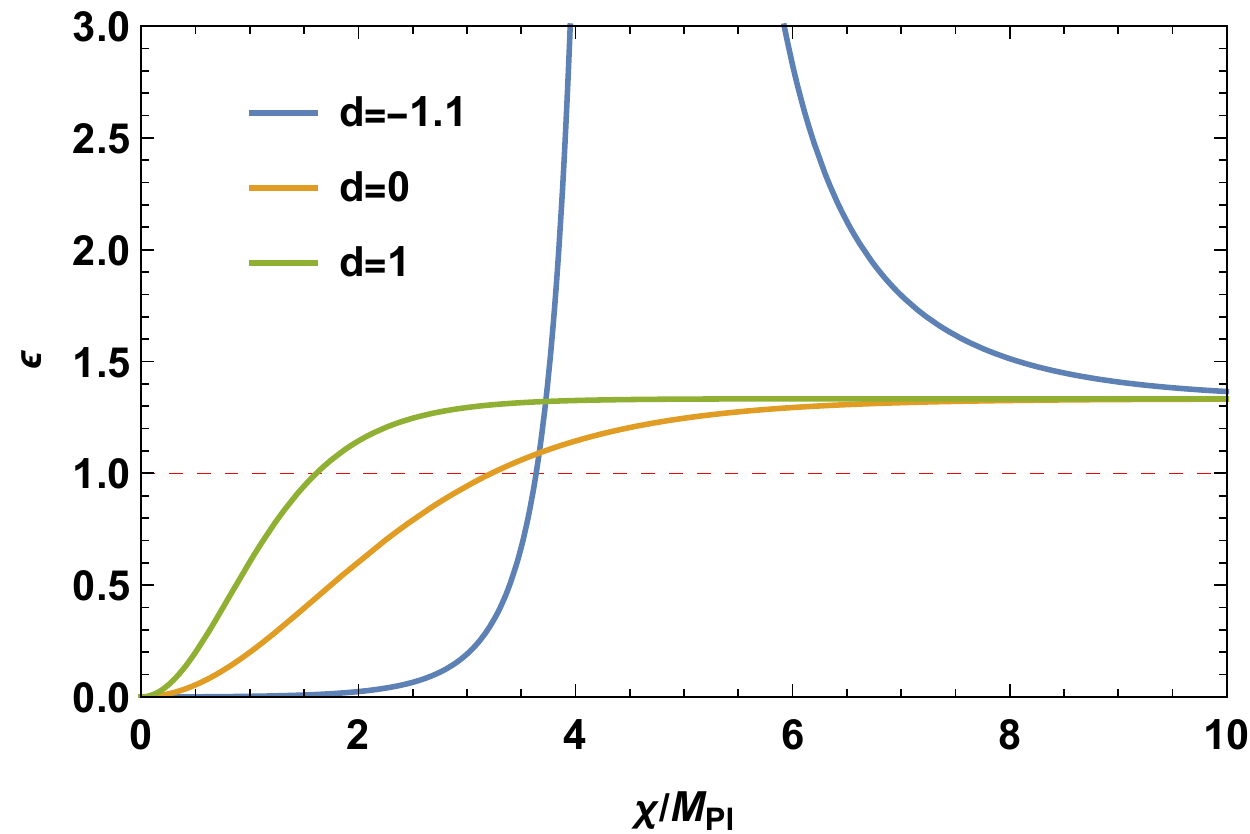}}
	\subfigure[]{\includegraphics[width=0.4\linewidth]{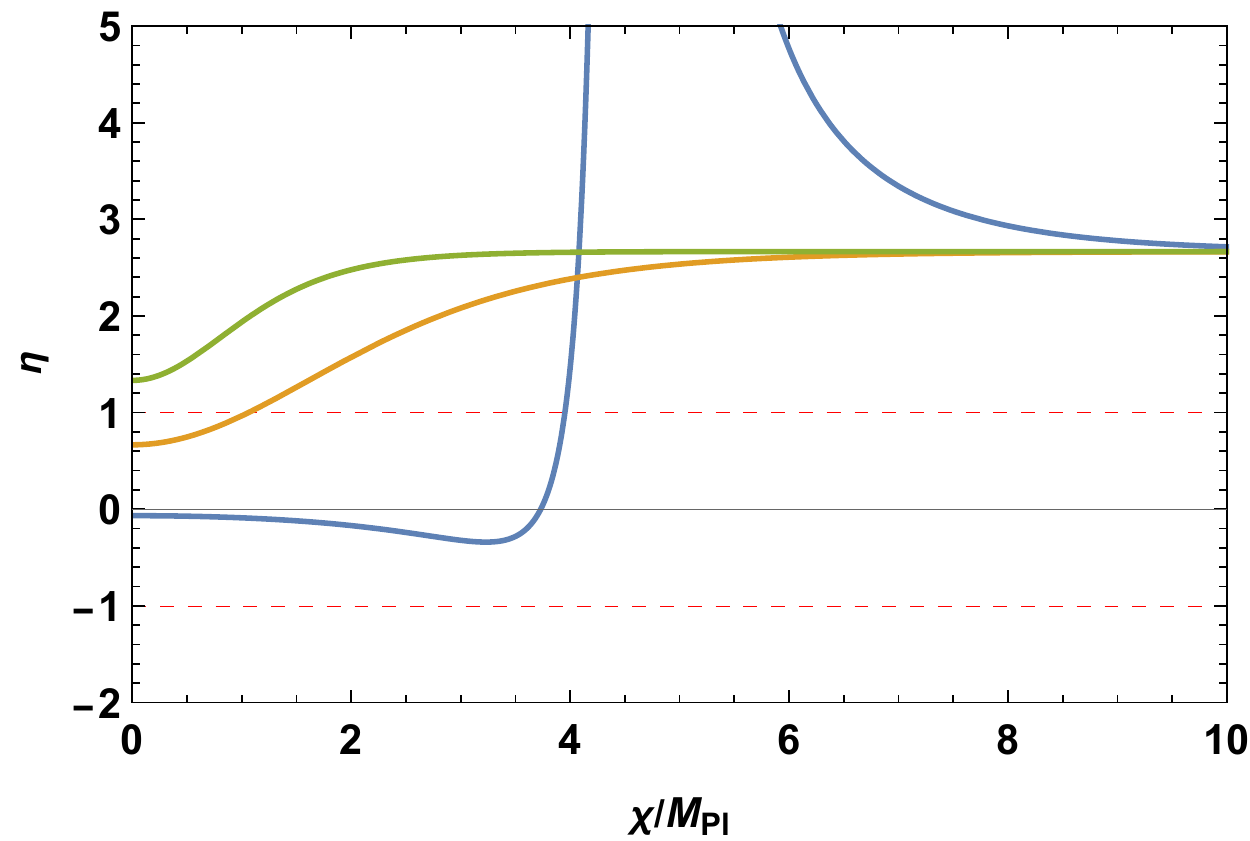}}
	\caption{The evolution of the slow-roll parameters (a) $\varepsilon$ and (b) $\eta$ for the inflaton potential in Eq. \eqref{Eq:V_oneE}.}\label{fig:slowroll-oneE}
\end{figure}
The potential becomes
\begin{equation}
\begin{split}
	V&=V_0e^{-2 \sqrt{\frac{2}{3}} \chi } \left( \left(e^{\sqrt{\frac{2}{3}} \chi }+1\right)^2+d \left(e^{\sqrt{\frac{2}{3}} \chi }-1\right)^2\right)^2\label{Eq:V_oneE}\\
\end{split}
\end{equation}
with $V_0=a_1^2/48c_1^2$ and $d=3a_3c_1/a_1$. The potential is an even function in terms of inflaton $\chi$, shown in Fig. \ref{fig:pot-oneE}. So we will consider inflation in the positive field regime. When $d>-1$, there is only one minimum at $\chi_m=0$ for the potential and the slow-roll inflation is not allowed due to $\varepsilon,\eta>1$. When $d<-1$, there are two minima at $\chi_{m1,m2}=\sqrt{ {3}/{2}} \log \left(- {(1\pm \sqrt{-d})^2}/{(1+d)}\right)$ and one maximum at $\chi_M=0$ for the potential. From Fig. \ref{fig:slowroll-oneE}, the slow-roll parameters are $\varepsilon\to 4/3$ and $\eta\to 8/3$ in the limit $\chi\to\infty$, thus inflation only happens in the region $[\phi_M,\phi_{m1}]$. The predicted observations $n_s$ and $r$ are also shown in Fig. \ref{fig:pot-oneE}(b).

\subsection{$a_3=0$ case}
\begin{figure}[!h]
	\subfigure[]{\includegraphics[width=0.45\linewidth]{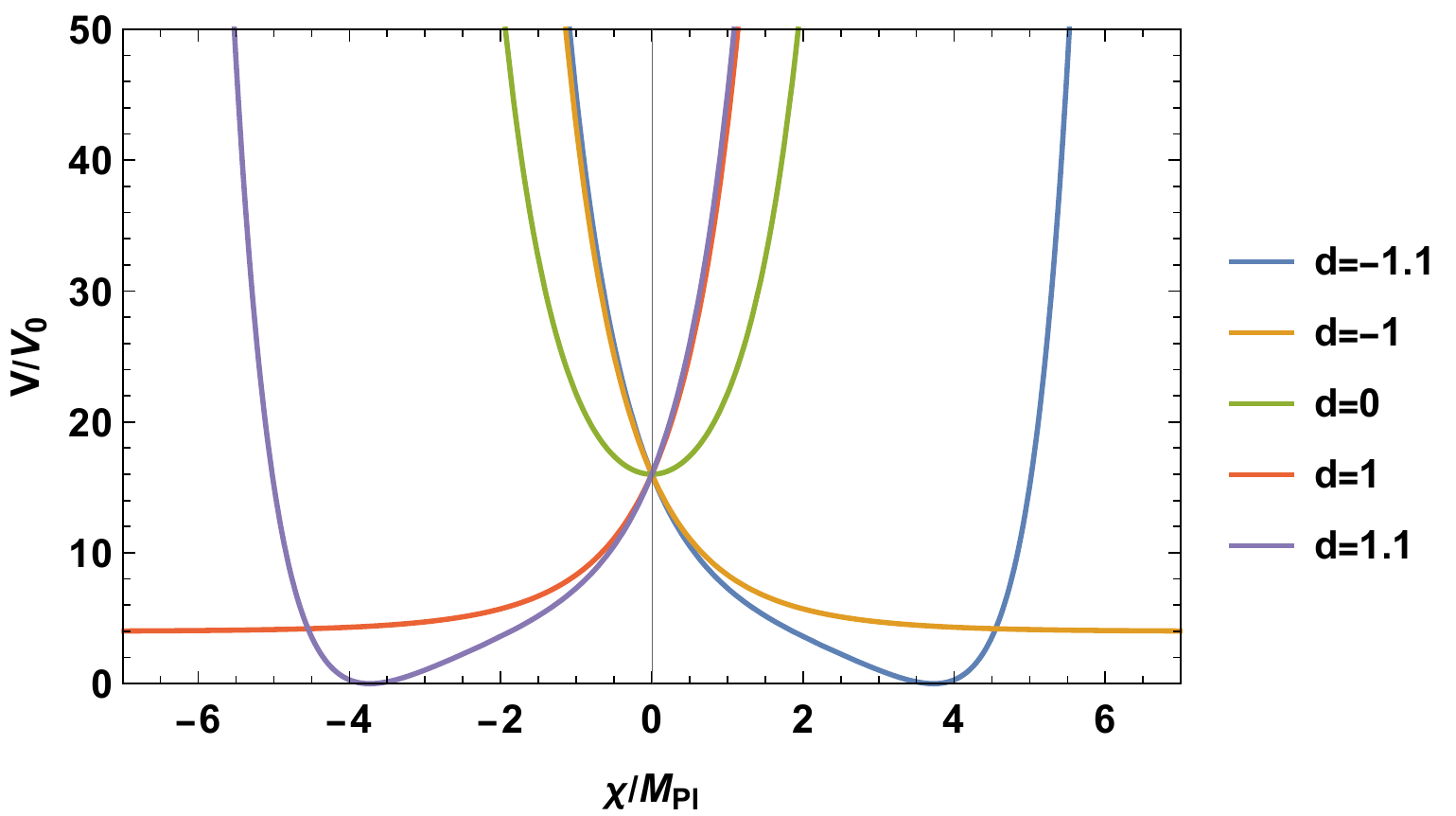}\label{fig:pot-oneF}}
	\subfigure[]{\includegraphics[width=0.4\linewidth]{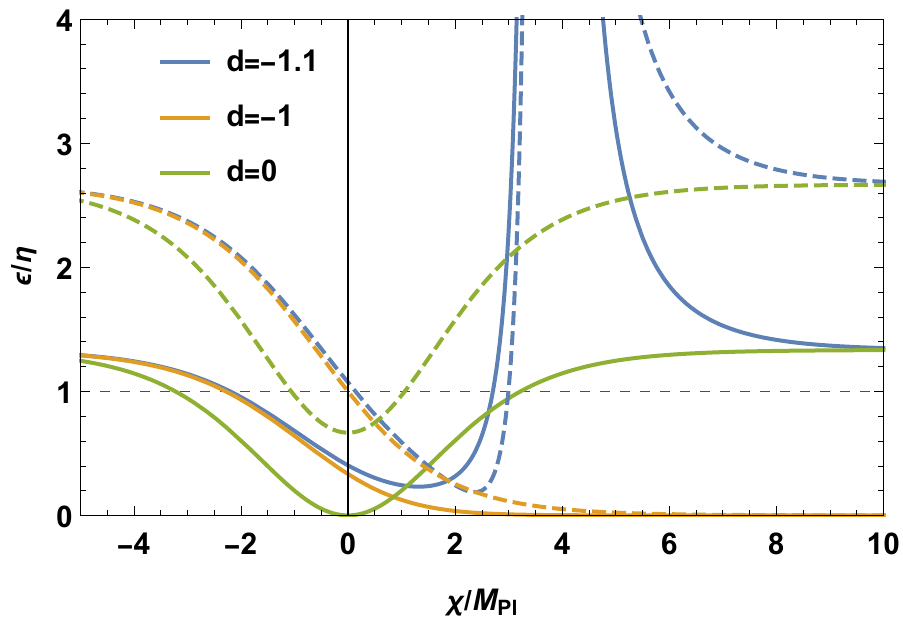}\label{fig:slowroll-oneF}}
	\caption{Inflationary potential (a) and the evolution of the slow-roll parameters (b) 
for the inflaton potential in Eq. \eqref{Eq:V_oneF}. The solid and dashed lines 
correspond to $\varepsilon$ and $\eta$, respectively.}
\end{figure}

The potential becomes 
\begin{equation}
\begin{split}
	V&=V_0 \left(1+e^{-\sqrt{\frac{2}{3}}\chi}\right)^2 \left( \left(e^{\sqrt{\frac{2}{3}} \chi }+1\right)+d  \left(e^{\sqrt{\frac{2}{3}} \chi }-1\right)\right)^2\label{Eq:V_oneF}\\
\end{split}
\end{equation}
with $V_0=a_1^2/48c_1^2$ and $d=2a_2\sqrt{c_1}/a_1$. The potential is shown in Fig. \ref{fig:pot-oneF}, and remains the same when both parameter $d$ and field $\chi$ become negative. Thus, we will seek the  inflation trajectory with $d\leq0$. There is a minimum at $\chi_m= \sqrt{\frac{3}{8}} \log \left(\frac{2}{d+1}-1\right)$ for potential with $-1<d\leq0$ and a minimum at $\sqrt{\frac{3}{2}} \log \left(\frac{d-1}{d+1}\right)$ with $d<-1$. From the evolution of slow-roll parameters in Fig. \ref{fig:slowroll-oneF}, we find that inflation impossibly happens for the model in Eq. \eqref{Eq:V_oneF}.

\subsection{General case}
\begin{figure}[t]
	\subfigure[]{\includegraphics[width=0.5\linewidth]{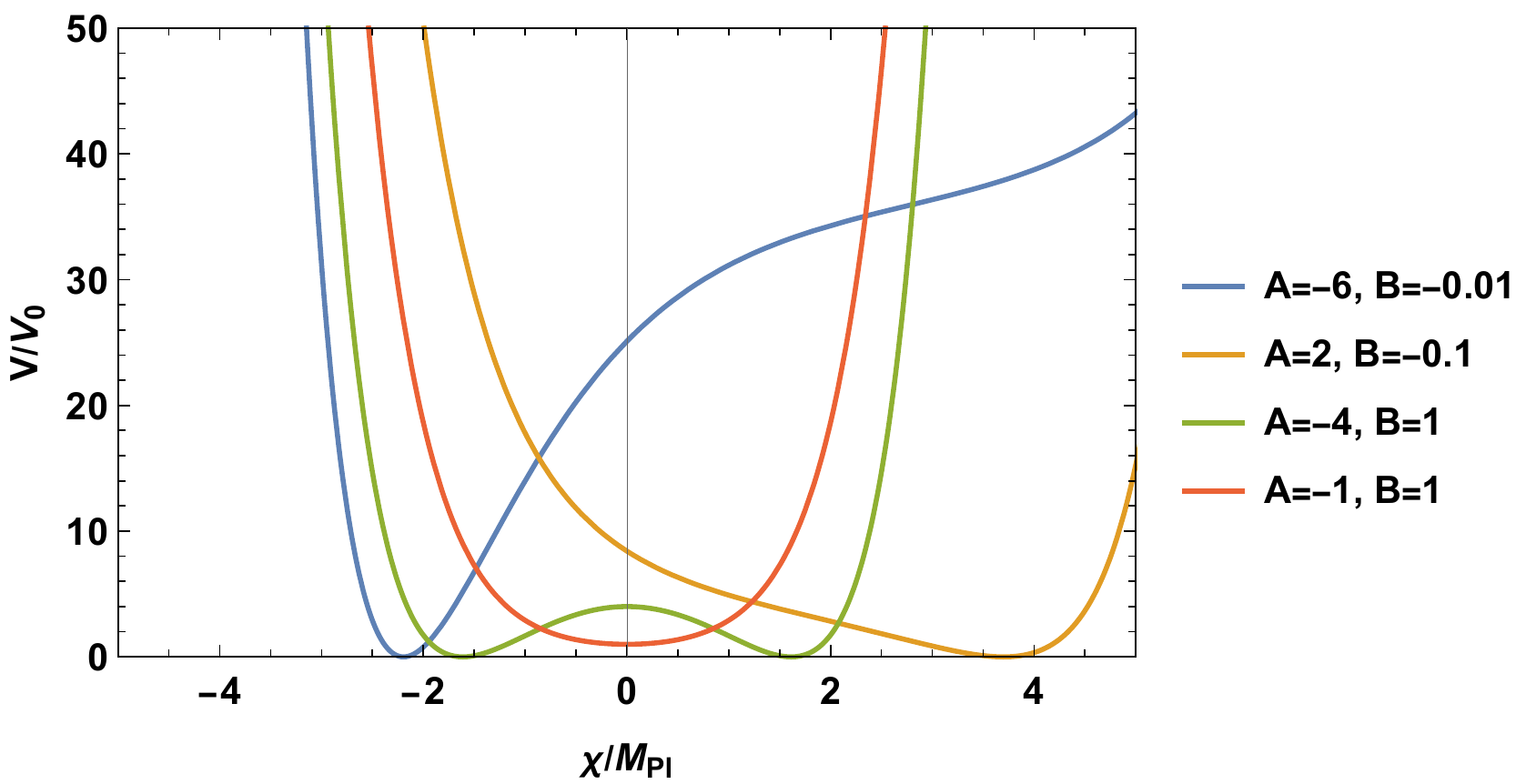}\label{fig:pot-oneG}}
	\subfigure[]{\includegraphics[width=0.4\linewidth]{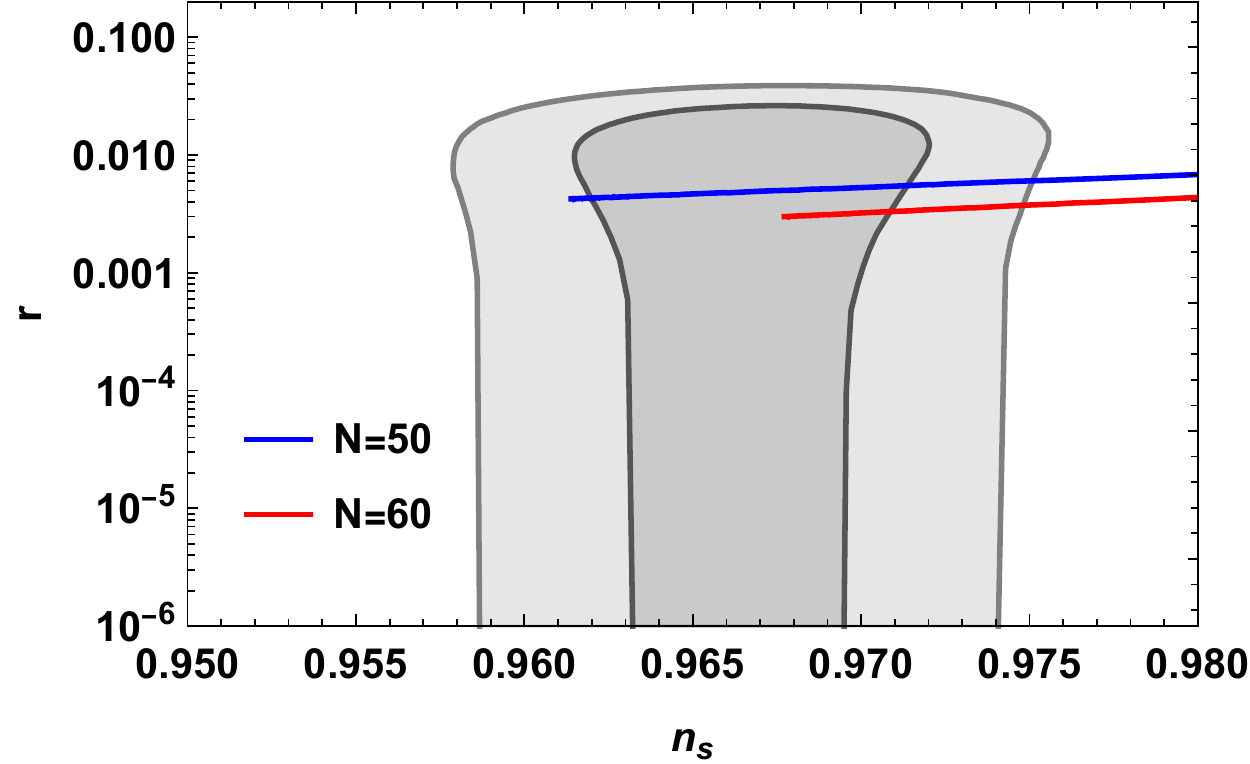}\label{fig:nsr-oneG1}}\subfigure[]{\includegraphics[width=0.4\linewidth]{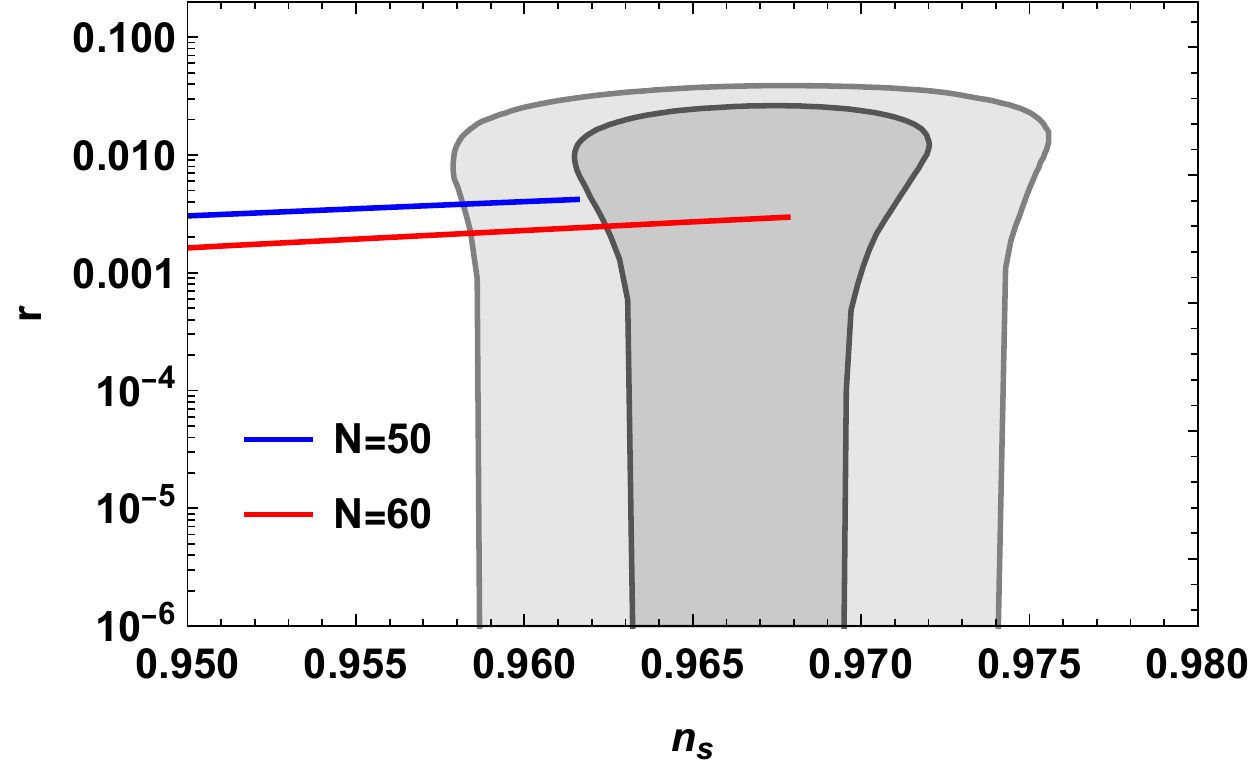}\label{fig:nsr-oneG2}}
	\caption{Inflationary potential (a) and the $r$ versus $n_s$ 
 predictions (b)-(c) for the inflaton potential
 in Eq. \eqref{Eq:V_simple}. Also,  (b) has $B_1<0$, and (c) has $A_1<0,~ B_1>0$ and $A_1^2>4B_1$.}\label{fig:nsr-oneG}
\end{figure}
\begin{figure}
	\includegraphics[width=0.4\linewidth]{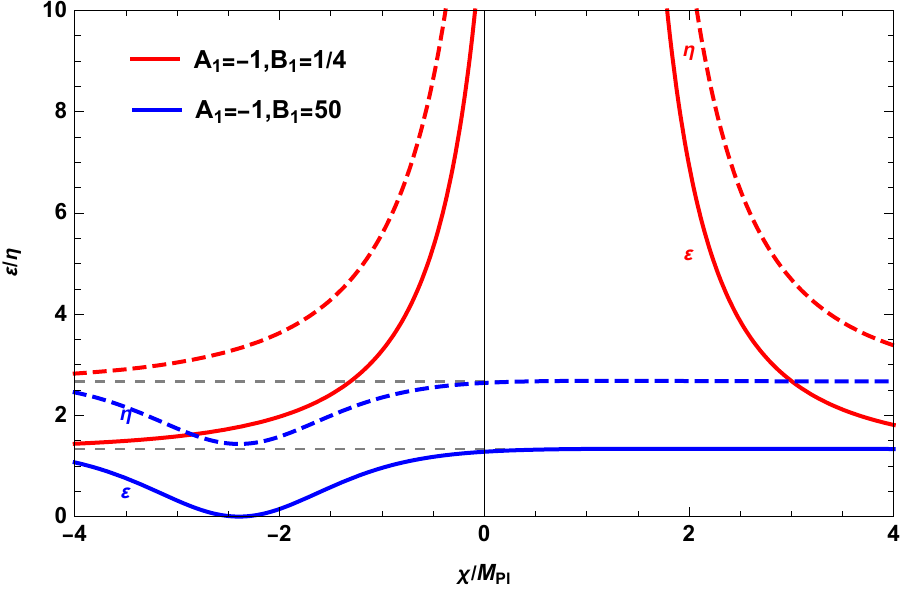}
	\caption{The evolution of the slow-roll parameters. Solid and dashed lines are corresponding to $\varepsilon$ and $\eta$, respectively. }\label{fig:slowroll-oneG}
\end{figure}

To discuss the general case for the inflation model with one modulus, we will start again with the potential in Eq. \eqref{Eq:V_simple}. The slow-roll parameters are 
\begin{equation}
	\begin{split}
		\varepsilon(\chi)&=\frac{4 \left(B_1 e^{2 \sqrt{\frac{2}{3}} \chi }-1\right)^2}{3 \left(1+A_1 e^{\sqrt{\frac{2}{3}} \chi }+B_1 e^{2 \sqrt{\frac{2}{3}} \chi } \right)^2},\\
		\eta(\chi)&=\frac{4 \left(2+A_1 e^{\sqrt{\frac{2}{3}}\chi }+A_1  B_1 e^{\sqrt{6} \chi }  +2 B_1^2 e^{4 \sqrt{\frac{2}{3}} \chi } \right)}{3 \left(1+A_1 e^{\sqrt{\frac{2}{3}} \chi }+B_1 e^{2 \sqrt{\frac{2}{3}} \chi } \right)^2}.\label{Eq:slowroll-oneG}
	\end{split}
\end{equation} 
In the limit $\chi\to \pm\infty$, $\varepsilon\to 4/3$ and $\eta\to8/3$, inflation cannot be realized
 since the slow-roll conditions cannot be satisfied.
By solving $V_{\chi}=0$, three extreme points are obtained as 
\begin{equation}
	\begin{split}
		\chi_1&=-\frac{1}{2} \sqrt{\frac{3}{2}} \log (B_1),\\
		\chi_2&=\sqrt{\frac{3}{2}} \log \left(-\frac{\sqrt{A_1^2-4 B_1}+A_1}{2 B_1}\right),\\
		\chi_3&=\sqrt{\frac{3}{2}} \log \left(\frac{ \text{sgn}[B_1]\sqrt{A_1^2-4 B_1}-A_1}{2 B_1}\right).
	\end{split}
\end{equation} 
The potential with different parameters is shown in Fig. \ref{fig:pot-oneG}. When $B_1<0$, there is only one minimum for the potential at $\chi_m=\chi_3$, and the CMB predictions are shown in Fig. \ref{fig:nsr-oneG1}. The spectral index $n_s$ is increasing as $|A_1|$ decreasing or $|B_1|$ increasing.
When $A_1<0,~ B_1>0$ and $A_1^2>4B_1$, there is a maximum at $\chi_M=\chi_1$, and two minima at $\chi_{m1,m2}=\chi_{2,3}$. Thus, the possible inflationary trajectories are from $\chi_M$ to $\chi_{m1}$ or to $\chi_{m2}$, where the CMB predictions are shown in Fig. \ref{fig:nsr-oneG2}. As $|A_1|\gg B_1$, the potential goes back to the case with $a_2=0$ and the inflation gives 
the proper CMB observed value; while for other parameter spaces, there is only one minimum for the potential at $\chi_m=\chi_1$, i.e, $V(\chi_1)=\left(A_1+2 \sqrt{B_1}\right)^2$. To avoid the cosmological constant problem, the two parameters are related as $A_1=-2\sqrt{B_1}$. Thus, the slow-roll parameters in Eq. \eqref{Eq:slowroll-oneG} become
\begin{equation*}
	\begin{split}
		\varepsilon(\chi)&=\frac{4}{3}\left(\frac{  1+\sqrt{B_1} e^{\sqrt{\frac{2}{3}} \chi } }{ 1-\sqrt{B_1} e^{\sqrt{\frac{2}{3}} \chi } }\right)^2\geq\frac{4}{3}~,~\\
		\eta(\chi)&=\frac{8}{3}\frac{ \left(1+\sqrt{B_1} e^{\sqrt{\frac{2}{3}} \chi }+B_1 e^{2\sqrt{\frac{2}{3}}\chi}\right)}{\left(1-\sqrt{B_1} e^{\sqrt{\frac{2}{3}} \chi }\right)^2}\geq\frac{8}{3}~.~
	\end{split}
\end{equation*}
If we ignore the cosmological constant problem, {\it i.e.}, $A_1\neq-2\sqrt{B_1}$, the slow-roll parameters in Eq. \eqref{Eq:slowroll-oneG} are in the ranges of 
\begin{equation*}
	0\leq \varepsilon(\chi)<\frac{4}{3},~~\frac{4}{3}\leq \eta(\chi)<\frac{8}{3}.
\end{equation*}
The slow-roll conditions are violated due to the steep slope. We also show the evolution of slow-roll parameters for models with $A_1=-1, B_1=1/4$ and $A_1=-1, B_1=50$ in Fig. \ref{fig:slowroll-oneG}.

\section{Two Moduli Inflationary Model}\label{sec:M2}

In this section, we  consider the no-scale inflation model realized via the orbifold compactification of M-theory on $T^6/Z_{12}$
by keeping singlets under $SU(2)\times U(1)$
symmetry, and then the compactification on $S^1/Z_2$~\cite{Li:1998sq}, where the K\"ahler potential 
with two moduli $T_{1,2}$ and one chiral superfield $\varphi$ is
\begin{equation}
	K=-2\log({T_1+\overline{T}_1}-2|\varphi|^2)-\log({T_2+\overline{T}_2})~.~
\end{equation}
Here, for simplicity we neglect the irrelevant scalar fields.  The scalar potential in Jordan frame is
\begin{equation}
	V=\frac{|W_{\varphi}|^2}{4c_2(c_1-2|\varphi|^2)}.
\end{equation}
After field transformation with $\varphi=\sqrt{c_1/2} \tanh \left(\chi /2\right)$, it becomes
\begin{equation}
	\begin{split}
		V=V_0\frac{1+A_2e^{\chi }+B_2e^{2\chi }}{e^{\chi }\left(1+e^{\chi } \right)^2},
	\end{split}
\end{equation}
with $V_0=\frac{\left(a_1-2 a_2 \sqrt{c_1}+3 a_3 c_1\right)^2}{8 c_1 c_2}$, $A_2=\frac{2 (a_1-3 a_3 c_1)}{a_1-2 a_2 \sqrt{c_1}+3 a_3 c_1}$ and $B_2=\frac{a_1+2 a_2 \sqrt{c_1}+3 a_3 c_1}{a_1-2 a_2 \sqrt{c_1}+3 a_3 c_1}$.

\subsection{$a_1=0$ case}

\begin{figure}[!h]
	\subfigure[]{\includegraphics[height=4cm]{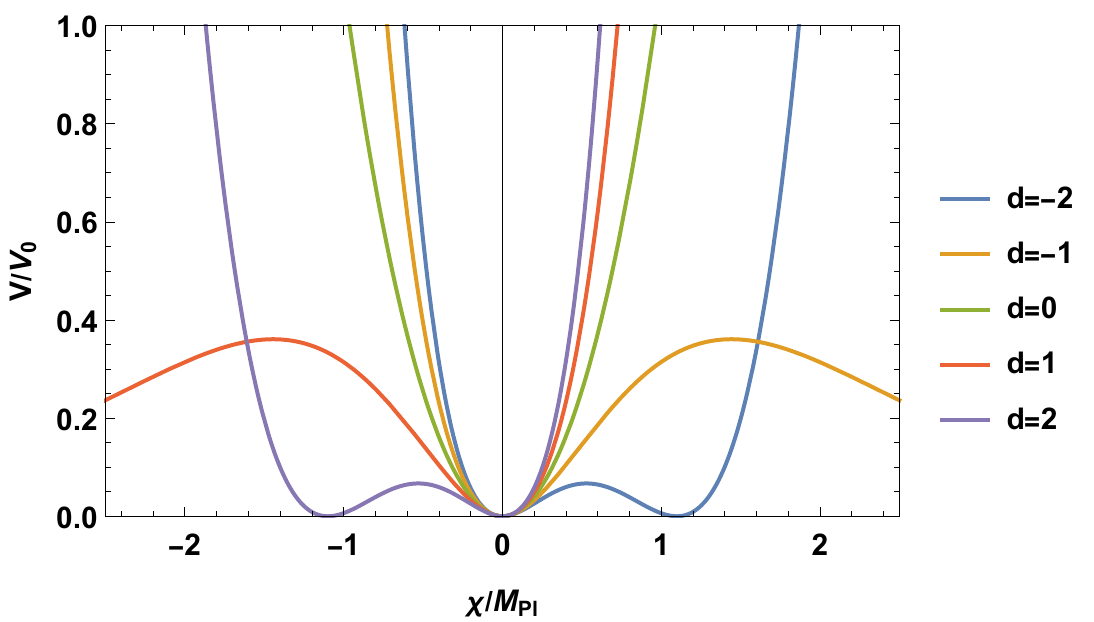}}
	\subfigure[]{\includegraphics[height=4cm]{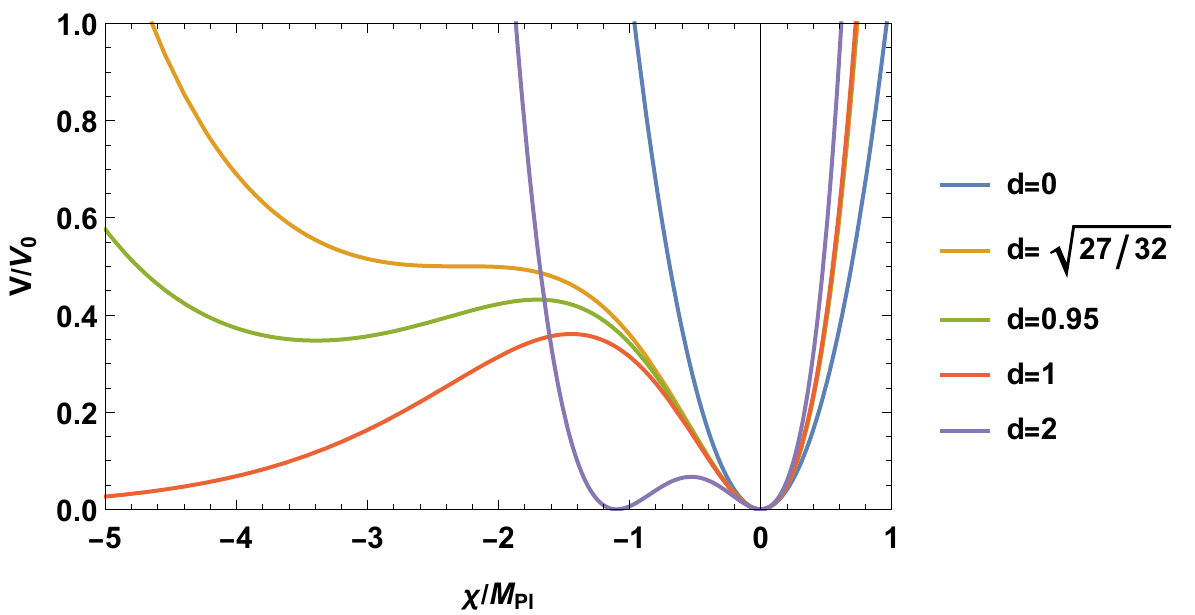}}
	\caption{The inflaton potential in Eq. \eqref{Eq:V_twoA}.}\label{fig:pot-twoA}
\end{figure}

The scalar potential is 
\begin{equation}
	\begin{split}
		V&=V_0\frac{ \left(1-e^{\chi }\right)^2 \left(1+\frac{1+d}{1-d}e^{\chi }\right)^2}{e^{\chi }\left(1+e^{\chi }\right)^2} \label{Eq:V_twoA}
	\end{split}
\end{equation}
where $V_0=\left(2 a_2-3 a_3 \sqrt{c_1}\right)^2/8 c_2$ and $d= 3a_3\sqrt{c_1}/2a_2$. From Fig. \ref{fig:pot-twoA}, the potential remains the same after both parameter $d$ and field $\chi$ become negative. Therefore, in the following we will only discuss inflation with $d\geq0$.

\begin{figure}[!h]
	\subfigure[]{\includegraphics[width=0.4\linewidth]{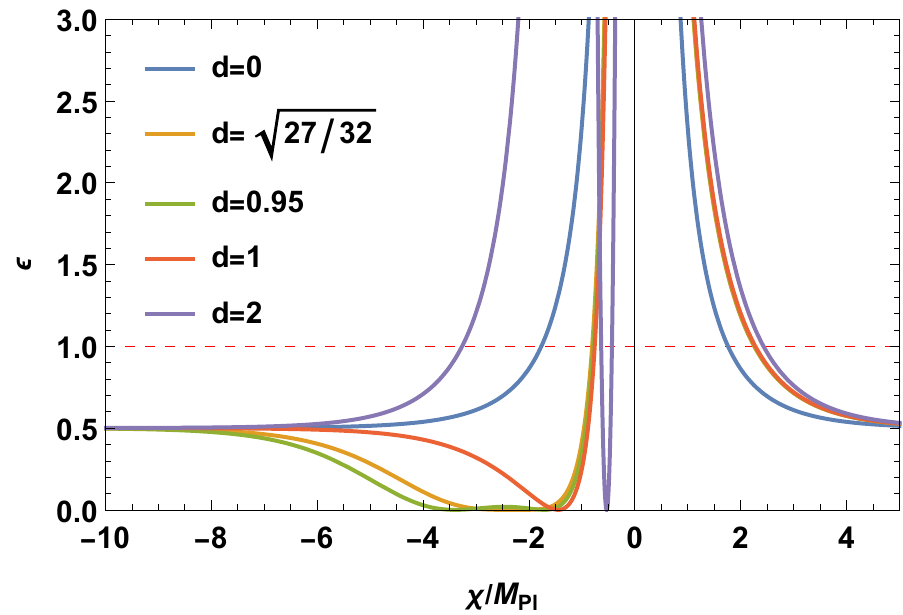}}
	\subfigure[]{\includegraphics[width=0.4\linewidth]{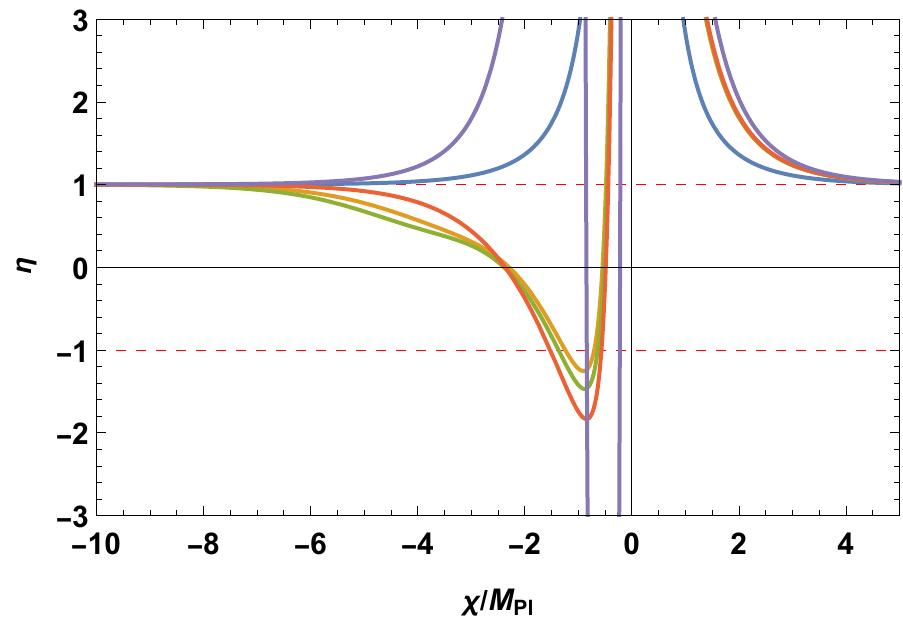}}
	\caption{The evolution of the slow-roll parameters (a) $\varepsilon$ and (b) $\eta$ for the inflaton potential in Eq. \eqref{Eq:V_twoA}.}\label{fig:slowroll-twoA1}
\end{figure}

When $d>\sqrt{27/32}$, the potential has three extreme points: $\chi_{m1}<\chi_M<\chi_{m2}=0$. There are four trajectories for inflation: on the left side of the minimum $\chi_{m1}$, from the maximum $\chi_M$ to its minima $\chi_{m1,m2}$ and on the right sides of the minimum $\chi_{m2}$.  However, slow-roll inflation cannot occur on the left side of $\chi_{m1}$ and on the right side of $\chi_{m2}$, since the slow-roll parameters $\varepsilon\to 1/2$ and $\eta\to1$ when $|\chi|>5M_{\text{Pl}}$, as shown in Fig. \ref{fig:slowroll-twoA1}. Moreover, we find that inflation cannot end on the trajectory from $\chi_M$ to $\chi_{m1}$,
 and slow-roll inflation is not permitted from $\chi_M$ to $\chi_{m2}$ because of $\eta(\chi_M)$ greater than 1. 

When $0\leq d\leq \sqrt{27/32}$, the potential has a minimum at $\chi_{m2}$ and the potential has an inflection point when $d=\sqrt{27/32}$. Here, we will discuss the impossibility for the inflation along the right side of the minimum. For the sake of convenience, we choose $d=0$, where the potential becomes 
\begin{equation}
	V=\frac{2a_2^2}{c_2} \sinh^2\left(\frac{\chi }{2}\right).
\end{equation}
In this case, the slow-roll parameters are 
\begin{equation}
	\begin{split}
		\varepsilon(\chi)&=\frac{1}{2} \coth ^2\left(\frac{\chi }{2}\right),\\
		\eta(\chi)&=\frac{1}{2} \cosh (\chi ) \text{csch}^2\left(\frac{\chi }{2}\right).
	\end{split}
\end{equation}
The parameter $\eta$ is larger than 1 on the whole trajectory. Therefore, slow-roll inflation cannot be obtained
 when $d=0$. Similarly, inflation from the right side of $\chi_{m2}=0$ for all $d$ and 
from the left side of $\chi_{m1}$ for $d>\sqrt{27/32}$ is not allowable as well. 

\begin{figure}[!h]
	\includegraphics[width=0.4\linewidth]{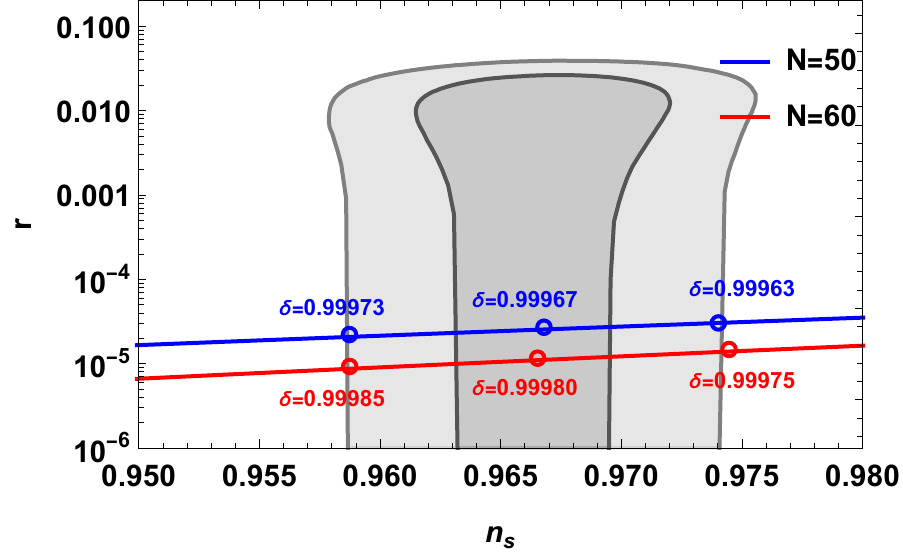}
	\caption{The $r$ versus $n_s$ predictions for the inflaton potential
 in Eq. \eqref{Eq:V_twoA} with parameter $d=\sqrt{27/32}\delta$.} \label{fig:nsr-twoA}
\end{figure}

Thus, the feasible inflationary trajectory is on the left side of $\chi_{m2}=0$ for $0< d \leq \sqrt{27/32}$. 
Similar to the Starobinsky-like E-model with one modulus, the predicted $r$ that satisfy the experimental limits 
are obtained from a tiny region for the parameter $d\lesssim\sqrt{27/32}$. The results are shown in Fig. \ref{fig:nsr-twoA}, where we also show triple-pack benchmark points corresponding to the central and $2\sigma$ bound values for observed $n_s$. Thus, the spectrum index is increasing as the parameter $d$ is decreasing. From the numerical calculations, one notes that inflation initials at $\chi_*\simeq -2.26 M_{\rm Pl}$, the higher order term $\left(\frac{1+d}{1-d}e^{\chi}\right)^2$ of the  numerator in Eq.\eqref{Eq:V_twoA} is not small and  cannot be ignored. When $d=\sqrt{27/32}$, the tensor-to-scalar ratio $r$ and $e$-folding number are approximate to be
\begin{equation*}
\begin{split}
    r&\simeq\frac{8 \left(5-3 \left(8 \sqrt{6}+13\right) e^{\chi }+3 \left(32 \sqrt{6}+77\right) e^{2 \chi } \right)^2}{ \left(5+\left(24 \sqrt{6}+59\right) e^{\chi } \right)^2}\\
    N&\simeq  \frac{824-336 \sqrt{6}}{\left(11 \sqrt{6}-24\right) e^{\chi }-103 \sqrt{6}+252}
\end{split}
\end{equation*}
and 
\begin{equation}
    \begin{split}
     r  \simeq\frac{8192 \left(49-20 \sqrt{6}\right)}{N^4}
       \simeq \frac{83.60}{N^4}   \\
    \end{split}
\end{equation}
For $N=50-60$, the predicted $r$ is about $10^{-5}$ and consistent with numerical calculations.

\subsection{$a_2=0$ case} 
\begin{figure}[t]
	\includegraphics[width=0.45\linewidth]{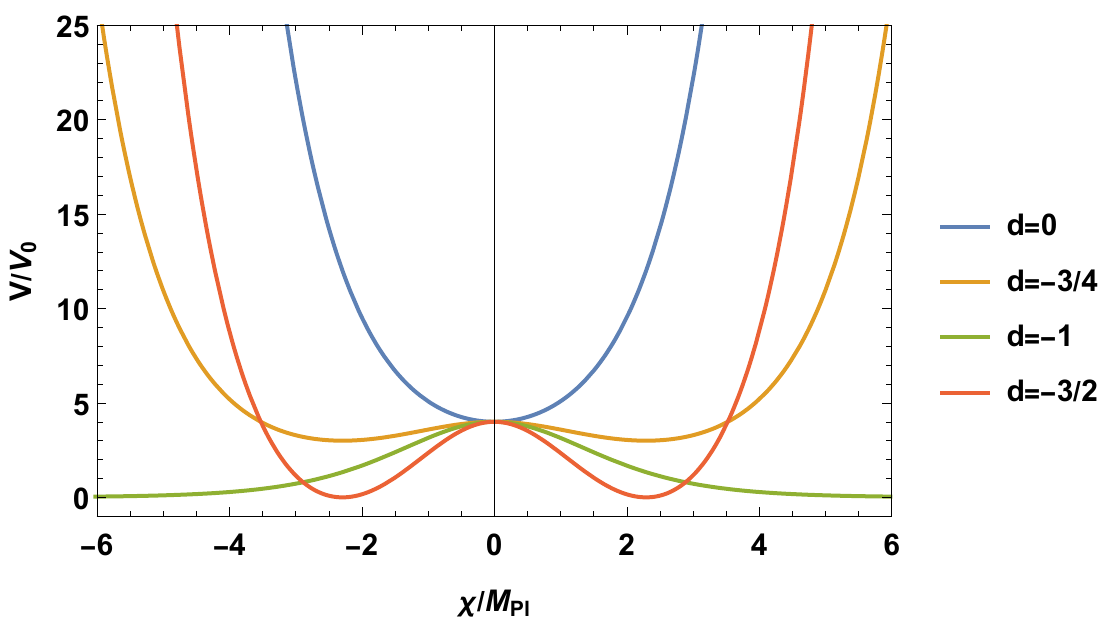}
	\caption{Inflaton potential in Eq. \eqref{Eq:V_twoB}. }\label{fig:pot-twoB}
\end{figure}
The scalar potential  becomes
\begin{equation}
	\begin{split}
		V&=V_0\frac{ \left(\left(e^{\chi }+1\right)^2+d \left(e^{\chi }-1\right)^2\right)^2}{e^{-\chi }\left(e^{\chi }+1\right)^2} \label{Eq:V_twoB}
	\end{split}
\end{equation}
with $V_0=a_1^2/(8c_1c_2)$ and $d=3a_3c_1/a_1 $. It is an even function in terms of $\chi$. Thus, for simplicity, 
we will study inflation in the positive region $\chi> 0$, as shown in Fig. \ref{fig:pot-twoB}.
There is a minimum and maximum at $\chi=0$ for the potential 
with $d\geq-1/2$ and $d=-1$, respectively.
Also, there is a maximum at $\chi_M=0$ and two minima at $\chi_m=\log \left(-\left(1+3 d\pm2 \sqrt{d (2 d+1)}\right)/\left(d+1\right)\right)$ and $\chi_m=\log \left(-\left(1-d\pm2 \sqrt{-d}\right)/\left(d+1\right)\right)$ 
respectively for $-1<d<-1/2$ and $d<-1$. However, 
the slow-roll inflation cannot end. 
To understand this behavior, we choose $d=-1$, and the corresponding slow-roll parameters in terms of the field $\chi$ are
\begin{equation}
	\varepsilon(\chi)=\frac{1}{2}-
\frac{2e^{\chi}}{\left(e^{\chi }+1\right)^2}
~,~~\eta(\chi)=1- \frac{6e^{\chi}}{\left(e^{\chi }+1\right)^2}.
\end{equation} 
Thus, we have $\varepsilon(0)=0$, $\eta(0)=-1/2$. For $\chi > 0$, we have $0< \varepsilon < 1/2$,
 $-1/2<\eta<1$, and $\eta\to 1$ as $\chi\to\infty$.  Thus, we cannot exit the slow-roll inflation.

\subsection{$a_3=0$ case}
\begin{figure}[!h]
	\includegraphics[width=0.45\linewidth]{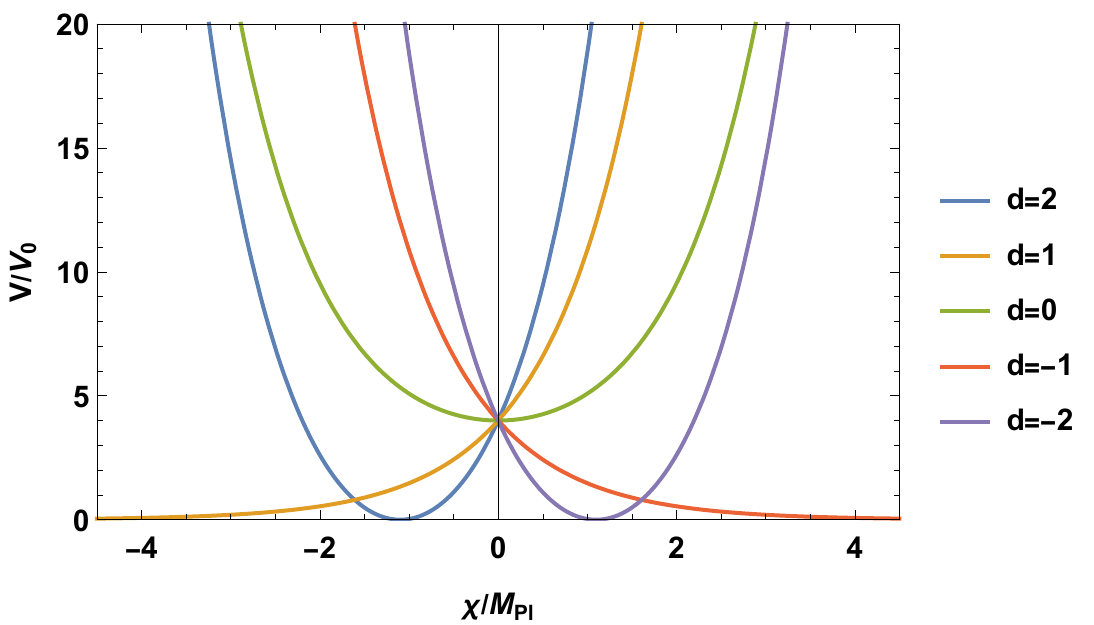}
	\caption{Inflaton potential in Eq. \eqref{Eq:V_twoC}.}\label{fig:pot-twoC}
\end{figure}
The scalar potential with $a_3=0$ becomes
\begin{equation}
	\begin{split}
		V=V_0 e^{-\chi } \left( \left(e^{\chi }+1\right)+d  \left(e^{\chi }-1\right)\right)^2\label{Eq:V_twoC}
	\end{split}
\end{equation}
where $V_0=a_1^2/8c_1c_2$ and $d =2a_2\sqrt{c_1}/a_1$. 
Because the potential is invariant as $d\to -d$ and $\chi\to -\chi$ (see Fig. \ref{fig:pot-twoC}), 
 we only need to study the cosmological predictions with $d\geq0$. There is a minimum at $\chi_m=\log\left(|1-d|/(1+d)\right)$. Note that $V(\chi_m)$ is larger than 0 with $d<1$, 
we have the cosmological constant problem. 
However, it still does not have slow-roll inflation with $d\ge 1$. The slow-roll parameters are 
\begin{equation}
	\begin{split}
		\varepsilon(\chi)=\frac{\left( e^{\chi }+\xi\right)^2}{ 2 \left(e^{\chi}-\xi\right)^2},~
		\eta(\chi) 
		=1+\frac{2\xi e^{\chi}}{\left(e^{\chi}-\xi\right)^2} 
	\end{split}
\end{equation}
with $\xi=(d-1)/(d+1)$. For $d\ge 1$, we obtain $\eta \ge 1$, and thus the slow-roll conditions cannot be satisfied. 

\subsection{General case}
\begin{figure}[t]
	\subfigure[]{\includegraphics[height=0.26\linewidth]{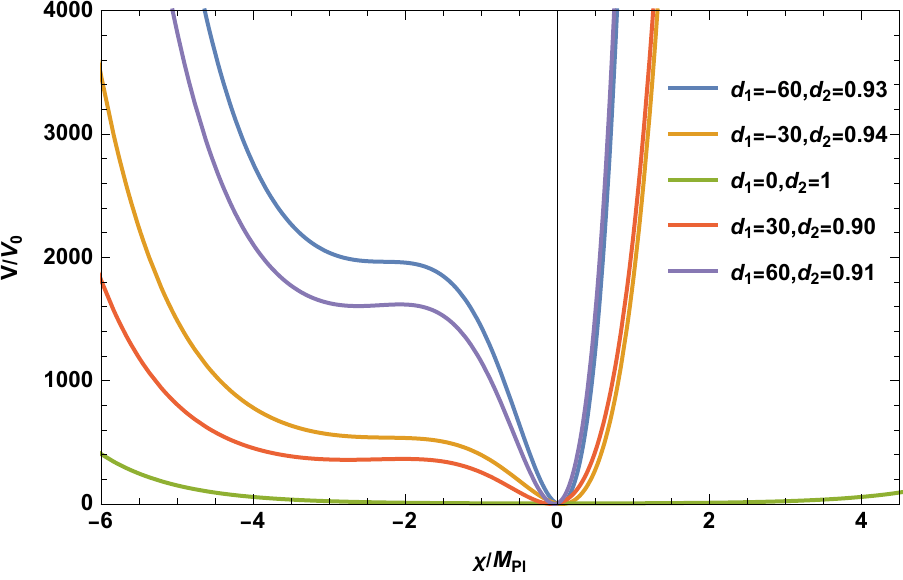}}\\
	\subfigure[]{\includegraphics[height=0.26\linewidth]{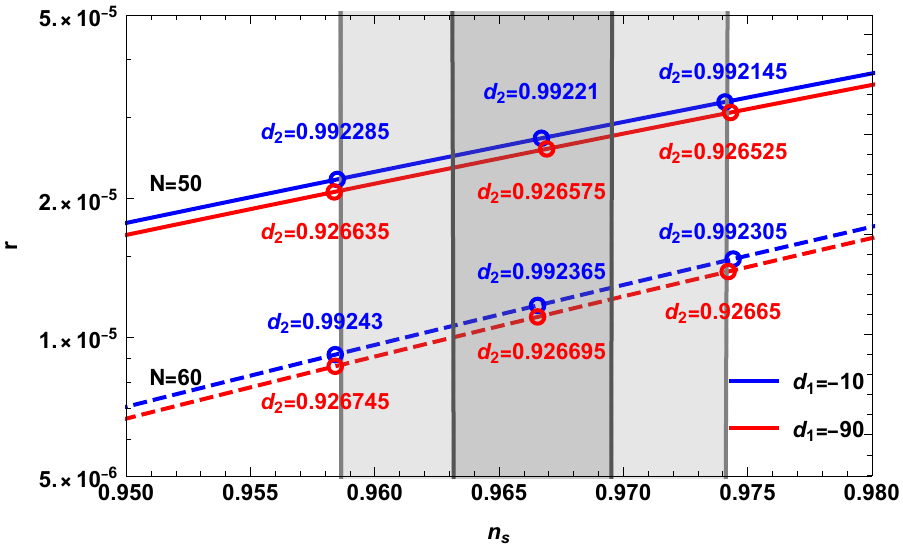}}
	\subfigure[]{\includegraphics[height=0.26\linewidth]{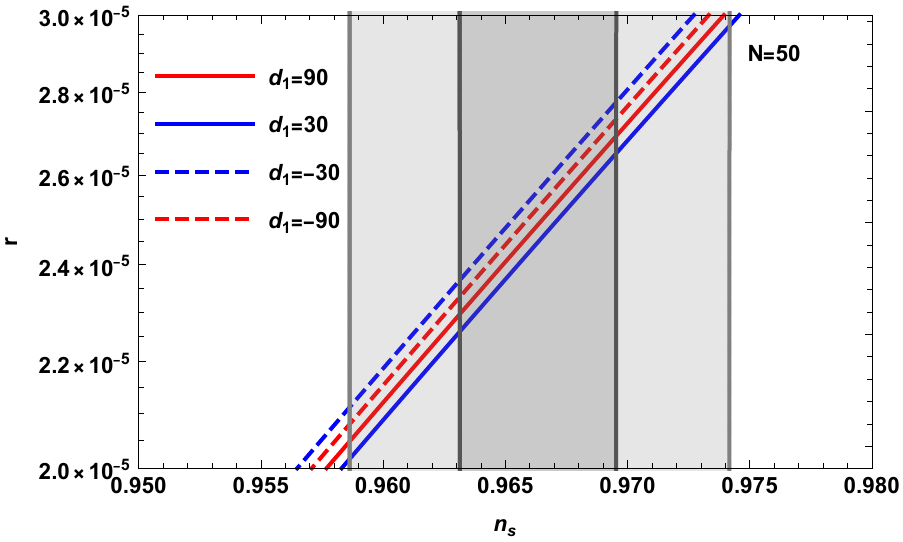}}
	\caption{The $r$ versus $n_s$ predictions (b),(c) for the inflaton potential (a) in Eq. \eqref{Eq:V_twoG}.}\label{fig:nsr-twoG}
\end{figure}
Now, we discuss the general case whose inflaton potential is 
\begin{equation}
	\begin{split}
		V=V_0\frac{ \left((e^{\chi } +1)^2+d_1(e^{\chi } -1)\left((e^{\chi } +1)+d_2(e^{\chi } -1)\right)\right)^2}{e^{\chi }(e^{\chi } +1)^2}\label{Eq:V_twoG}
	\end{split}
\end{equation}
with $V_0=a_1^2/8c_1c_2$, $d_1=2a_2\sqrt{c_1}/a_1$ and $d_2=3a_3\sqrt{c_1}/2a_2$.  The potential remains the same after both parameters $d_1,d_2$ and field $\chi$ become negative. Thus, we will only discuss inflation with $d_2\geq0$. The potential and $r$ versus $n_s$ predictions are shown in Fig. \ref{fig:nsr-twoG}. The possible inflation is along the flat trajectory $\chi<0$. Similar to the $a_1=0$ case, the inflation is sensitive to the parameter $d_2$ and the benchmark points corresponding to the central and $2\sigma$ bound values for observation $n_s$ are also shown in Fig. \ref{fig:nsr-twoG}(b). The solid and dashed lines are corresponding to $N=50$ and $N=60$, respectively. From the numerical results, we know that the spectrum index increases as the parameter $d_2$ decreases. Moreover, the tensor-to-scalar ratio $r$ predicted from models with $d_1<0$ is larger than that with $d_1>0$. We plot $r$ versus $n_s$ for models with $d_1=\pm30,~\pm 90$ in Fig. \ref{fig:nsr-twoG}(c).

\section{Three Moduli Inflationary Model}\label{sec:M3}

In this section, we consider no-scale inflation realized by the orbifold compactifications of M-theory on  $T^6/Z_{12}$ as well as $S^1/Z_2$~\cite{Li:1998sq}, which has three moduli and the K\"ahler potential is given by Eq. \eqref{eq:genK} with $N_1=N_2=N_3=1$.
Stabilizing the muduli with $T_{i}=\overline{T}_{i}=c_i/2$, the scalar potential along the real part of field $\varphi$ is given by
\begin{equation}
	V =\frac{|W_{\varphi}|^2}{2c_2c_3}.
	\label{Eq:pot_threeG}
\end{equation}
The three moduli inflationary model is similar to that with
 the global supersymmetry, and has a non-negative scalar potential, but the K\"ahler potential is different.
Redefining the canonical field $\chi$ with $\varphi=\sqrt{c_1/2} \tanh \left(\chi /\sqrt{2}\right)$, the potential becomes 
\begin{equation}
	V = V_0\frac{\left(1+A_3 e^{\sqrt{2} \chi } +B_3 e^{2 \sqrt{2} \chi }\right)^2}{\left(1+e^{\sqrt{2} \chi }\right)^4} \label{Eq:V_threeMG}
\end{equation}
with $V_0=\frac{\left(a_1-2 a_2 \sqrt{c_1}+3 a_3 c_1\right)^2}{c_2 c_3}$, $A_3=\frac{2 (a_1-3 a_3 c_1)}{a_1-2 a_2 \sqrt{c_1}+3 a_3  c_1}$, and $B_3=\frac{a_1+2 a_2 \sqrt{ c_1}+3 a_3  c_1}{a_1-2 a_2 \sqrt{ c_1}+3 a_3  c_1}$.

\subsection{T-model realization: $a_{1,3}=0$ and $a_{1,2}=0$ cases}
When $a_{1,3}=0$, the inflation model in Eq. \eqref{Eq:pot_threeG} reduces to  $V\propto\varphi^2$. We know that the chaotic inflation \cite{Linde:1983gd,Creminelli:2014nqa} predicts
\begin{equation}\label{eq:nsr_chaotic}
n_s\simeq1-\frac{2}{N},~~r\simeq\frac{8}{N},
\end{equation} 
which will give a larger $r\sim 0.14$ for $N\sim 55$ and the inflation with a quadratic potential is ruled out by the Planck and BICEP/Keck experiments \cite{Akrami:2018odb,BICEP:2021xfz}. After field transformation, the T-model  \cite{Kallosh:2013yoa} for the $\varphi^2$ inflation model is achieved with potential as
\begin{equation}
	V=V_0\tanh^2\left(\frac{\chi}{\sqrt{2}}\right).
\end{equation} 
Similar, the T-model \cite{Kallosh:2013yoa} for $\varphi^4$  is realized with $a_{1,2}=0$ and the potential is given by $V=V_0\tanh^4\left(\chi/\sqrt{2}\right)$. We list the realization of the T-model in the three moduli model in Table~\ref{tab:T-Emodel}. If the factor $N_1$ is generalized to $\alpha N_1$, the T-model of the $\alpha$-attractor can be obtained.

For the T-model for $\varphi^{2n}$, the slow-roll parameters are 
\begin{equation}
\begin{split}
	\varepsilon&=4 n^2 \text{csch}^2\left(\sqrt{2} \chi\right),\\
	\eta&=4 n  \left(2n-\cosh \left(\sqrt{2} \chi\right)\right)\text{csch}^2\left(\sqrt{2} \chi\right). 
\end{split}
\end{equation}
The inflation ends at $\chi_e=\sinh^{-1}(2n)/\sqrt{2}$ with $\varepsilon=1$. Then the \textit{e}-folding number can be rewritten 
in terms of  the inflaton at horizon crossing $\chi_*$
\begin{equation}
\begin{split}
	N&=\left.\frac{1}{4 n}\cosh \left(\sqrt{2} \chi\right)\right|_{\chi_e}^{\chi_*}\\
	&=\frac{1}{4 n}\left(\cosh \left(\sqrt{2} \chi_*\right)-\sqrt{4 n^2+1}\right)\\
\end{split}
\end{equation} 
The cosmological predictions are given by 
\begin{equation}
	\begin{split}
		n_s&=1-\frac{2 \left(n+\sqrt{4 n^2+1}+4 n N\right)}{n+2 N \sqrt{4 n^2+1}+4 n N^2}\\
		r&=\frac{16 n}{n+2 N \sqrt{4 n^2+1} +4 n N^2}
	\end{split}
\end{equation}
Comparing with Eq. \eqref{eq:nsr_chaotic}, the spectral index changes a little, whereas the tensor-to-scalar ratio is effectively suppressed by the factor $4/N^2$:
$n_s=0.9602,~r=0.0016$ for $N=50$, and $n_s=0.9668,~r=0.0011$ for $N=60$.

\subsection{$a_1=0$ case}
The effective potential of scalar field $\varphi$ is
\begin{equation}
	V=\frac{4}{c_2c_3}\left|\sqrt{2}a_2\varphi+3a_3\varphi^2\right|^2,~
\end{equation}
and can be rewritten as a three-term polynomial inflation model $V=V_0|\phi+\alpha\phi^2|^2$ which predicts a large tensor-to-scalar ratio $r\sim [0.025,0.249]$ \cite{Li:2014zfa} and the parameter space is partially ruled out by the latest limits. However, the field $\varphi$ is noncanonical in the no-scale SUGRA models. In the following discussion, one will find that tensor-to-scalar ratio $r$ is suppressed by $1/N^2$ rather than $1/N$.

\begin{figure}[!h]
	\includegraphics[width=0.4\linewidth]{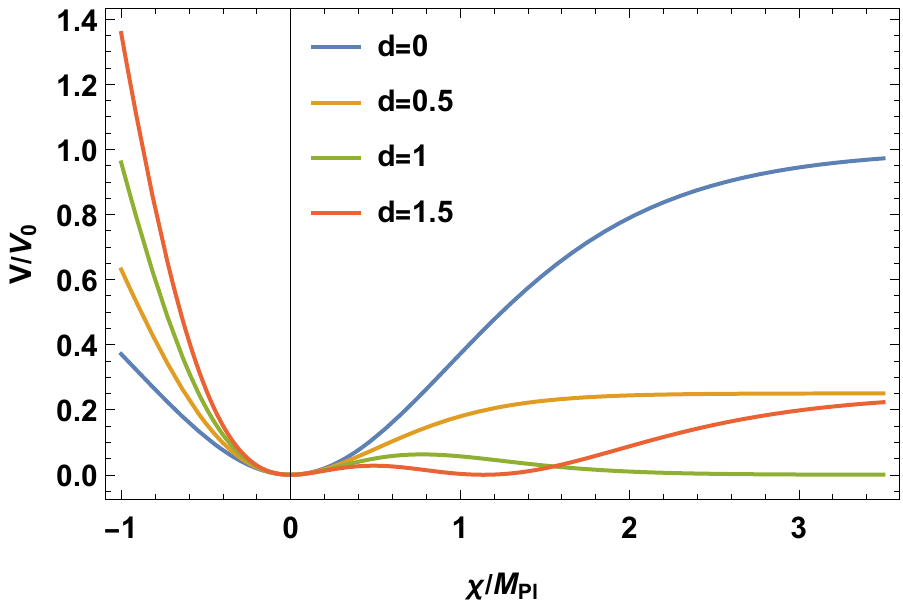}
	\caption{Inflaton potential in Eq. \eqref{Eq:V_threeMB}.}\label{fig:pot-threeB}
\end{figure}

Using the field transformation, the potential with canonical field $\chi$ can be rewritten in the general form Eq. \eqref{Eq:V_threeMG}
with $V_0=\frac{\left(2 a_2 \sqrt{c_1}-3 a_3 c_1\right)^2}{c_2 c_3}$, $A_3=\frac{ 6 a_3 c_1}{2 a_2 \sqrt{c_1}-3 a_3  c_1}$, $B_3=-\frac{2 a_2 \sqrt{ c_1}+3 a_3  c_1}{2 a_2 \sqrt{ c_1}-3 a_3  c_1}$. The two parameters $A_3$ and $B_3$ are interrelated, $A_3+B_3=-1$. After defining a new parameter $d=-3a_3\sqrt{c_1}/2a_2$, they become $A_3=-{2d}/{ (d+1)}$ and $B_3={ d-1}/{(d+1)}$. The potential can be rewritten with two parameters $V_0$ and $d$ as follows:
\begin{equation}
	V=V_0  \frac{\left(d \left(e^{\sqrt{2} \chi }-1\right)^2-e^{2 \sqrt{2} \chi }+1\right)^2}{ \left(1+e^{\sqrt{2} \chi }\right)^4}.\label{Eq:V_threeMB}
\end{equation}
The parameter $V_0=4a_2^2c_1/c_2c_3$ can be fixed by the power spectrum $A_s=2.10\times10^{-9}$ \cite{Akrami:2018odb} at horizon crossing. One can find that the potential $V(-d,-\chi)$ equals to $V(d,\chi)$, which indicates that the negative trajectory $\chi<0$ with $d<0$ 
will give the same observations with positive trajectory $\chi>0$ with $d>0$, and vice versa. For convenience, we study the inflation by setting $d>0$. Inflation with $d=0$ was discussed in the previous subsection. 

\begin{figure}[!h]
	\subfigure[]{\includegraphics[width=0.4\linewidth]{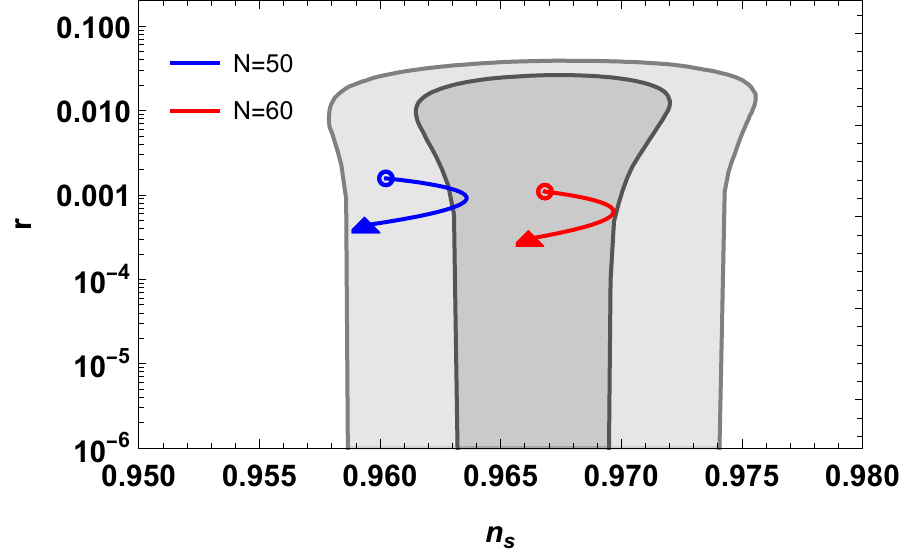}}
	\subfigure[]{\includegraphics[width=0.4\linewidth]{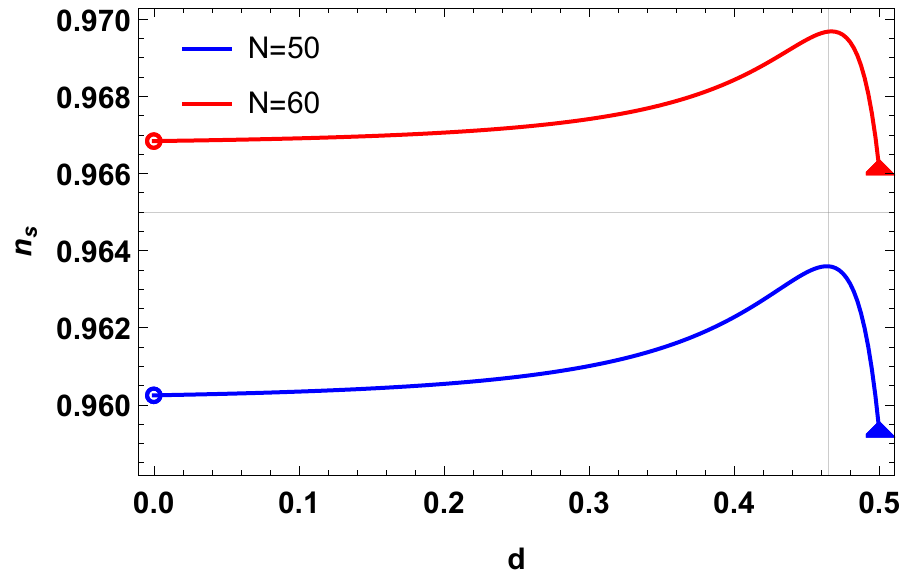}}\subfigure[]{\includegraphics[width=0.4\linewidth]{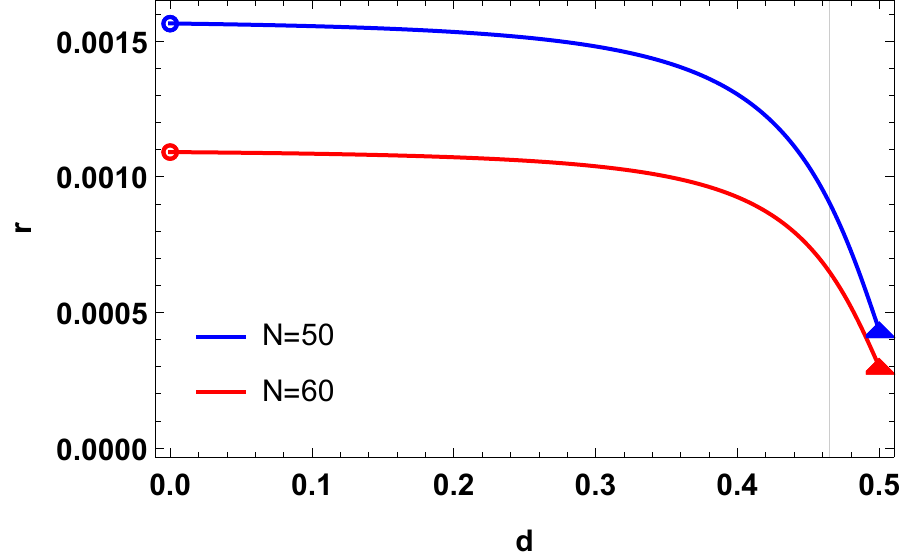}}
	\caption{(a) The $r$ versus $n_s$ predictions for the inflaton potential
 in Eq. \eqref{Eq:V_threeMB} with $d\leq1/2$; (b) $n_s$ versus the parameter  $d$,
and (c)  $r$ versus the parameter  $d$. The circles and triangles  
correspond to $d=0$ (T-model for $\varphi^2$) and $d=1/2$, respectively.}\label{fig:ns-threeB-d1}
\end{figure}

\begin{figure}[!h]
	\subfigure[]{\includegraphics[width=0.4\linewidth]{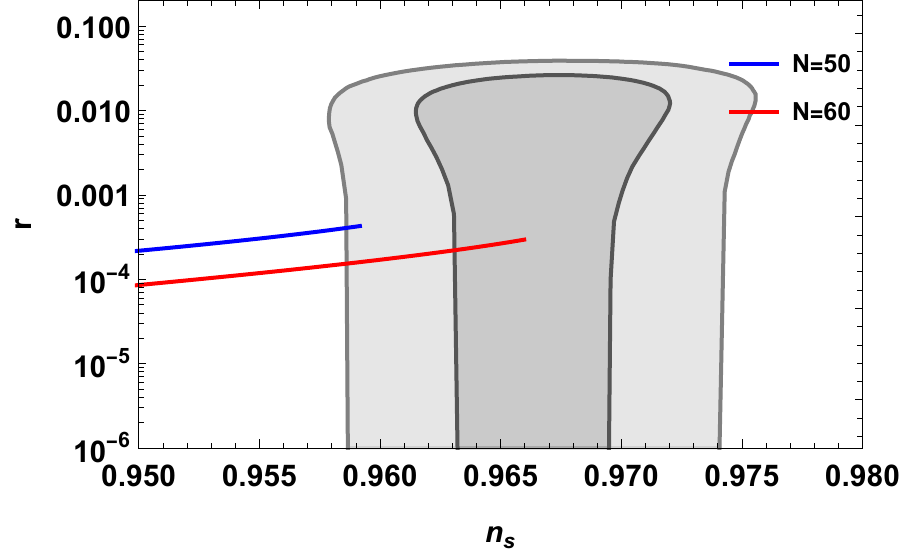}\label{fig:ns-threeB-d2}}
	\subfigure[]{\includegraphics[width=0.4\linewidth]{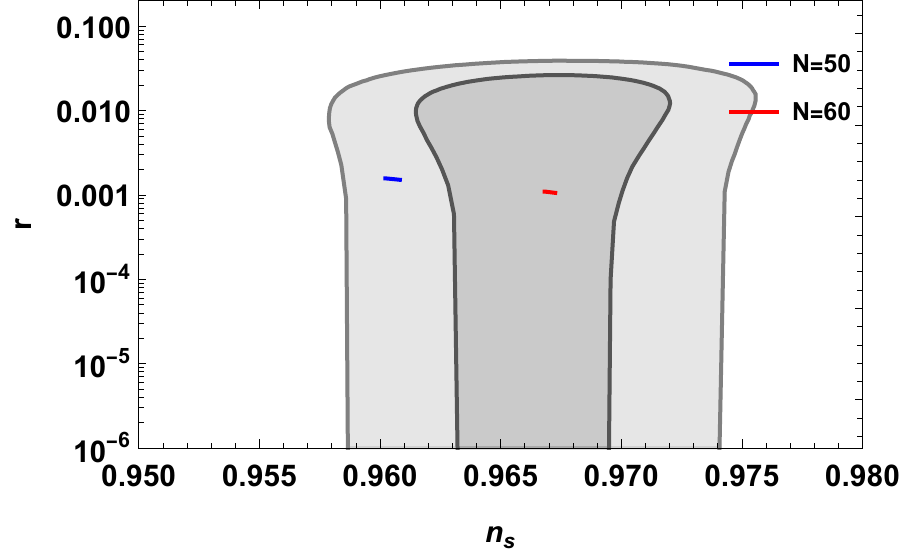}\label{fig:ns-threeB-d3}}
	\caption{The $r$ versus $n_s$ predictions for the inflaton potential
 in Eq. \eqref{Eq:V_threeMB} with (a) $1/2<d\leq1$ and (b) $d>1$.}
\end{figure}

\begin{figure}[!h]
	\includegraphics[width=0.4\linewidth]{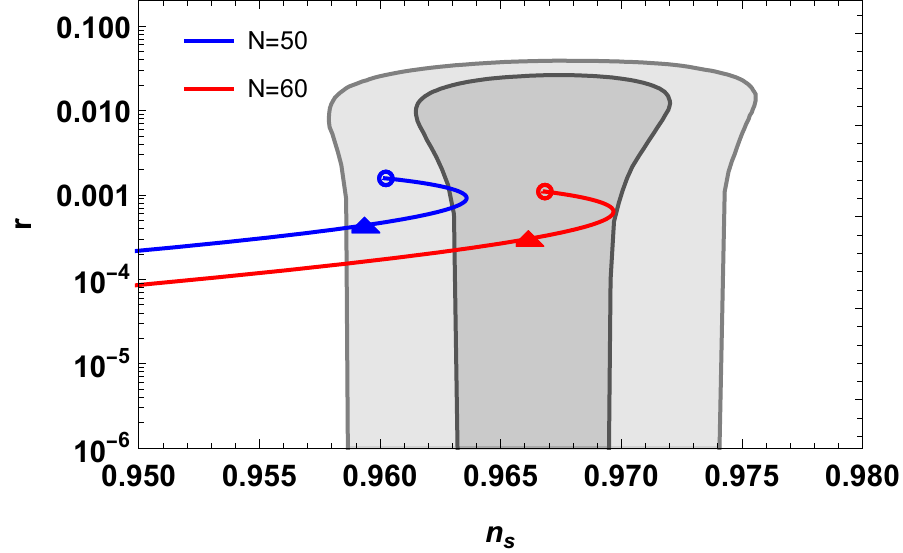}
	\caption{The $r$ versus $n_s$ predictions for the inflaton potential in Eq. \eqref{Eq:V_threeMB}. 
The circles and triangles correspond to $d=0$ (T-model for $\varphi^2$) and $d=1/2$, respectively. }\label{fig:ns-threeB}
\end{figure}

The potential in terms of inflaton field $\chi$ with different parameter $d$ are shown in Fig.~\ref{fig:pot-threeB}. There is a minimum at $\chi=0$, and the inflation will occur on the negative branch $R_1$ ($\chi<\chi_m$ region). Since the steep trajectory in the region of $\chi<0$ weakly depends on the parameter $d$, the potential goes back to that in the previous  $a_{1,3}=0$ case (T-model) and the predictions are $n_s\simeq0.9602$ and $r\simeq0.0016$ for $N=50$, as well as
 $n_s\simeq 0.9667$ and $r\simeq0.0011$ for $N=60$.  Next, we will study inflation on the positive field branch $\chi>0$. Similar
 to the $a_{1,3}=0$ case, the potential with $d<1/2$ only has one minimum at $\chi=0$. The inflation can occur at both branches $\chi>0$ and $\chi<0$, however, the two inflation trajectories are not symmetric.  When the inflation occurs at the positive branch $\chi>0$, $n_s$ and $r$ are very sensitive to $d$. The numerical results are shown in Fig. \ref{fig:ns-threeB-d1} and $n_s$ has a maximum when $d=0.465$: $n_s=0.9636( 0.9697)$ for \textit{e}-folders $N=50( 60)$. When $1/2<d\leq 1$, the potential has a minimum at $\chi_m=0$ and a maximum at $\chi_M=\sqrt{2}\tanh^{-1}\left( {1}/{2d}\right)$. There are three possible inflation trajectories, labeled as $R_1$ ($\chi<\chi_m$ region), $R_2$ ($\chi_m<\chi<\chi_M$ region), and $R_3$ ($\chi>\chi_M$ region). The parameter is restricted to a range $0.5<d<0.55$ for the $R_2$ trajectory due to small $n_s$  and lack of \textit{e}-folds. The scalar spectral index $n_s$ decreases as $d$ is increasing and the numerical results are shown in Fig. \ref{fig:ns-threeB-d2}. The last trajectory $R_3$ is ruled out since the potential is too flat where the inflation either cannot end or cannot last long enough. When $d>1$, the potential has
 two minima at $\chi_{m_1}=0$ and $\chi_{m_2}=\sqrt{2}\tanh^{-1}\left({1}/{d}\right)$ and one maximum at $\chi_M=\sqrt{2}\tanh^{-1}\left({1}/{2d}\right)$. Thus, there are four possible trajectories, labeled as $R_1$ ($\chi<\chi_m$ region), $R_2$ ($\chi_{m_1}<\chi<\chi_M$ region), $R_3$ ($\chi_M<\chi<\chi_{m_2}$ region), and $R_4$ ($\chi>\chi_{m2}$ region). As discussed above, 
the inflation in $R_2$ and $R_3$ is forbidden since the \textit{e}-folding numbers are not enough.
 The inflation on $R_4$ is similar to that on negative branch $R_1$ and the results are shown in Fig. \ref{fig:ns-threeB-d3}. At last, the above results are summarized in Fig. \ref{fig:ns-threeB}.


As discussed above, $e^{-\sqrt{2}\chi_*}$ is small as the mode crossing the horizon, then potential \eqref{Eq:V_threeMB} is expanded up to the leading order in $e^{-2b\chi}$ as
\begin{equation}
	V\propto \left(1-a e^{-2b\chi}+\mathcal{O}(e^{-4b\chi})\right)
\end{equation}
where $a=4(1-2d)/(1-d)$ and $b=1/\sqrt{2}$. 
The spectrum index, tensor-to-scalar ratio, and \textit{e}-folding number are
\begin{align}
	n_s &\simeq 1-8ab^2 e^{-2b\chi}+\mathcal{O}(e^{-4b\chi}) \\
	r&\simeq 32 a^2 b^2 e^{-4b\chi}+\mathcal{O}(e^{-8b\chi}) \\
	N&\simeq \frac{e^{2b\chi}}{4ab^2}+\mathcal{O}(e^{-4b\chi}) 
\end{align}
and 
\begin{equation}
	n_s\simeq 1-\frac{2}{N},~~r\simeq \frac{2}{b^2N^2}.
\end{equation}
Thus, the predictions $n_s=[0.960,0.967]$ and $r=[0.0016,0.0011]$ for $N=[50,60]$ are consistent with 
the Planck 2018 results \cite{Akrami:2018odb}.

\subsection{$a_2=0$ case}

\begin{figure}
	\includegraphics[width=0.4\linewidth]{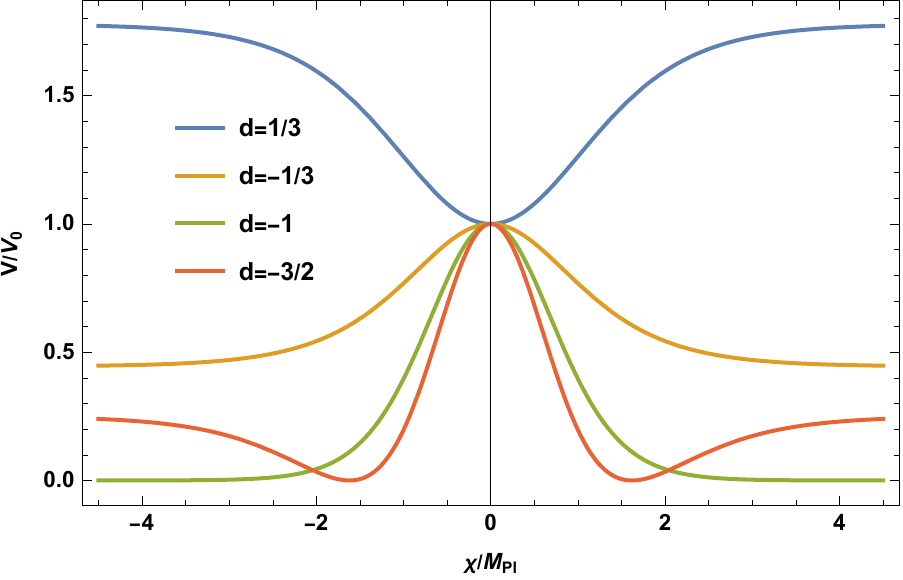}
	\caption{Inflaton potential in  Eq. \eqref{Eq:V_threeMC}.}\label{fig:pot-threeC}
\end{figure}

\begin{figure}[t]
	\includegraphics[width=0.4\linewidth]{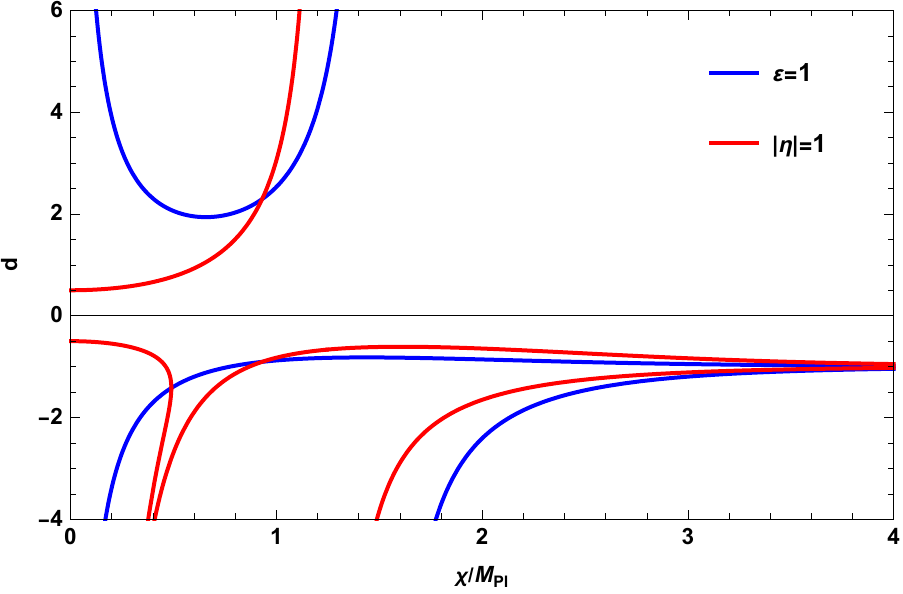}
	\caption{The parameter $d$ with respect to inflaton $\chi/M_{\text{Pl}}$. }\label{fig:d-threeC}
\end{figure}

\begin{figure}[!h]
	\subfigure{\includegraphics[width=0.3\linewidth]{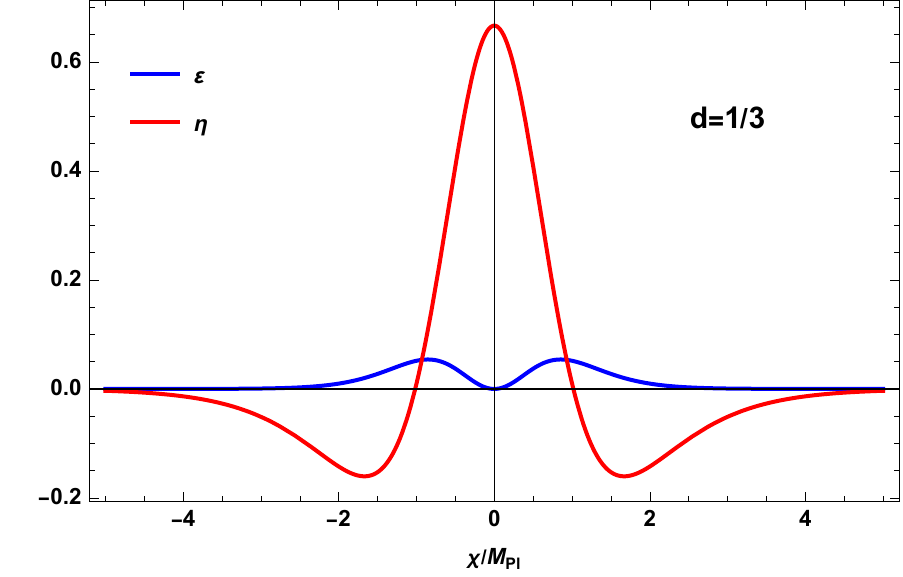}}
	\subfigure{\includegraphics[width=0.3\linewidth]{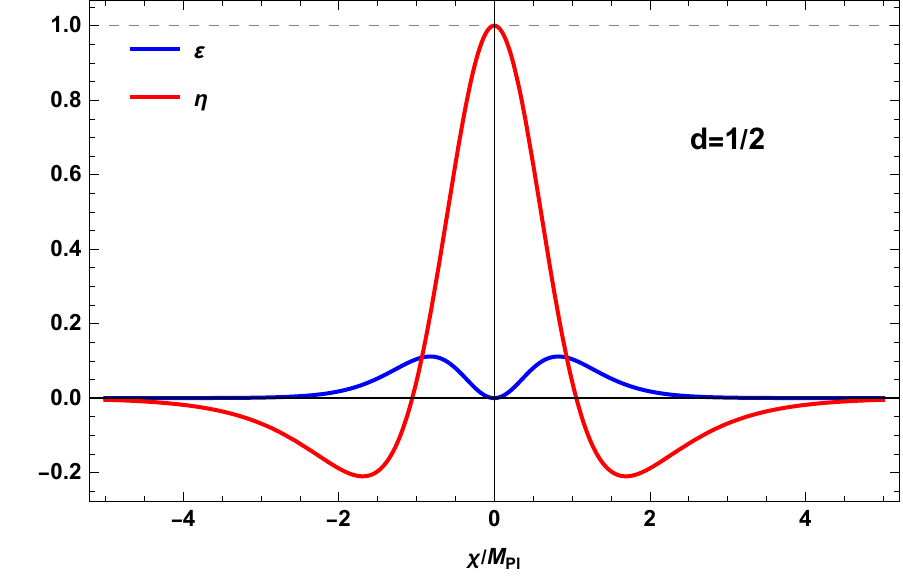}}
	\subfigure{\includegraphics[width=0.3\linewidth]{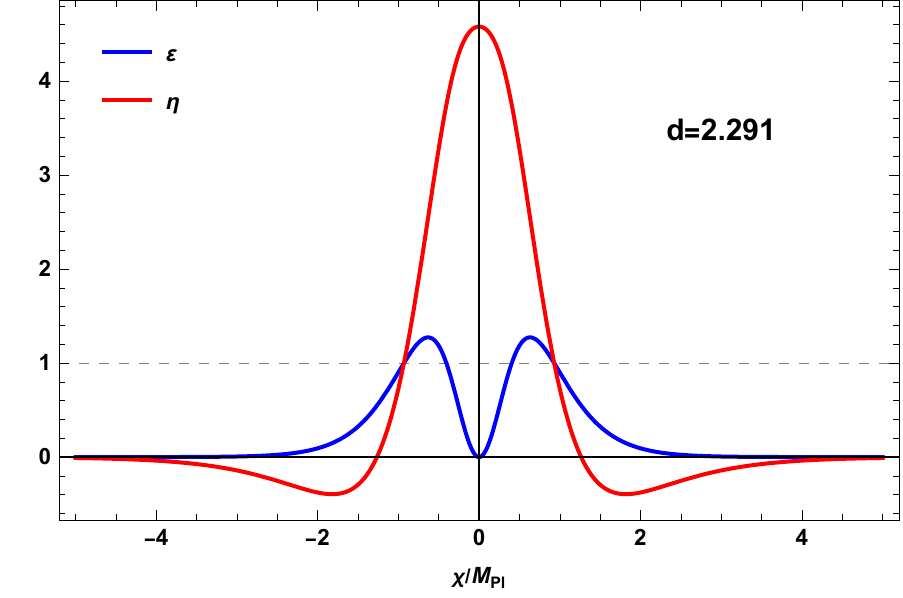}}
	\subfigure{\includegraphics[width=0.3\linewidth]{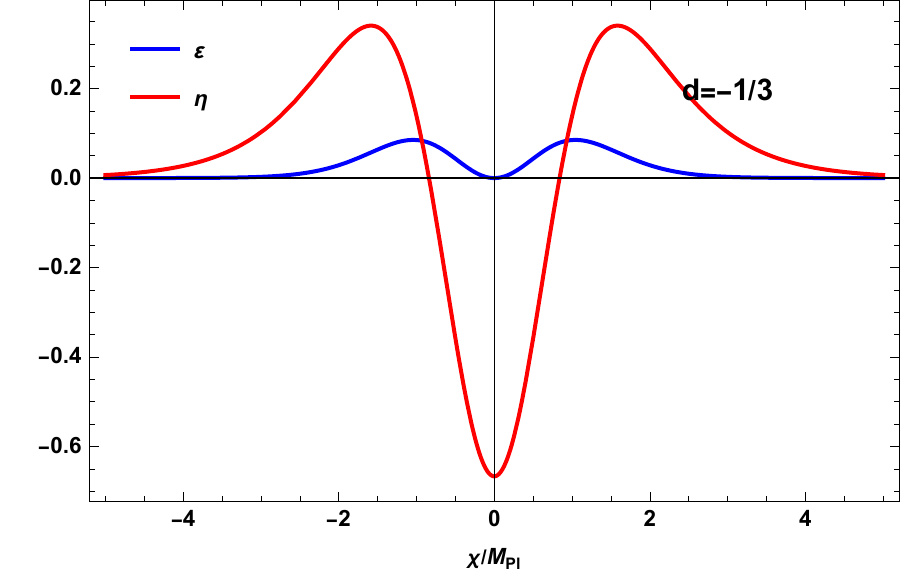}}
	\subfigure{\includegraphics[width=0.3\linewidth]{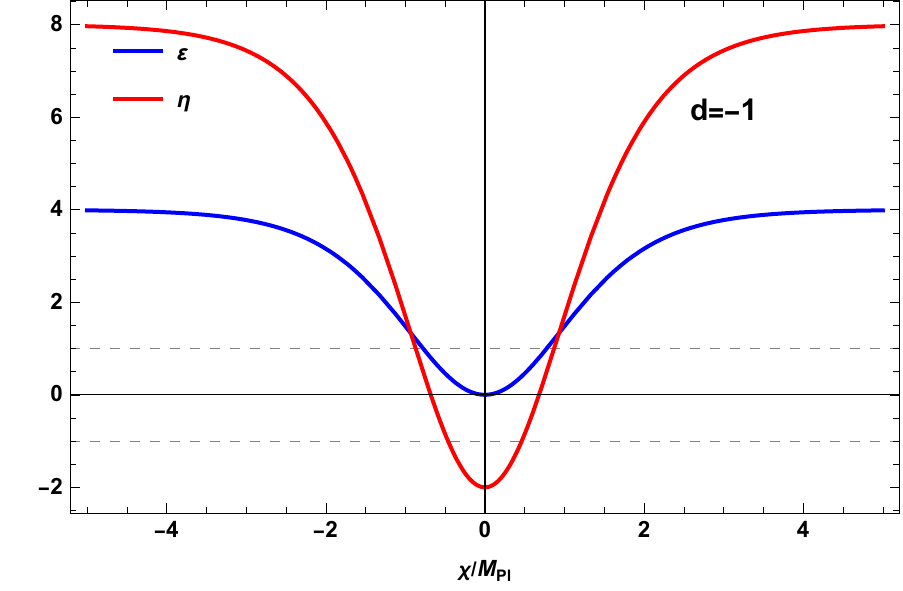}}
	\subfigure{\includegraphics[width=0.3\linewidth]{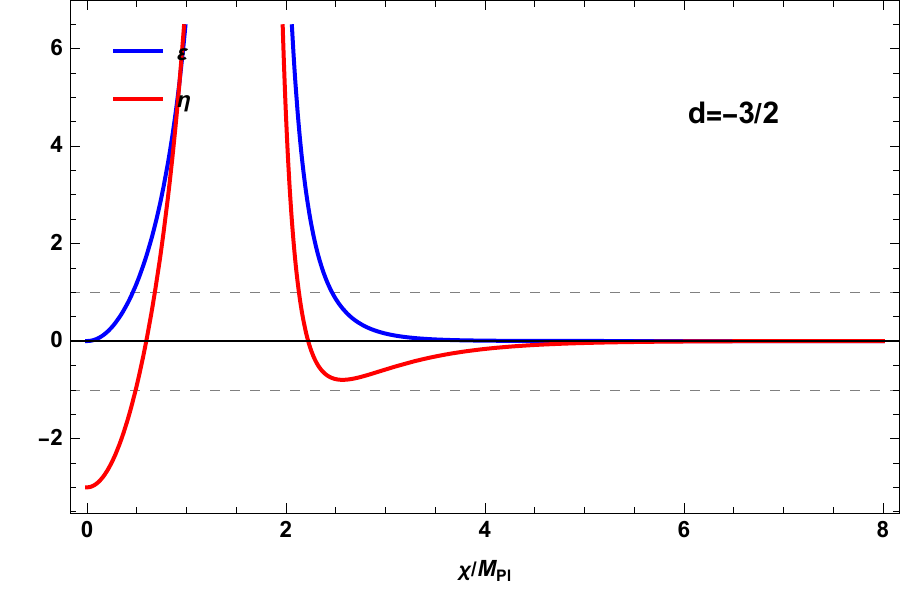}}
	\caption{The evolutions of the slow-roll parameters $\varepsilon,~\eta$ for 
the inflaton potential in Eq.~\eqref{Eq:V_threeMC} .}\label{fig:slowroll-threeC}
\end{figure}

When $a_2=0$, the potential with canonical field $\chi$ can be rewritten as
\begin{equation}
	\begin{split}
		V &=V_0\left(1+d\tanh^2\left(\frac{\chi}{\sqrt{2}}\right)\right)^2\label{Eq:V_threeMC}\\
	\end{split}
\end{equation}
with $V_0=a_1^2/c_2c_3$ and $d=3a_3c_1/a_1$, and we show the potential with different $d$ in Fig.~\ref{fig:pot-threeC}. From the plots, one notes that the positive and negative branches are symmetric, and here we will only discuss the inflation for 
the positive branch. Thus, we need to find the possible parameter space of $d$ for inflation. The slow-roll parameters 
are given by the following analytic form 
\begin{equation*}
	\begin{split}
		\varepsilon&=\frac{16 d^2 \tanh ^2\left(b \chi \right)}{\left((d+1) \cosh \left(2b \chi \right)-d+1\right)^2}, \\
		\eta&= \frac{d\left(1-2 \xi_1 \cosh \left(2b \chi \right)+\xi_2 \cosh \left(4b \chi \right)\right)\text{sech}^6\left( b\chi  \right) }{6(3 d-1) \left(d \tanh ^2\left( b\chi \right)+1\right)^2}.
	\end{split}
\end{equation*}
with $\xi_1=\left(5d+1\right)/(9d-3)$ and $\xi_2=(d+1)/(9d-3)$. 
Solving the equations of $\varepsilon=1 $ and $|\eta|=1$, we get 
\begin{equation}
\begin{split}
	\varepsilon=1 &\to	d_{1}=\frac{2 \cosh ^2\left(b\chi \right) \coth \left(b\chi \right)}{4 \mp\sinh \left(2b \chi \right)}~,~\\
	|\eta|=1 &\to	d_{2}=\left\{	
		\begin{array}{l}
			\frac{\left(\Delta_1\pm\Delta_2\right)\coth ^2\left(b\chi\right)}{\cosh \left(4b\chi \right)+16 \cosh \left(2b\chi \right)-65}\\
			\frac{\left(\Delta_3\pm \Delta_4\right)\coth ^2\left(b\chi\right)}{\cosh \left(4b\chi\right)-16 \cosh \left(2b\chi\right)+63}
		\end{array} 
	\right. ~,~
\end{split}\label{Eq:d-threeC}
\end{equation}
where  $\Delta_1=8 \sqrt{\cosh \left(4b\chi \right)-4 \cosh \left(2b\chi \right)+4}$, $\Delta_2= \cosh \left(4b\chi \right)+8 \cosh \left(2b\chi \right)-17$, $\Delta_3=8 \sqrt{5-4 \cosh \left(2b\chi \right)}$, 
and $\Delta_4=\cosh \left(4b\chi \right)-8 \cosh \left(2b\chi \right)+15$.
Due to the complicated formula, we plot  $d$ versus $\chi/M_{\text{Pl}}$ in Fig. \ref{fig:d-threeC} and find that the inflation will occur at the regime $|d|>1/2$ since inflaton will go down to the vacuum of potential without truncation when $|d|<1/2$. The evolution of slow-roll parameters in terms of inflaton field are shown in Fig. \ref{fig:slowroll-threeC}. 

\begin{figure}[t]
	\includegraphics[width=0.4\linewidth]{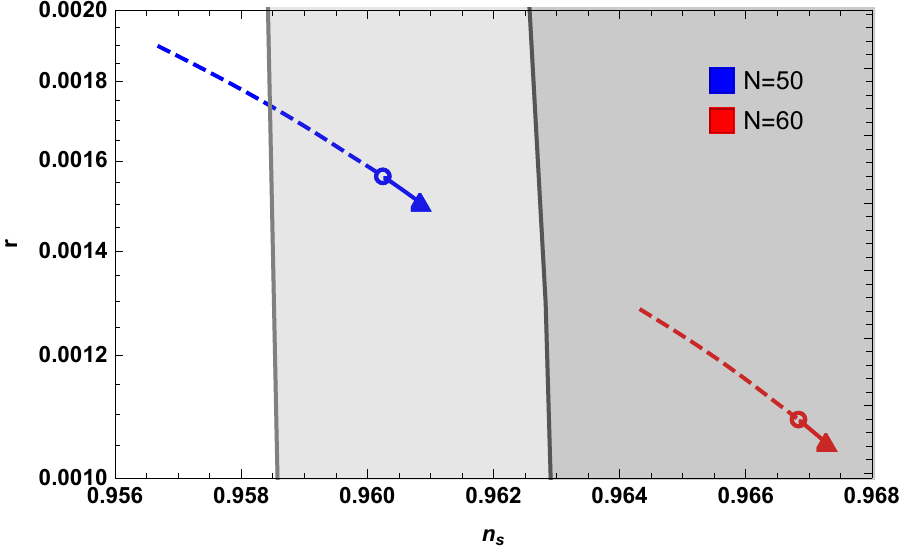}
	\caption{The CMB predictions for the model in Eq. \eqref{Eq:V_threeMC}. The circles and triangles are corresponding to T- and E-model, respectively.}  \label{fig:nsr-threeC}
\end{figure}

When $d>1/2$,  there is a minimum for the potential at $\chi_m=0$, and the gradient increases as $d$ increases, which indicates
 that inflation can be truncated only with large $d$. Based on the evolution of slow-roll parameters from Fig. \ref{fig:slowroll-threeC}, one can find that inflation  will  end if the condition  $\varepsilon=1$ is satisfied when $d>1.937$, whereas inflation will frist  end  at $\eta=1$ until $d>2.291$. The predictions $n_s$ and $r$ for $N=50$ and $60$ are shown with dashed lines in Fig. \ref{fig:nsr-threeC}. As $d$ is rising, the scalar spectral index $n_s$ is increasing. In the limit $d\to \infty$, the constant term $1$ can be ignored in the potential \eqref{Eq:V_threeMC} and then the $\alpha$-attractor T-model for $\phi^4$ \cite{Kallosh:2013yoa} can be achieved. As the parameter $d$ locates in the range $-1<d<-1/2$, there is only a maximum for the potential at $\chi=0$. However, slow-roll inflation cannot happen due to the large $|\eta|$. While $d<-1$, there is a maximum at $\chi=0$ and two minima at $\chi_m=\pm \sqrt{2} \tanh ^{-1}\left(\sqrt{{1}/{|d|}}\right)$. From the evolution of the slow-roll parameters, inflation only occurs at the region $\chi>\chi_m$ and the predictons are shown with solid lines in Fig. \ref{fig:nsr-threeC}. The spectral index $n_s$ and the tensor-to-scalar ratio $r$  slightly depend on the parameter $d$, and  $n_s$ improves when the parameter $d$ goes from $-\infty$ to $-1$. In the limit $d\to -1$, the inflaton $\chi_*\sim 6 M_{\text{Pl}}$ and $e^{-\sqrt{2}\chi_*}$ is small when the pivot scale leaves the horizon, so the potential \eqref{Eq:V_threeMC} can be expanded as
\begin{equation}
	\begin{split}
		V&\simeq V_0(1+d)^2\left(1-\frac{4d}{1+d}e^{-\sqrt{2}\chi}\right)^2.\\
	\end{split}
\end{equation} 
The E-model \cite{Kallosh:2013yoa,Kallosh:2021mnu} is carried out and  the predicted observations are
\begin{equation}
	n_s\simeq1-\frac{8 (4 N+1)}{(1-4N)^2},~~r\simeq\frac{64}{(1-4 N)^2}.
\end{equation}
The numerical results of the predictions for T- and E-models are also shown as circles and triangles in Fig. \ref{fig:nsr-threeC}, respectively. One notes that the zeroth and first order in $e^{-\sqrt{2}\chi}$ expansion 
for the potential of T- and E-models are identical, however, the precise predictions are different due to the distinction at the higher orders when $\sqrt{2}\chi\simeq \mathcal{O}(1)$. We conclude the possible parameter space for $d$ in Table \ref{tab:d-threeC}. 

\begin{table*}[t]
	\caption{The possible parameter space for $d$. } \label{tab:d-threeC}
	\begin{tabular}{ccc}
			\hline
			{\multirow{2}{*}{Condition}} &  \multicolumn{2}{c}{$d$}\\
			\cline{2-3}
			& $a_2=0$ case & $a_3=0$ case\\
			\hline
			$\varepsilon=1$
			& $[-\infty,-1)~\cup~[2.2911,\infty]$~~&~~ $[-\infty,-1]~\cup~[1,\infty]$\\
			$|\eta|=1$ 
			 &	$[0.5,2.2911]$& $[-1,-0.5774]~\cup~[0.5774,1]$\\
			No truncation 
			& $[-0.5,0.5]$& $[-0.5774,0.5774]$\\
			No slow-roll 
			& $[-1,-0.5]$&\\
			\hline
	\end{tabular}
\end{table*}

\subsection{$a_3=0$ case}

\begin{figure}[t]
	\subfigure[]{\includegraphics[width=0.4\linewidth]{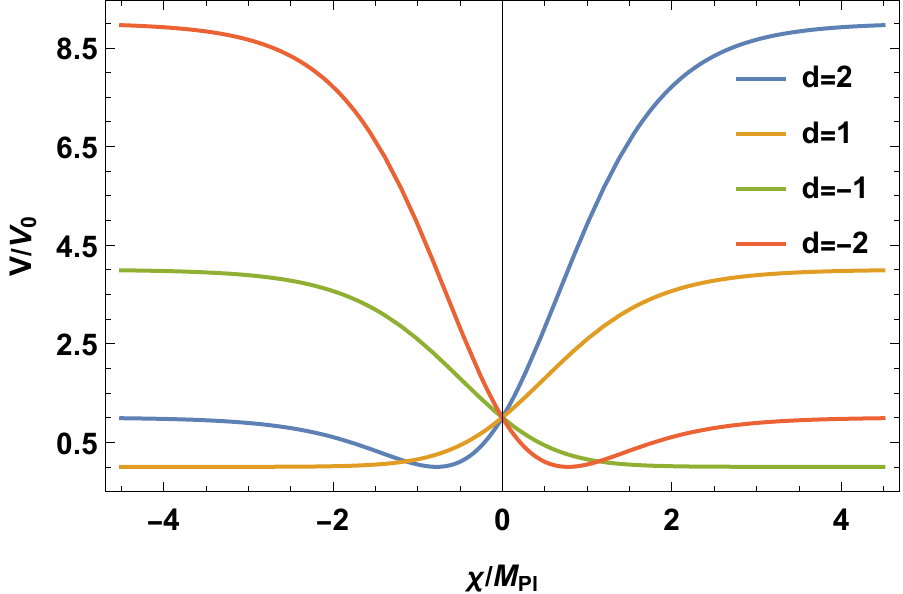}\label{fig:pot-threeD}}
	\subfigure[]{\includegraphics[width=0.4\linewidth]{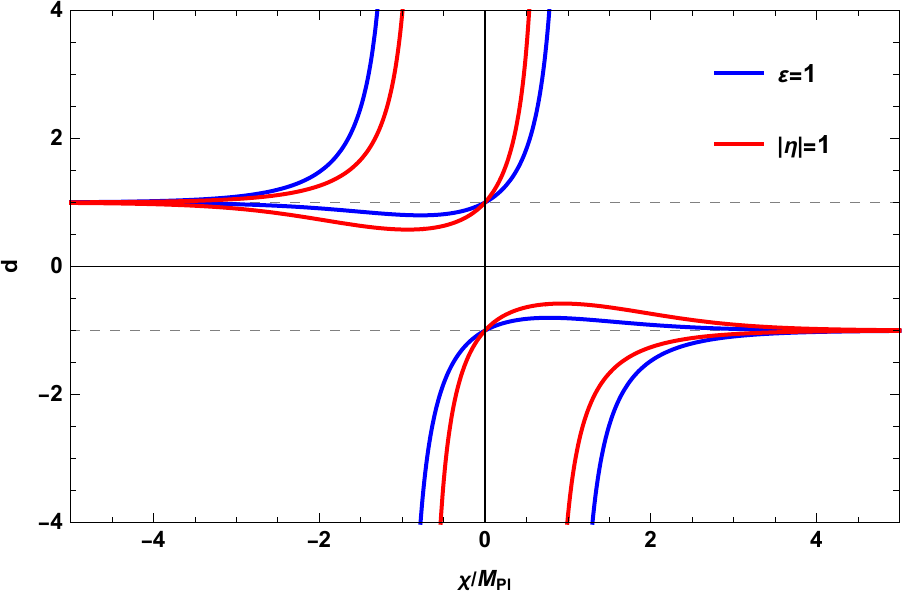}\label{fig:d-threeD}}
	\caption{Inflaton potential in Eq. \eqref{Eq:V_threeMD} (a) and its possible parameter space for $d$ (b).}
\end{figure}

\begin{figure}[t]
	\subfigure{\includegraphics[width=0.4\linewidth]{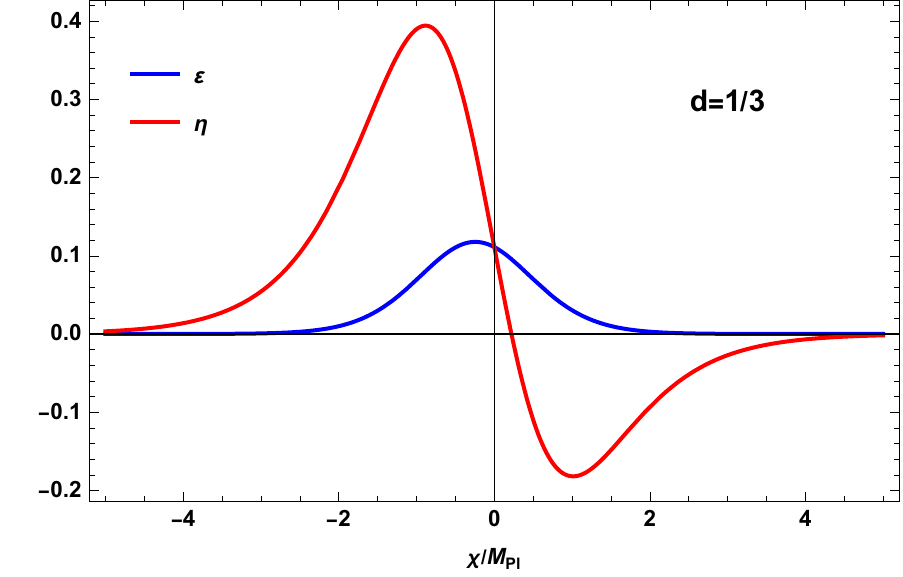}}
	\subfigure{\includegraphics[width=0.4\linewidth]{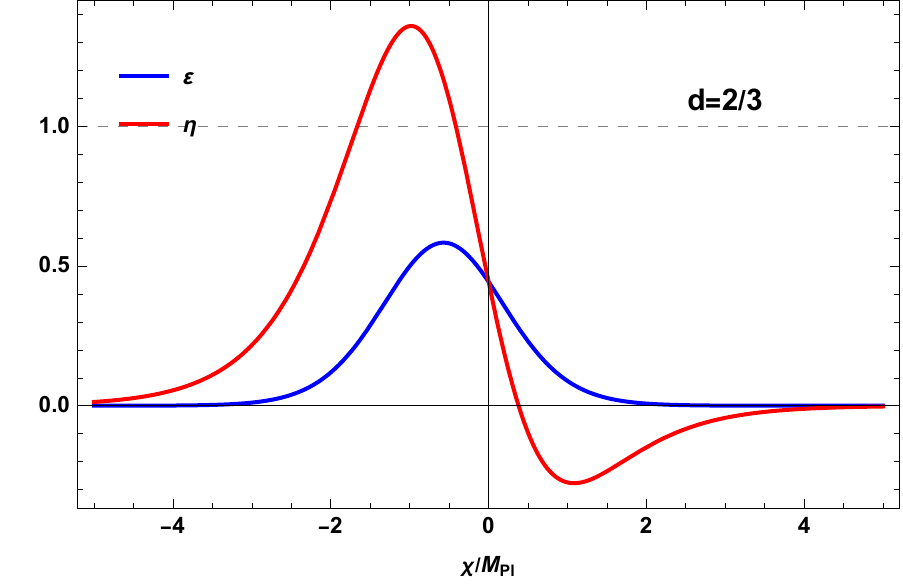}}\\
	\subfigure{\includegraphics[width=0.4\linewidth]{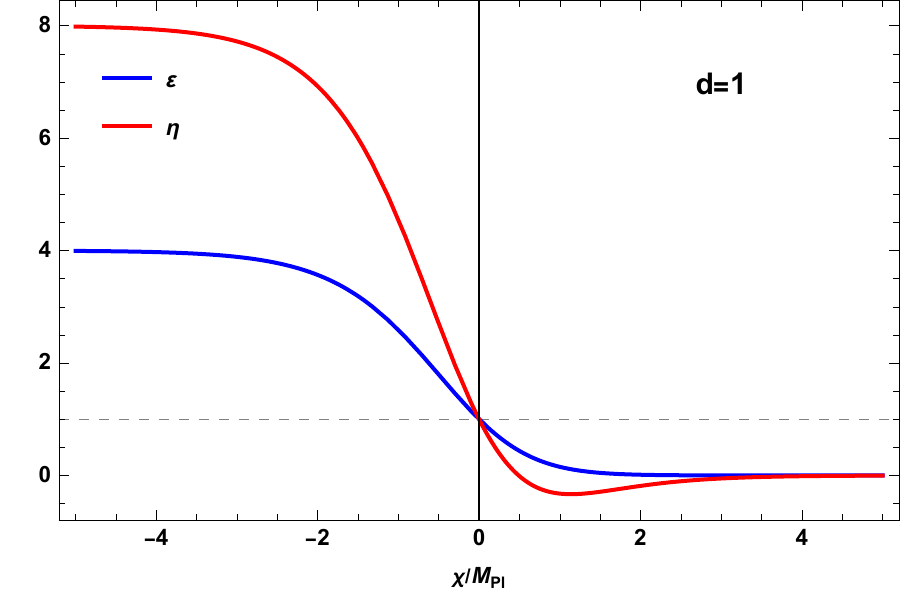}}
	\subfigure{\includegraphics[width=0.4\linewidth]{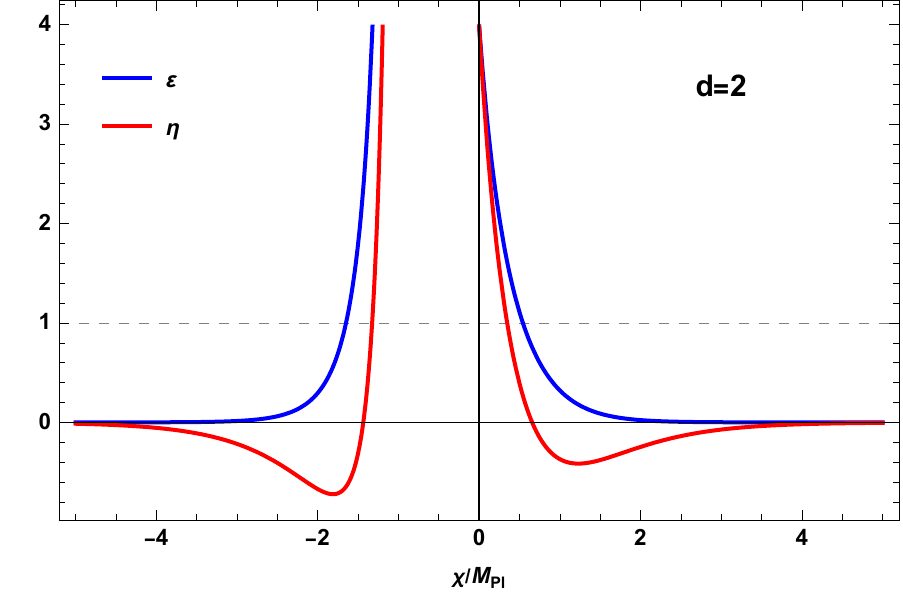}}
	\caption{The evolution of the slow-roll parameters $\varepsilon$ and $\eta$.}\label{fig:slowroll-threeD}
\end{figure}

\begin{figure}[t]
	{\includegraphics[width=0.4\linewidth]{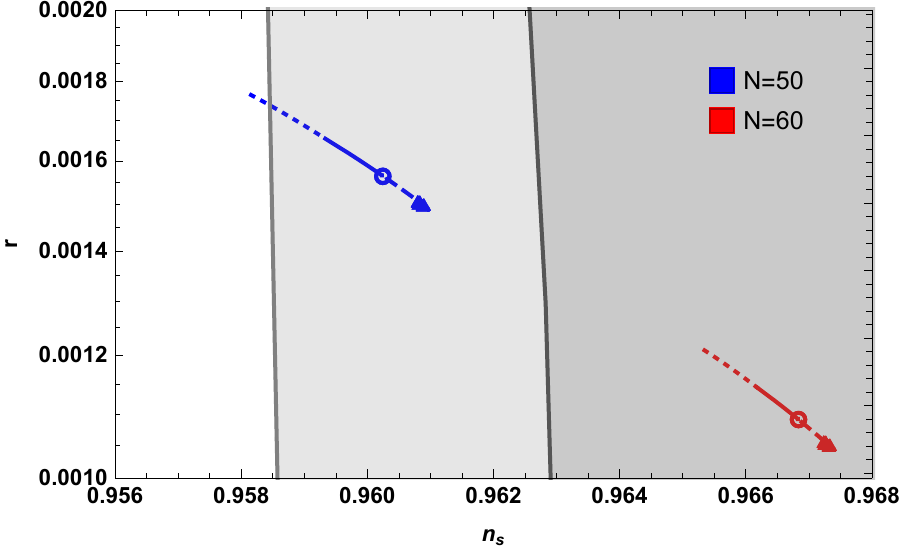}}
	\caption{ The $r$ versus $n_s$  predictions for the  inflaton potential 
in Eq. \eqref{Eq:V_threeMD}. The circles and triangles correspond to T- and E-models, respectively.}\label{fig:nsr-threeD}

\end{figure}

The inflaton potential in terms of  canoical field $\chi$ is
\begin{equation}
	\begin{split}
		V=V_0 \left(1+d\tanh\left(\frac{\chi}{\sqrt{2}}\right)\right)^2,\label{Eq:V_threeMD}
	\end{split}
\end{equation}
with $V_0=a_1^2/c_2c_3$ and $d=2a_2\sqrt{c_1}/a_1$. 
The inflaton potential is shown in Fig. \ref{fig:pot-threeD}, and we will only consider inflation with $d>0$, 
given the equivalent potential under the exchange $(d,\chi)\leftrightarrow (-d,-\chi)$. The slow-roll parameters are
\begin{equation}
	\begin{split}
		\varepsilon&=\frac{d^2 \text{sech}^4\left(b\chi\right)}{\left(1+d \tanh \left(b\chi\right)\right)^2},\\
		\eta&=\frac{d \text{sech}^4\left(b\chi\right) \left(2 d-d \cosh \left(2b\chi \right)-\sinh \left(2b\chi \right)\right)}{\left(1+d \tanh \left(b\chi\right)\right)^2}.
	\end{split}
\end{equation}
Under the slow-roll conditions, the solutions for the parameter $d$ with respect to inflaton $\chi$ are
\begin{equation}
	\begin{split}
		\varepsilon=1 &\to d_1=\frac{2 \cosh ^2\left(b\chi\right)}{\sinh \left(2b\chi \right)\pm 2}~,~\\
		|\eta|=1 &\to d_2=\left\{	
		\begin{array}{l}
			-\frac{2 \cosh \left(b\chi\right)\left(\Delta_1\pm\Delta_2\right)}{8 \cosh \left(2b\chi \right)+\cosh \left(4b\chi \right)-17} \\
			\frac{-2 i \cosh \left(b\chi\right)\pm \sinh \left(2b\chi \right)}{2 i \sinh \left(b\chi\right)\pm \left(3-\cosh \left(2b\chi \right)\right)}
		\end{array} 
		\right.~,~\,
	\end{split}\label{Eq:d-threeD}
\end{equation}
where $\Delta_1=5 \sinh \left(b\chi\right)+\sinh \left(3b\chi \right)$, and $\Delta_2=4 \sqrt{\cosh \left(2b\chi \right)}$. 
From Figs. \ref{fig:d-threeD} and  \ref{fig:slowroll-threeD}, one can find that it is hard to end inflation when $|d|<0.5774$ due to the too flattened potential. When $d\in[0.5774,1]$, inflation ends first as the condition $|\eta|=1$ is satisfied, even the condition $\varepsilon=1$ has a solution as $d>0.8$. There is no extreme point for the potential and the inflaton rolls from the positive to the negative region. The numerical predictions are given by dotted lines in Fig. \ref{fig:nsr-threeD} and the spectral index $n_s$ increases with $d$ increasing. With other $d\geq 1$, there is a minimum for the potential and inflation can happen in the two trajectories, right and left sides of the minimum. In both branches, inflation ends when the condition $\varepsilon=1$ is satisfied. The predicted  $n_s$ and $r$ for inflation on the right and left trajectories are given by solid and dashed lines, respectively, in Fig. \ref{fig:nsr-threeD}. As parameter $d$ is increasing, the index $n_s$ on the right trajectory increases, while $n_s$ on the left trajectory decreases. The possible choices for parameter $d$ are also shown in Table \ref{tab:d-threeC}.

As discussed before, in the limit $d\to \infty$, the constant term 1 can be ignored and the potential in Eq. \eqref{Eq:V_threeMD} returns to $\tanh^2(\chi/\sqrt{2})$, which is identical to the potential of the T-model for $\varphi^2$ \cite{Kallosh:2013yoa}. In this way, the left and right trajectories are symmetric and give the same cosmological predictions, which makes the connection of  solid lines and dashed lines in Fig. \ref{fig:nsr-threeD}.  While in the limit $d\to 1$, the E-model can be realized by expanding the potential in Eq. \eqref{Eq:V_threeMD}  on the left trajectory as $V\propto\left(1-\frac{2d}{1+d}e^{-\sqrt{2}\chi}\right)^2$. That is why our models connect the T-model and the E-model in a different approximation.

\subsection{General case}

\begin{figure}[!h]
	\subfigure[]{\includegraphics[width=0.48\linewidth]{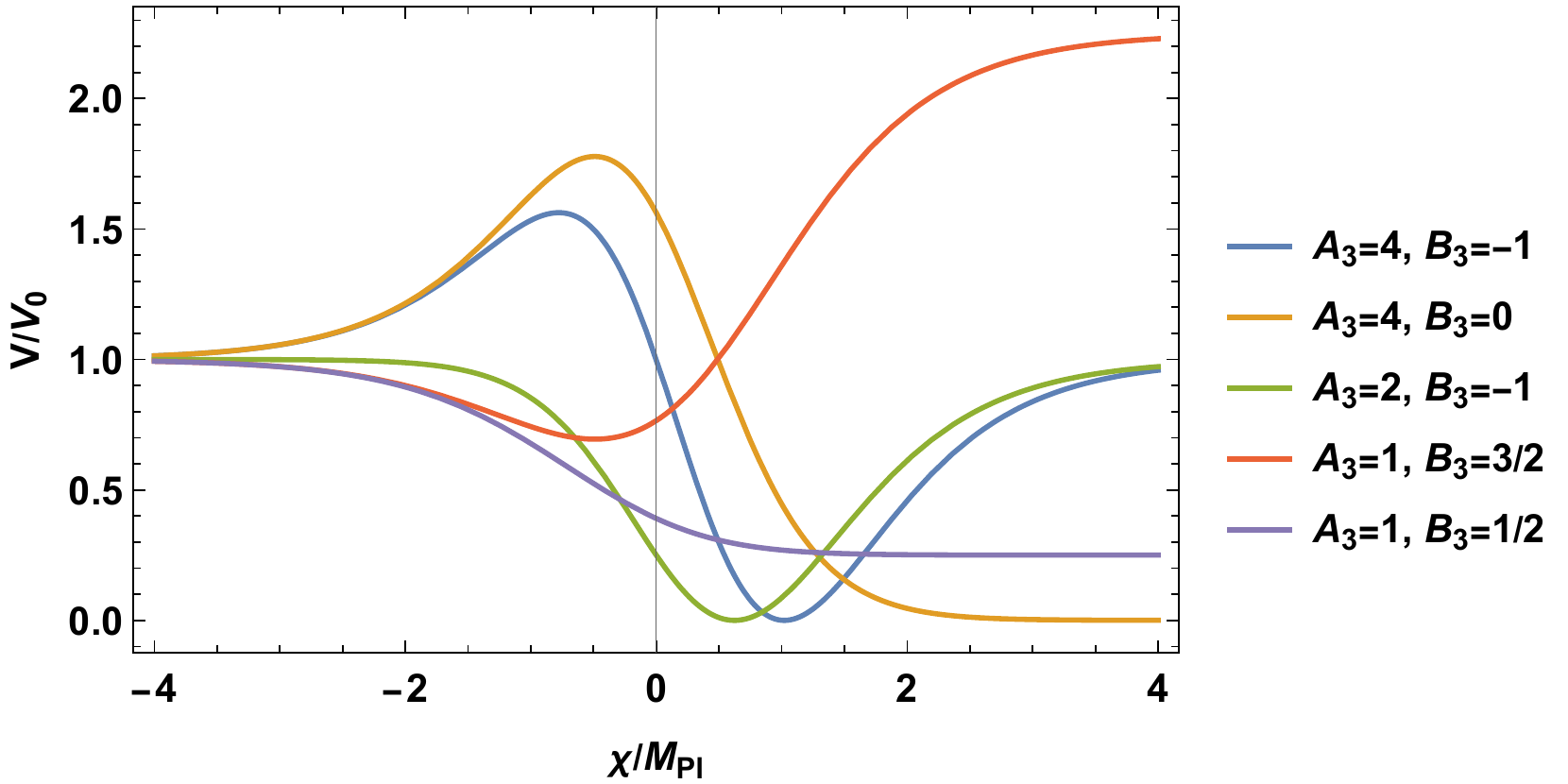}}
	\subfigure[]{\includegraphics[width=0.48\linewidth]{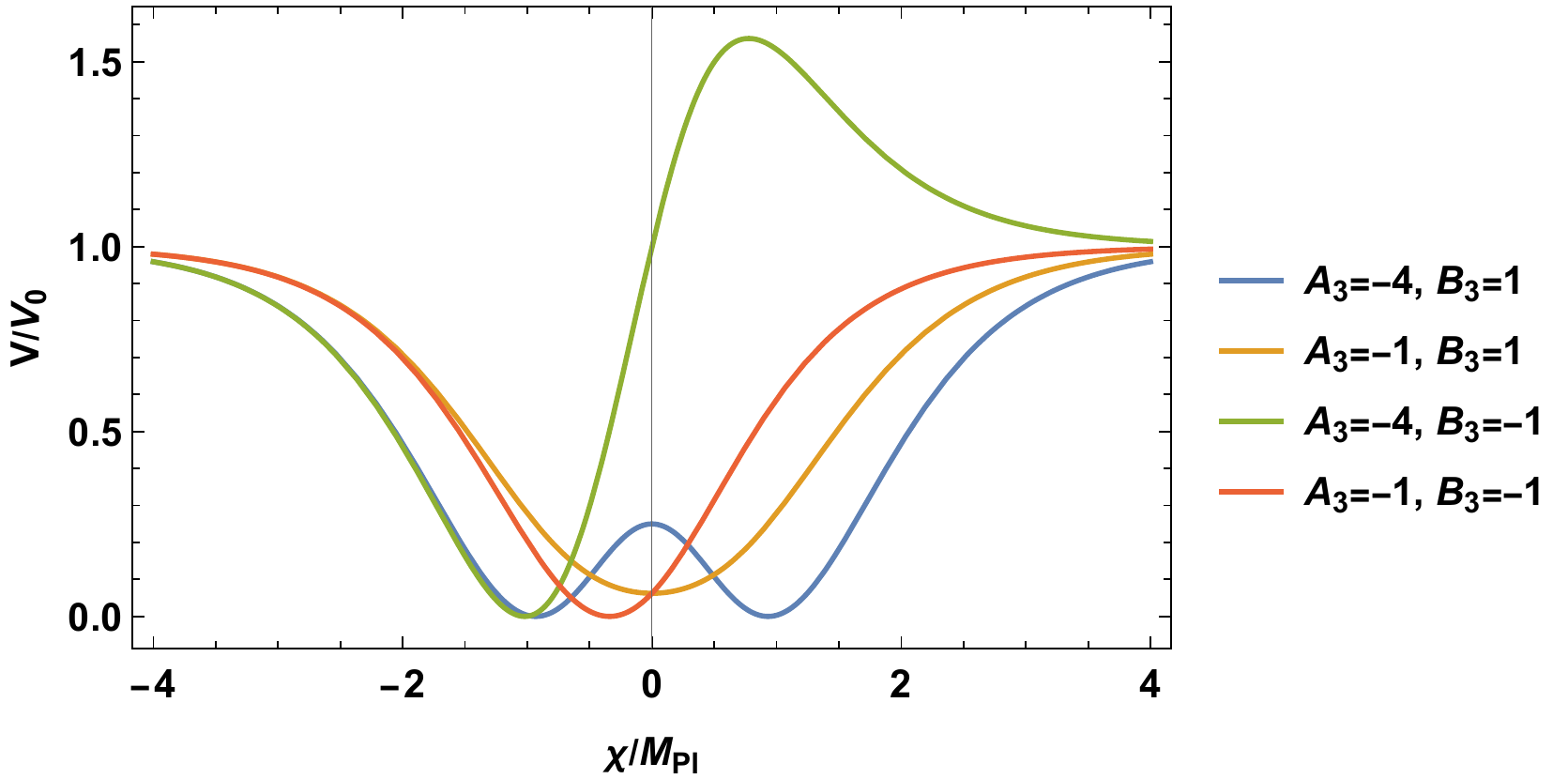}}
	\caption{Inflaton potential in Eq. \eqref{Eq:V_threeMG} with $A>0$ (a) and $A<0$ (b).}\label{fig:pot_threeMG}
\end{figure}

\begin{figure}[!h]
	\subfigure[]{\includegraphics[width=0.4\linewidth]{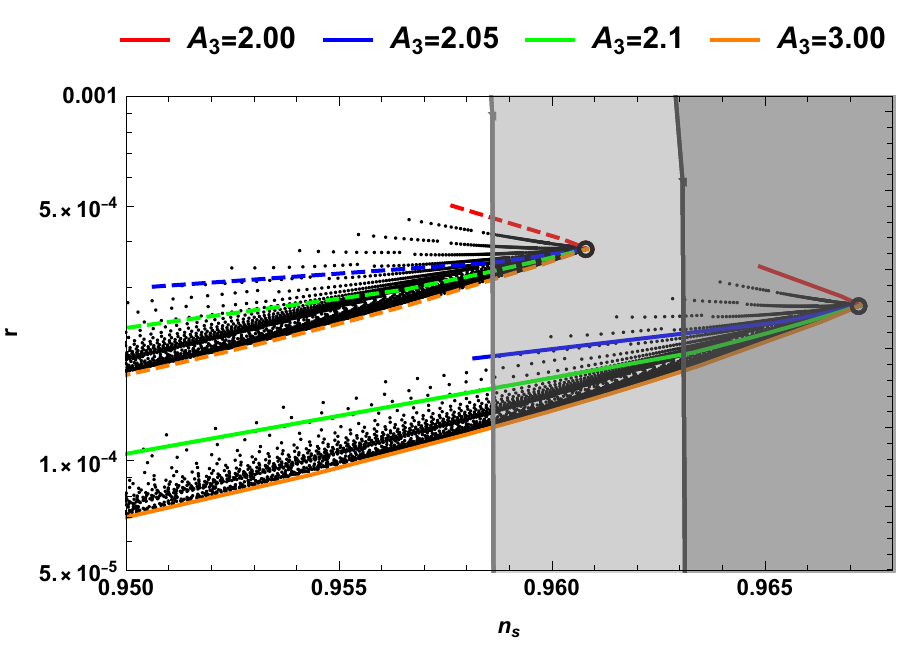}}
	\subfigure[]{\includegraphics[width=0.4\linewidth]{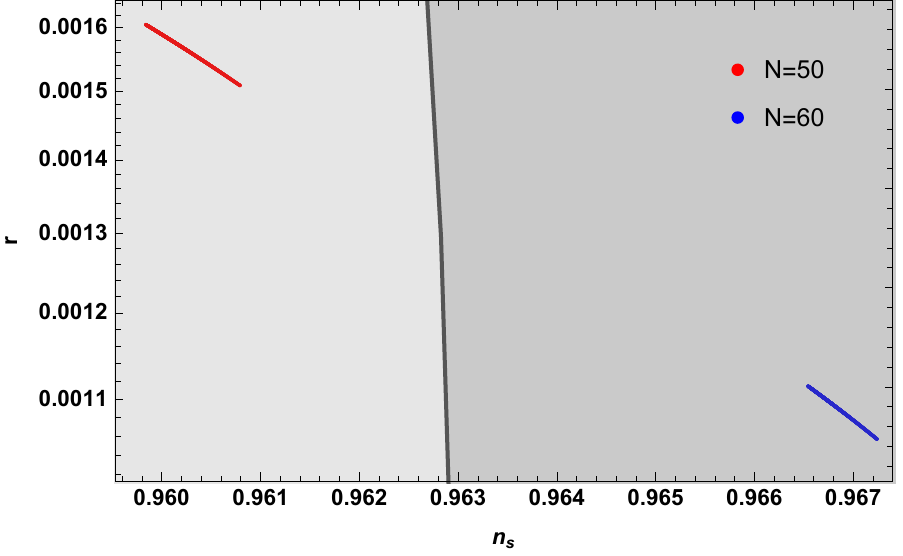}}
	\caption{The  $r$ versus $n_s$ predictions for the inflaton potential in Eq. \eqref{Eq:V_threeMG} 
with $A_3\geq2$ and $B_3<0$. (a) In the region $[\chi_M,\chi_m]$; the circles are in the limit $B_3\to -\infty$.  The dashed and solid lines are corresponding to $N=50$ and $N=60$. (b) In the region $[\chi_m,\infty]$.}\label{fig:nsr-threeG1}
\end{figure}

Finally, we discuss the following generic potential in Eq. \eqref{Eq:V_threeMG}. 
Similarly, the parameter $V_0$ can be determined by the constraint $A_s=2.1\times10^{-9}$ at horizon crossing.
There are three extreme points for the potential
\begin{equation}
	\begin{split}
		\chi_1&=\frac{1}{\sqrt{2}}\log \left(\frac{A_3-2}{A_3-2 B_3}\right),\\ \chi_2&=\frac{1}{\sqrt{2}}\log \left(-\frac{\sqrt{A_3^2-4 B_3}+A_3}{2 B_3}\right),\\
		\chi_3&=\frac{1}{\sqrt{2}}\log \left(-\frac{A_3-\text{sgn}(B_3)\sqrt{A_3^2-4 B_3} }{2 B_3}\right).
	\end{split}
\end{equation} 
The potential with different parameters is shown in Fig. \ref{fig:pot_threeMG}. For $A_3>2$ and $B_3<0$, the potential has a maximum at $\chi_M=\chi_1$ and a minimum at $\chi_m=\chi_3$. Similarly, inflation occurs in the regions $[\chi_M,\chi_m]$ and $[\chi_m,\infty]$ since the slow-roll conditions are violated in the region $[-\infty,\chi_M]$. The cosmological predictions in both regions are shown in Fig. \ref{fig:nsr-threeG1}. 
The slow-roll parameters and number of $e$-folding are
\begin{equation}
	\begin{split}
		\varepsilon&=\frac{4 e^{2 \sqrt{2} \chi } \left(\delta_0-\delta_4 e^{\sqrt{2} \chi } \right)^2}{\left(e^{\sqrt{2} \chi }+1\right)^2 \left(1+A_3 e^{\sqrt{2} \chi }+B_3 e^{2 \sqrt{2} \chi }\right)^2}~,~\\
		\eta&=\frac{4 e^{\sqrt{2} \chi } \left(\delta_0+2 \delta_1 e^{\sqrt{2} \chi } -3 \delta_2 e^{2 \sqrt{2} \chi } +2 \delta_3e^{3 \sqrt{2} \chi } +B_3 \delta_4 e^{4 \sqrt{2} \chi } \right)}{\left(e^{\sqrt{2} \chi }+1\right)^2 \left(1+A_3 e^{\sqrt{2} \chi }+B_3 e^{2 \sqrt{2} \chi }\right)^2}~,~\\
		N&=\frac{1}{4} \left(-\frac{1}{\delta_0}e^{-\sqrt{2} \chi }-\frac{B_3 }{\delta_4}e^{\sqrt{2} \chi }+\frac{\sqrt{2} \gamma_1 }{\delta_0^2}\chi-\frac{2 \gamma_2 }{\delta_0^2 \delta_4^2}\log \left(e^{\sqrt{2} \chi } \delta_4-\delta_0\right)\right)~,~
	\end{split}
\end{equation}
where $\delta_0=A_3-2$, $\delta_1=A_3^2-5 A_3+2 B_3+4$, $\delta_2=2 A_3^2-3 A_3 (B_3+1)+4 B_3$, $\delta_3=A_3^2-5 A_3 B_3+2 B_3 (2 B_3+1)$, $\delta_4=A_3-2 B_3$, $\gamma_1=A_3^2-2 B_3-2$, and $\gamma_2=\left(A_3^2-4 B_3\right) (-A_3+B_3+1)^2$.
In order to obtain a prediction within the $2\sigma$ region of the Planck results, we need $B_3\lessapprox-450$. Then these new parameters are approximated to be
\begin{equation*}
	\begin{split}
		\delta_1\sim 2B_3,~\delta_2\sim (4-3A_3)B_3,
		~\delta_3\sim 4B_3^2, ~\delta_4\sim -2B_3.
	\end{split}
\end{equation*}
Thus, the tensor-to-scalar ratio approximately is
\begin{equation*}
	r=16\varepsilon\simeq\frac{4\delta_4^2}{(B_3e^{\sqrt{2}\chi})^2}\sim256e^{-2\sqrt{2}\chi}.
\end{equation*}
In the region $[\chi_M,\chi_m]$, the pivot scale leaves the horizon at $\chi_*\sim -5M_{\text{Pl}}$, and the tensor-to-scalar ratio is $10^{-4}$, which is consistent with numerical calculations. 
Moreover, in the limit $B_3\to -\infty$, the slow-roll parameters are independent on the parameter $A_3$. Thus, the predictions are $n_s\simeq 0.9608,r\simeq 3.8\times10^{-4} $ for $N=50$ and $n_s\simeq 0.9672,r\simeq 2.7\times10^{-4}$ for $N=60$, which locate in the $2\sigma$ region of the Planck results and are shown as circles in Fig.  \ref{fig:nsr-threeG1}(a). In the figure, the black scatters of the numerical predictions form a fan where the two boundaries and vertex of the fan come from the limits $A_3\to 2$, $A_3\to \infty$ and $B_3\to -\infty$, respectively. 
There is a minimum at $\chi_m=\chi_3$ for the potential as setting $A_3=2$ and $B_3<0$, so there are two inflationary 
trajectories for which the predicted $n_s$ and $r$ are also shown in Fig. \ref{fig:nsr-threeG1}. 

\begin{figure}[!h]
	\subfigure[]{\includegraphics[width=0.4\linewidth]{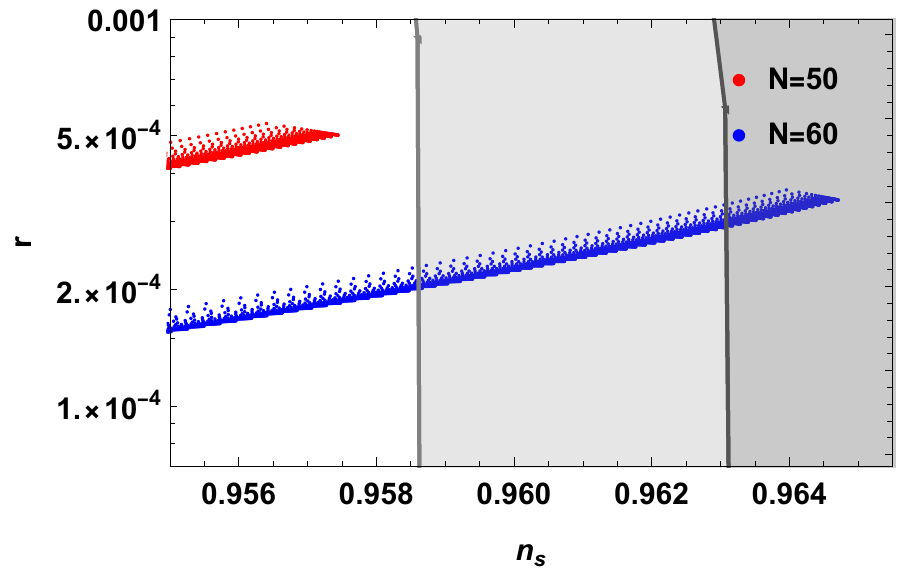}}
	\subfigure[]{\includegraphics[width=0.4\linewidth]{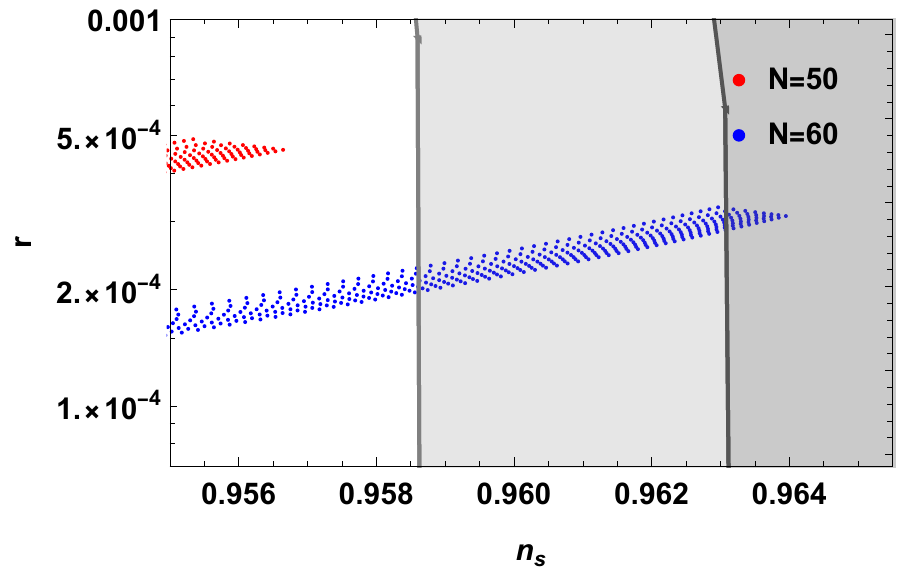}}
	\caption{The $r$ versus $n_s$ predictions for the inflaton potential  in Eq. \eqref{Eq:V_threeMG} 
		with $A_3>2,~B_3>0$, and $A_3>2B_3$. (a) In the region $[-\infty,\chi_M]$. (b) In the region $[\chi_M,\infty]$.}\label{fig:nsr-threeG3}
\end{figure}

\begin{figure}[!h]
	\includegraphics[width=0.4\linewidth]{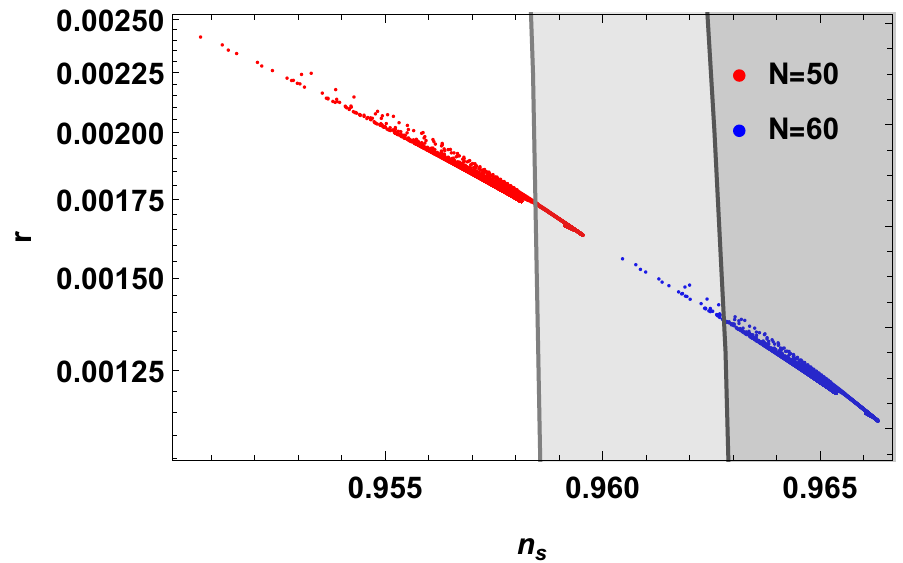}
	\caption{CMB observations for the model in Eq. \eqref{Eq:V_threeMG} with $0\leq  A_3<2, A_3<2B_3$. Inflation occurs on the right side of $\chi_m$.}\label{fig:nsr-threeG4}
\end{figure}

When the parameters are in the region $A_3>2, ~B_3>0$, and $A_3>2B_3$, the potential has a maximum at $\chi_M=\chi_1$. Inflation can occur on both sides of the maximum where the predicted $n_s$ and $r$ are shown in Fig. \ref{fig:nsr-threeG3}. In order to get predictions which are consistent with the experimental data, the parameters should be $A_3\simeq B_3$ or $A_3\simeq 2, B_3\sim 0$.		
While for $0\leq  A_3<2$, and $A_3<2B_3$, inflation can occur on the right side of the minimum $\chi_m=\chi_1$ for which the predicted $n_s$ and $r$ are  shown in Fig. \ref{fig:nsr-threeG4}.
 
\begin{figure}[!h]
	\subfigure[]{\includegraphics[width=0.4\linewidth]{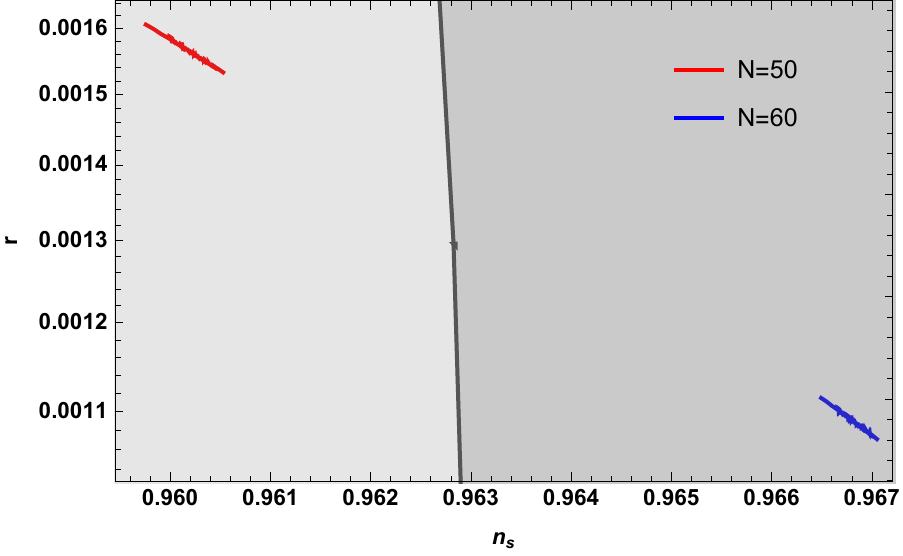}}
	\subfigure[]{\includegraphics[width=0.4\linewidth]{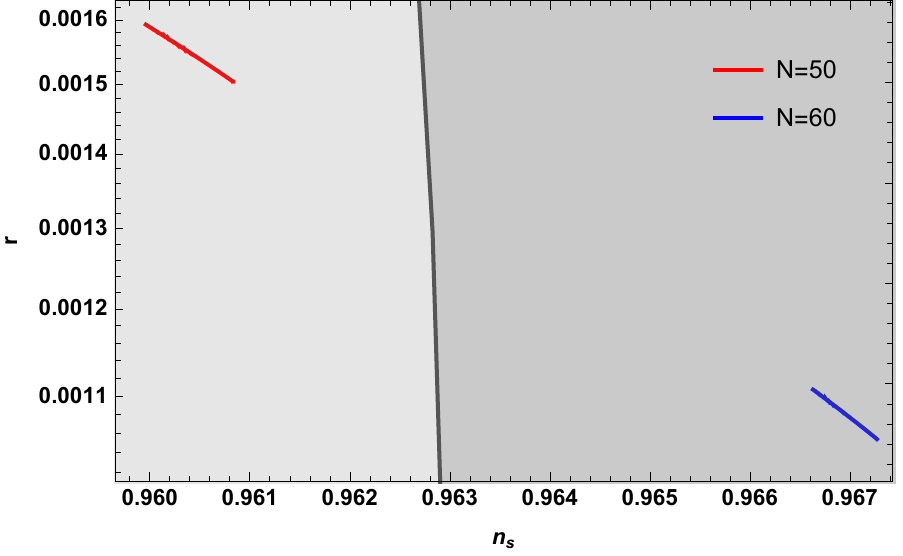}}
	\caption{The $r$ versus $n_s$ predictions for the inflaton potential in Eq. \eqref{Eq:V_threeMG} with $A_3<0,~B_3>0$, and $A_3^2>4B_3$. Inflation occurs in the regions (a) $[-\infty,\chi_{m1}]$ and (b) $[\chi_{m2},\infty]$.}\label{fig:nsr-threeG5}
\end{figure}

\begin{figure}[!h]
	\subfigure[]{\includegraphics[width=0.4\linewidth]{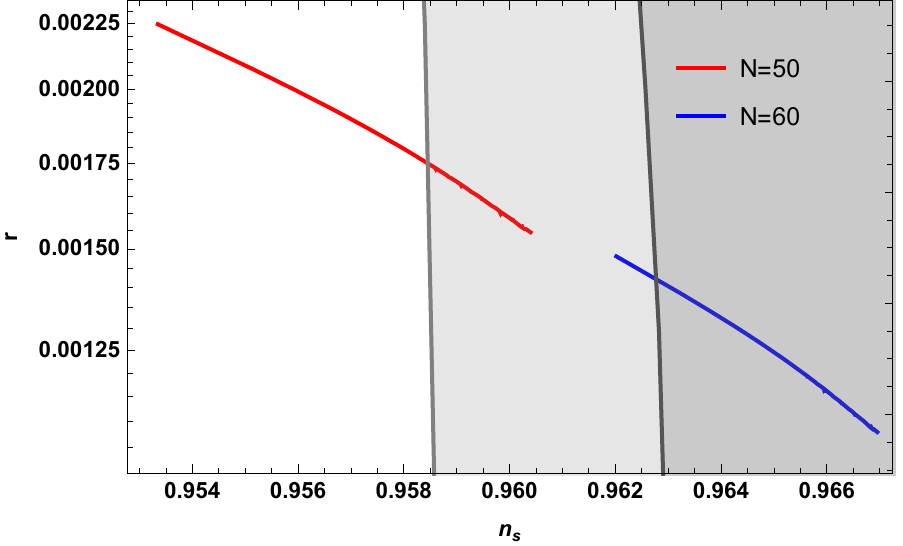}}
	\subfigure[]{\includegraphics[width=0.4\linewidth]{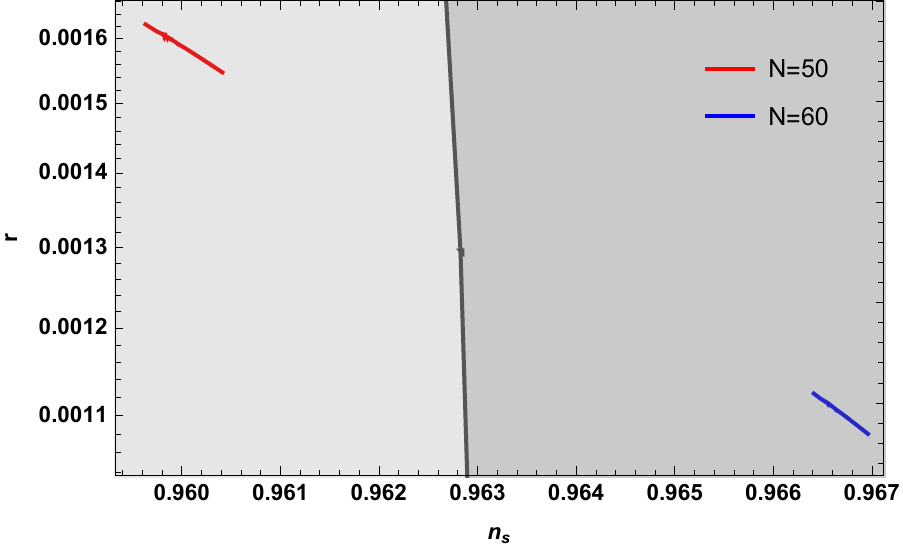}}
	\caption{The $r$ versus $n_s$ predictions for the inflaton potential 
		in Eq. \eqref{Eq:V_threeMG} with $A_3<0,~B_3>0$, and $A_3^2<4B_3$. Inflation occurs in the regions (a) $[-\infty,\chi_m]$ and (b) $[\chi_m,\infty]$.}\label{fig:nsr-threeG6}
\end{figure}

\begin{figure}[!h]
	\subfigure[]{\includegraphics[width=0.4\linewidth]{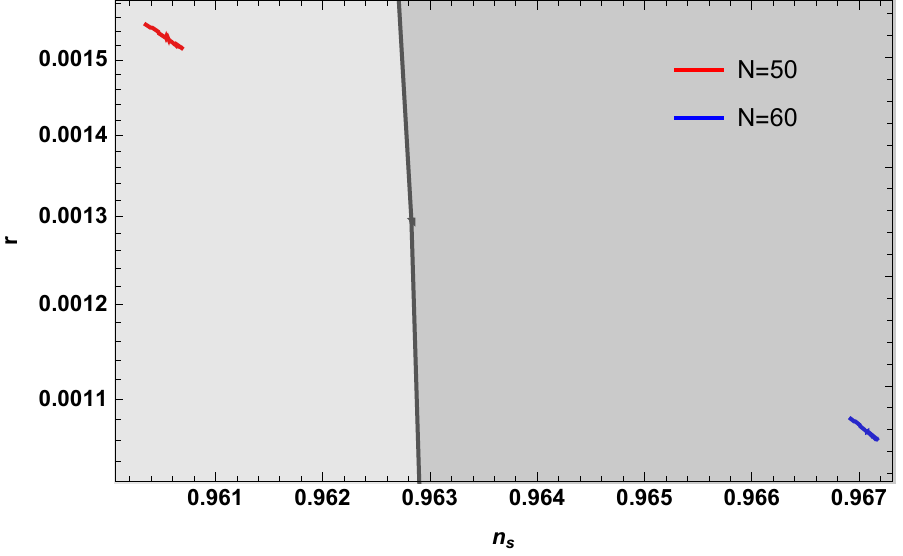}}
	\subfigure[]{\includegraphics[width=0.4\linewidth]{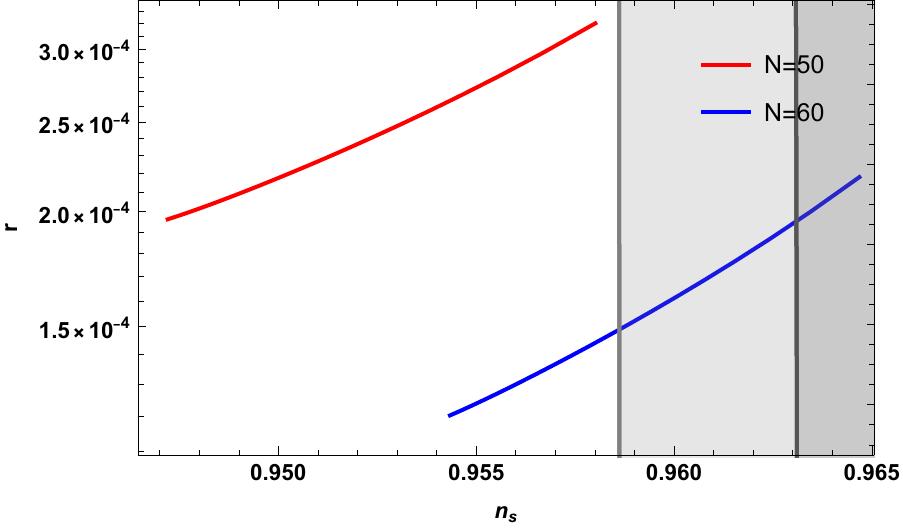}}
	\caption{ The $r$ versus $n_s$ predictions for the inflaton potential  in Eq. \eqref{Eq:V_threeMG} 
		with $A_3<0,~B_3<0$, and $A_3<2B_3$. Inflation occurs in the regions (a) $[-\infty,\chi_m]$ and (b) $[\chi_m,\chi_M]$.}\label{fig:nsr-threeG7}
\end{figure}

\begin{figure}[!h]
	\subfigure[]{\includegraphics[width=0.4\linewidth]{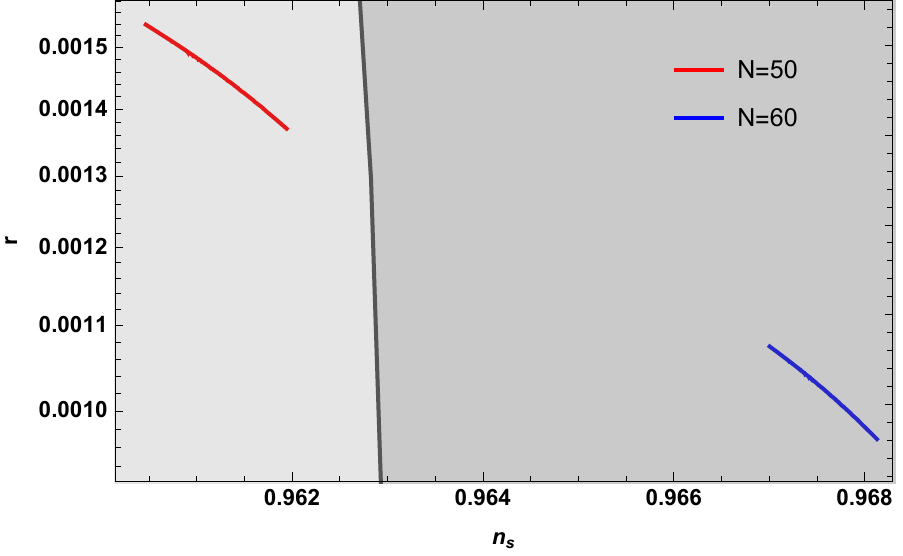}}
	\subfigure[]{\includegraphics[width=0.4\linewidth]{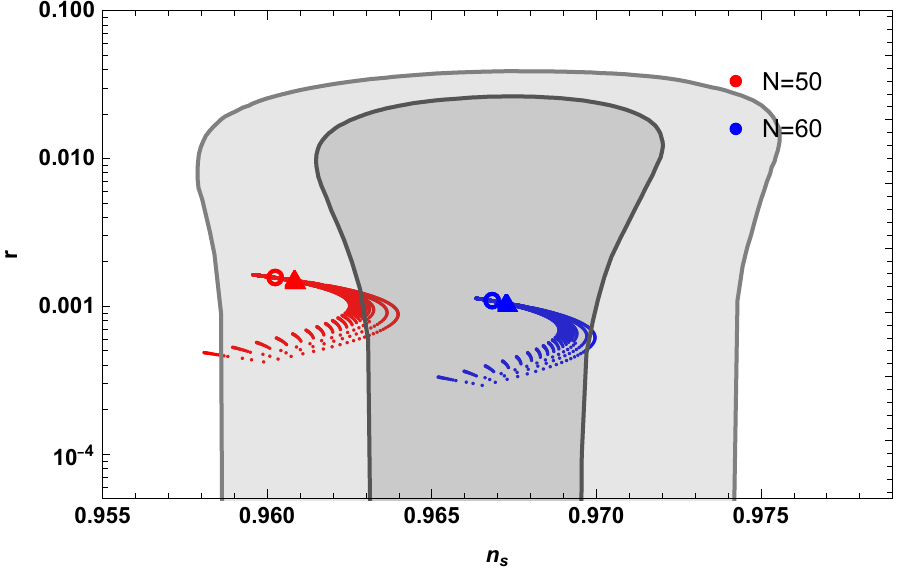}}
	\caption{The $r$ versus $n_s$ predictions for the  inflaton potential
		in Eq. \eqref{Eq:V_threeMG} with $A_3<0,~B_3<0$, and $A_3>2B_3$. Inflation occurs in the regions (a) $[-\infty,\chi_m]$ and (b) $[\chi_m,\infty]$. The circles and triangles correspond to T- and E-models, respectively.}\label{fig:nsr-threeG8}
\end{figure}
There are two minima at $\chi_{m1}=\chi_2$ and $\chi_{m2}=\chi_3$ and one maximum at $\chi_M=\chi_1$ 
for the potential with $A_3<0,~B_3>0$, and $A_3^2>4B_3$. Inflation occurs in the regions $[-\infty,\chi_{m1}]$ and $[\chi_{m2},\infty]$, except the regions $[\chi_{m1},\chi_M]$ and $[\chi_M,\chi_{m2}]$, where slow-roll inflation
 cannot be realized due to $\eta>1$ at $\chi_M$. The predicted  $n_s$ and $r$
are shown in Fig. \ref{fig:nsr-threeG5}. Moreover, inflation can happen on both the left and right sides of the minimum $\chi_m=\chi_1$ for the potential with  $A_3<0,~B_3>0,~A_3^2<4B_3$, where the predicted  $n_s$ and $r$  are shown in Fig. \ref{fig:nsr-threeG6}. while for $A_3<0,~B_3<0$, and $A_3<2B_3$, the potential is a maximum at $\chi_M=\chi_1$ and a minimum at $\chi_m=\chi_3$. Similarly, the possible inflationary trajectories are in the regions  $[-\infty,\chi_m]$ and $[\chi_m,\chi_M]$, 
and the predicted  $n_s$ and $r$ are shown in Fig. \ref{fig:nsr-threeG7}. Especially, 
the relationship for parameters is $A\sim2B$ in the region $[\chi_m,\chi_M]$. At last, for $A_3<0,~B_3<0$, and $A_3>2B_3$, the potential only has one minimum at $\chi_m=\chi_3$,
and the predicted $n_s$ and $r$ are shown in Fig. \ref{fig:nsr-threeG8}. On the right side  of the minimum, the predictions of T- and E-models can also be covered. For the rest of the parameter spaces,	there is no extreme point for the potential, and 
the slow-roll inflation is not possible on any trajectories.

\section{Conclusion}\label{sec:conclusion}

We have studied three classes of no-scale inflation models with one, two, and three moduli
which can be realized naturally via string compactifications.
Also, we considered the general renormalizable superpotential as a three-order polynomials of the inflaton field.   
The E-model and T-model for a fixed $\alpha$ are realized in the one modulus model and the three moduli model, respectively.
They are connected by the three moduli model in the limits $2a_2\sqrt{c_1}/a_1\to 1$ 
and $2a_2\sqrt{c_1}/a_1\to \infty$. The detailed analyses of the spectral indices and the tensor-to-scalar ratio 
have been preformed, and they are consistent with the Planck and BICEP/Keck experimental data 
on the cosmic microwave background.  The spectral index is  $n_s\simeq 1-2/N \sim 0.965$ 
for all models. Similar to the Starobinsky model, the 
tensor-to-scalar ratio $r$ in the one and three moduli models are $r \simeq 2/(b^2N^2)$, whereas  $r$ predicted in the two moduli models are  $r\simeq 83/N^4$ due to the non-negligible contribution from the higher order term in potential. The parameter $b$ is defined in terms of $N_1$ as $d=1/\sqrt{2N_1}$. Thus, we have $b=1/\sqrt{6}$ for one modulus models and $b=1/\sqrt{2}$ for three moduli models. 
The tensor-to-scalar ratio in models with quartic and quadratic potential 
is significantly suppressed in no-scale supergravity,  and then they can satisfy 
the strong bound from the current observations $r_{0.05}<0.036$. 
In other words, no-scale supergravity is a viable framework which makes the excluded models valid again. 
In the three moduli model, the scalar potential is similar to that in global supersymmetry, 
but the K\"ahler potential is different. 
Thus, such no-scale supergravity becomes a bridge between supergravity and global supersymmetry.

\begin{acknowledgments}

This work is supported in part by 
the National Key Research and Development Program of China Grant No. 2020YFC2201504, 
by the Projects No. 11875062, No. 11947302, and No. 12047503 supported 
by the National Natural Science Foundation of China,
by the Major Program of the National Natural Science Foundation of China under Grant No. 11690021, 
as well as by the Key Research Program 
of the Chinese Academy of Sciences, Grant No. XDPB15. 

\end{acknowledgments}


\begin{thebibliography}{69}%
	\makeatletter
	\providecommand \@ifxundefined [1]{%
		\@ifx{#1\undefined}
	}%
	\providecommand \@ifnum [1]{%
		\ifnum #1\expandafter \@firstoftwo
		\else \expandafter \@secondoftwo
		\fi
	}%
	\providecommand \@ifx [1]{%
		\ifx #1\expandafter \@firstoftwo
		\else \expandafter \@secondoftwo
		\fi
	}%
	\providecommand \natexlab [1]{#1}%
	\providecommand \enquote  [1]{``#1''}%
	\providecommand \bibnamefont  [1]{#1}%
	\providecommand \bibfnamefont [1]{#1}%
	\providecommand \citenamefont [1]{#1}%
	\providecommand \href@noop [0]{\@secondoftwo}%
	\providecommand \href [0]{\begingroup \@sanitize@url \@href}%
	\providecommand \@href[1]{\@@startlink{#1}\@@href}%
	\providecommand \@@href[1]{\endgroup#1\@@endlink}%
	\providecommand \@sanitize@url [0]{\catcode `\\12\catcode `\$12\catcode
		`\&12\catcode `\#12\catcode `\^12\catcode `\_12\catcode `\%12\relax}%
	\providecommand \@@startlink[1]{}%
	\providecommand \@@endlink[0]{}%
	\providecommand \url  [0]{\begingroup\@sanitize@url \@url }%
	\providecommand \@url [1]{\endgroup\@href {#1}{\urlprefix }}%
	\providecommand \urlprefix  [0]{URL }%
	\providecommand \Eprint [0]{\href }%
	\providecommand \doibase [0]{https://doi.org/}%
	\providecommand \selectlanguage [0]{\@gobble}%
	\providecommand \bibinfo  [0]{\@secondoftwo}%
	\providecommand \bibfield  [0]{\@secondoftwo}%
	\providecommand \translation [1]{[#1]}%
	\providecommand \BibitemOpen [0]{}%
	\providecommand \bibitemStop [0]{}%
	\providecommand \bibitemNoStop [0]{.\EOS\space}%
	\providecommand \EOS [0]{\spacefactor3000\relax}%
	\providecommand \BibitemShut  [1]{\csname bibitem#1\endcsname}%
	\let\auto@bib@innerbib\@empty
	\bibitem [{\citenamefont {Starobinsky}(1980)}]{Starobinsky:1980te}%
	\BibitemOpen
	\bibfield  {author} {\bibinfo {author} {\bibfnamefont {A.~A.}\ \bibnamefont
			{Starobinsky}},\ }\bibinfo {title} {{A New Type of Isotropic Cosmological
			Models Without Singularity}},\ \href
	{https://doi.org/10.1016/0370-2693(80)90670-X} {\bibfield  {journal}
		{\bibinfo  {journal} {Phys. Lett. B}\ }\textbf {\bibinfo {volume} {91}},\
		\bibinfo {pages} {99} (\bibinfo {year} {1980})}\BibitemShut {NoStop}%
	\bibitem [{\citenamefont {Guth}(1981)}]{Guth:1980zm}%
	\BibitemOpen
	\bibfield  {author} {\bibinfo {author} {\bibfnamefont {A.~H.}\ \bibnamefont
			{Guth}},\ }\bibinfo {title} {{The Inflationary Universe: A Possible Solution
			to the Horizon and Flatness Problems}},\ \href
	{https://doi.org/10.1103/PhysRevD.23.347} {\bibfield  {journal} {\bibinfo
			{journal} {Phys. Rev. D}\ }\textbf {\bibinfo {volume} {23}},\ \bibinfo
		{pages} {347} (\bibinfo {year} {1981})}\BibitemShut {NoStop}%
	\bibitem [{\citenamefont {Linde}(1982)}]{Linde:1981mu}%
	\BibitemOpen
	\bibfield  {author} {\bibinfo {author} {\bibfnamefont {A.~D.}\ \bibnamefont
			{Linde}},\ }\bibinfo {title} {{A New Inflationary Universe Scenario: A
			Possible Solution of the Horizon, Flatness, Homogeneity, Isotropy and
			Primordial Monopole Problems}},\ \href
	{https://doi.org/10.1016/0370-2693(82)91219-9} {\bibfield  {journal}
		{\bibinfo  {journal} {Phys. Lett. B}\ }\textbf {\bibinfo {volume} {108}},\
		\bibinfo {pages} {389} (\bibinfo {year} {1982})}\BibitemShut {NoStop}%
	\bibitem [{\citenamefont {Albrecht}\ and\ \citenamefont
		{Steinhardt}(1982)}]{Albrecht:1982wi}%
	\BibitemOpen
	\bibfield  {author} {\bibinfo {author} {\bibfnamefont {A.}~\bibnamefont
			{Albrecht}}\ and\ \bibinfo {author} {\bibfnamefont {P.~J.}\ \bibnamefont
			{Steinhardt}},\ }\bibinfo {title} {{Cosmology for Grand Unified Theories with
			Radiatively Induced Symmetry Breaking}},\ \href
	{https://doi.org/10.1103/PhysRevLett.48.1220} {\bibfield  {journal} {\bibinfo
			{journal} {Phys. Rev. Lett.}\ }\textbf {\bibinfo {volume} {48}},\ \bibinfo
		{pages} {1220} (\bibinfo {year} {1982})}\BibitemShut {NoStop}%
	\bibitem [{\citenamefont {Lyth}\ and\ \citenamefont
		{Riotto}(1999)}]{Lyth:1998xn}%
	\BibitemOpen
	\bibfield  {author} {\bibinfo {author} {\bibfnamefont {D.~H.}\ \bibnamefont
			{Lyth}}\ and\ \bibinfo {author} {\bibfnamefont {A.}~\bibnamefont {Riotto}},\
	}\bibinfo {title} {{Particle physics models of inflation and the cosmological
			density perturbation}},\ \href
	{https://doi.org/10.1016/S0370-1573(98)00128-8} {\bibfield  {journal}
		{\bibinfo  {journal} {Phys. Rept.}\ }\textbf {\bibinfo {volume} {314}},\
		\bibinfo {pages} {1} (\bibinfo {year} {1999})},\ \Eprint
	{https://arxiv.org/abs/hep-ph/9807278} {arXiv:hep-ph/9807278} \BibitemShut
	{NoStop}%
	\bibitem [{\citenamefont {Akrami}\ {\it et~al.}(2020)\citenamefont {Akrami}
		{\it et~al.}}]{Akrami:2018odb}%
	\BibitemOpen
	\bibfield  {author} {\bibinfo {author} {\bibfnamefont {Y.}~\bibnamefont
			{Akrami}} {\it et~al.} (\bibinfo {collaboration} {Planck Collaboration}),\
	}\bibinfo {title} {{Planck 2018 results. X. Constraints on inflation}},\
	\href {https://doi.org/10.1051/0004-6361/201833887} {\bibfield  {journal}
		{\bibinfo  {journal} {Astron. Astrophys.}\ }\textbf {\bibinfo {volume}
			{641}},\ \bibinfo {pages} {A10} (\bibinfo {year} {2020})},\ \Eprint
	{https://arxiv.org/abs/1807.06211} {arXiv:1807.06211 [astro-ph.CO]}
	\BibitemShut {NoStop}%
	\bibitem [{\citenamefont {Ade}\ {\it et~al.}(2021)\citenamefont {Ade} {\it
			et~al.}}]{BICEP:2021xfz}%
	\BibitemOpen
	\bibfield  {author} {\bibinfo {author} {\bibfnamefont {P.~A.~R.}\
			\bibnamefont {Ade}} {\it et~al.} (\bibinfo {collaboration} {BICEP and Keck
			Array Collaborations}),\ }\bibinfo {title} {{Improved Constraints on
			Primordial Gravitational Waves using Planck, WMAP, and BICEP/Keck
			Observations through the 2018 Observing Season}},\ \href
	{https://doi.org/10.1103/PhysRevLett.127.151301} {\bibfield  {journal}
		{\bibinfo  {journal} {Phys. Rev. Lett.}\ }\textbf {\bibinfo {volume} {127}},\
		\bibinfo {pages} {151301} (\bibinfo {year} {2021})},\ \Eprint
	{https://arxiv.org/abs/2110.00483} {arXiv:2110.00483 [astro-ph.CO]}
	\BibitemShut {NoStop}%
	\bibitem [{\citenamefont {Burgess}\ {\it et~al.}(2013)\citenamefont {Burgess},
		\citenamefont {Cicoli},\ and\ \citenamefont {Quevedo}}]{Burgess:2013sla}%
	\BibitemOpen
	\bibfield  {author} {\bibinfo {author} {\bibfnamefont {C.~P.}\ \bibnamefont
			{Burgess}}, \bibinfo {author} {\bibfnamefont {M.}~\bibnamefont {Cicoli}},\
		and\ \bibinfo {author} {\bibfnamefont {F.}~\bibnamefont {Quevedo}},\
	}\bibinfo {title} {{String Inflation After Planck 2013}},\ \href
	{https://doi.org/10.1088/1475-7516/2013/11/003} {J. Cosmol. Astropart. Phys.\
		\bibinfo {volume} {11}\bibfield  {year} {\bibinfo  {year} { (\textbf
				{2013})}\ }\bibfield  {pages} {\bibinfo  {pages} {003}},\ }\Eprint
	{https://arxiv.org/abs/1306.3512} {arXiv:1306.3512 [hep-th]} \BibitemShut
	{NoStop}%
	\bibitem [{\citenamefont {Bhattacharya}\ {\it et~al.}(2018)\citenamefont
		{Bhattacharya}, \citenamefont {Dutta}, \citenamefont {Gangopadhyay},\ and\
		\citenamefont {Maharana}}]{Bhattacharya:2017pws}%
	\BibitemOpen
	\bibfield  {author} {\bibinfo {author} {\bibfnamefont {S.}~\bibnamefont
			{Bhattacharya}}, \bibinfo {author} {\bibfnamefont {K.}~\bibnamefont {Dutta}},
		\bibinfo {author} {\bibfnamefont {M.~R.}\ \bibnamefont {Gangopadhyay}},\ and\
		\bibinfo {author} {\bibfnamefont {A.}~\bibnamefont {Maharana}},\ }\bibinfo
	{title} {{Confronting K\"ahler moduli inflation with CMB data}},\ \href
	{https://doi.org/10.1103/PhysRevD.97.123533} {\bibfield  {journal} {\bibinfo
			{journal} {Phys. Rev. D}\ }\textbf {\bibinfo {volume} {97}},\ \bibinfo
		{pages} {123533} (\bibinfo {year} {2018})},\ \Eprint
	{https://arxiv.org/abs/1711.04807} {arXiv:1711.04807 [astro-ph.CO]}
	\BibitemShut {NoStop}%
	\bibitem [{\citenamefont {Cicoli}\ {\it et~al.}(2018)\citenamefont {Cicoli},
		\citenamefont {Diaz},\ and\ \citenamefont {Pedro}}]{Cicoli:2018asa}%
	\BibitemOpen
	\bibfield  {author} {\bibinfo {author} {\bibfnamefont {M.}~\bibnamefont
			{Cicoli}}, \bibinfo {author} {\bibfnamefont {V.~A.}\ \bibnamefont {Diaz}},\
		and\ \bibinfo {author} {\bibfnamefont {F.~G.}\ \bibnamefont {Pedro}},\
	}\bibinfo {title} {{Primordial Black Holes from String Inflation}},\ \href
	{https://doi.org/10.1088/1475-7516/2018/06/034} {J. Cosmol. Astropart. Phys.\
		\bibinfo {volume} {06}\bibfield  {year} {\bibinfo  {year} { (\textbf
				{2018})}\ }\bibfield  {pages} {\bibinfo  {pages} {034}},\ }\Eprint
	{https://arxiv.org/abs/1803.02837} {arXiv:1803.02837 [hep-th]} \BibitemShut
	{NoStop}%
	\bibitem [{\citenamefont {Kallosh}\ and\ \citenamefont
		{Linde}(2021)}]{Kallosh:2021mnu}%
	\BibitemOpen
	\bibfield  {author} {\bibinfo {author} {\bibfnamefont {R.}~\bibnamefont
			{Kallosh}}\ and\ \bibinfo {author} {\bibfnamefont {A.}~\bibnamefont
			{Linde}},\ }\bibinfo {title} {{BICEP/Keck and cosmological attractors}},\
	\href {https://doi.org/10.1088/1475-7516/2021/12/008} {J. Cosmol. Astropart.
		Phys.\ \bibinfo {volume} {12}\bibfield  {year} {\bibinfo  {year} { (\textbf
				{2021})}\ }\bibfield  {pages} {\bibinfo  {pages} {008}},\ }\Eprint
	{https://arxiv.org/abs/2110.10902} {arXiv:2110.10902 [astro-ph.CO]}
	\BibitemShut {NoStop}%
	\bibitem [{\citenamefont {Ellis}\ {\it et~al.}(2022)\citenamefont {Ellis},
		\citenamefont {Garcia}, \citenamefont {Nanopoulos}, \citenamefont {Olive},\
		and\ \citenamefont {Verner}}]{Ellis:2021kad}%
	\BibitemOpen
	\bibfield  {author} {\bibinfo {author} {\bibfnamefont {J.}~\bibnamefont
			{Ellis}}, \bibinfo {author} {\bibfnamefont {M.~A.~G.}\ \bibnamefont
			{Garcia}}, \bibinfo {author} {\bibfnamefont {D.~V.}\ \bibnamefont
			{Nanopoulos}}, \bibinfo {author} {\bibfnamefont {K.~A.}\ \bibnamefont
			{Olive}},\ and\ \bibinfo {author} {\bibfnamefont {S.}~\bibnamefont
			{Verner}},\ }\bibinfo {title} {{BICEP/Keck constraints on attractor models of
			inflation and reheating}},\ \href
	{https://doi.org/10.1103/PhysRevD.105.043504} {\bibfield  {journal} {\bibinfo
			{journal} {Phys. Rev. D}\ }\textbf {\bibinfo {volume} {105}},\ \bibinfo
		{pages} {043504} (\bibinfo {year} {2022})},\ \Eprint
	{https://arxiv.org/abs/2112.04466} {arXiv:2112.04466 [hep-ph]} \BibitemShut
	{NoStop}%
	\bibitem [{\citenamefont {Copeland}\ {\it et~al.}(1994)\citenamefont
		{Copeland}, \citenamefont {Liddle}, \citenamefont {Lyth}, \citenamefont
		{Stewart},\ and\ \citenamefont {Wands}}]{Copeland:1994vg}%
	\BibitemOpen
	\bibfield  {author} {\bibinfo {author} {\bibfnamefont {E.~J.}\ \bibnamefont
			{Copeland}}, \bibinfo {author} {\bibfnamefont {A.~R.}\ \bibnamefont
			{Liddle}}, \bibinfo {author} {\bibfnamefont {D.~H.}\ \bibnamefont {Lyth}},
		\bibinfo {author} {\bibfnamefont {E.~D.}\ \bibnamefont {Stewart}},\ and\
		\bibinfo {author} {\bibfnamefont {D.}~\bibnamefont {Wands}},\ }\bibinfo
	{title} {{False vacuum inflation with Einstein gravity}},\ \href
	{https://doi.org/10.1103/PhysRevD.49.6410} {\bibfield  {journal} {\bibinfo
			{journal} {Phys. Rev. D}\ }\textbf {\bibinfo {volume} {49}},\ \bibinfo
		{pages} {6410} (\bibinfo {year} {1994})},\ \Eprint
	{https://arxiv.org/abs/astro-ph/9401011} {arXiv:astro-ph/9401011}
	\BibitemShut {NoStop}%
	\bibitem [{\citenamefont {Stewart}(1995)}]{Stewart:1994ts}%
	\BibitemOpen
	\bibfield  {author} {\bibinfo {author} {\bibfnamefont {E.~D.}\ \bibnamefont
			{Stewart}},\ }\bibinfo {title} {{Inflation, supergravity and superstrings}},\
	\href {https://doi.org/10.1103/PhysRevD.51.6847} {\bibfield  {journal}
		{\bibinfo  {journal} {Phys. Rev. D}\ }\textbf {\bibinfo {volume} {51}},\
		\bibinfo {pages} {6847} (\bibinfo {year} {1995})},\ \Eprint
	{https://arxiv.org/abs/hep-ph/9405389} {arXiv:hep-ph/9405389} \BibitemShut
	{NoStop}%
	\bibitem [{\citenamefont {Dine}\ {\it et~al.}(1995)\citenamefont {Dine},
		\citenamefont {Randall},\ and\ \citenamefont {Thomas}}]{Dine:1995uk}%
	\BibitemOpen
	\bibfield  {author} {\bibinfo {author} {\bibfnamefont {M.}~\bibnamefont
			{Dine}}, \bibinfo {author} {\bibfnamefont {L.}~\bibnamefont {Randall}},\ and\
		\bibinfo {author} {\bibfnamefont {S.~D.}\ \bibnamefont {Thomas}},\ }\bibinfo
	{title} {{Supersymmetry breaking in the early universe}},\ \href
	{https://doi.org/10.1103/PhysRevLett.75.398} {\bibfield  {journal} {\bibinfo
			{journal} {Phys. Rev. Lett.}\ }\textbf {\bibinfo {volume} {75}},\ \bibinfo
		{pages} {398} (\bibinfo {year} {1995})},\ \Eprint
	{https://arxiv.org/abs/hep-ph/9503303} {arXiv:hep-ph/9503303} \BibitemShut
	{NoStop}%
	\bibitem [{\citenamefont {Cremmer}\ {\it et~al.}(1983)\citenamefont {Cremmer},
		\citenamefont {Ferrara}, \citenamefont {Kounnas},\ and\ \citenamefont
		{Nanopoulos}}]{Cremmer:1983bf}%
	\BibitemOpen
	\bibfield  {author} {\bibinfo {author} {\bibfnamefont {E.}~\bibnamefont
			{Cremmer}}, \bibinfo {author} {\bibfnamefont {S.}~\bibnamefont {Ferrara}},
		\bibinfo {author} {\bibfnamefont {C.}~\bibnamefont {Kounnas}},\ and\ \bibinfo
		{author} {\bibfnamefont {D.~V.}\ \bibnamefont {Nanopoulos}},\ }\bibinfo
	{title} {{Naturally Vanishing Cosmological Constant in N=1 Supergravity}},\
	\href {https://doi.org/10.1016/0370-2693(83)90106-5} {\bibfield  {journal}
		{\bibinfo  {journal} {Phys. Lett. B}\ }\textbf {\bibinfo {volume} {133}},\
		\bibinfo {pages} {61} (\bibinfo {year} {1983})}\BibitemShut {NoStop}%
	\bibitem [{\citenamefont {Ellis}\ {\it et~al.}(1984)\citenamefont {Ellis},
		\citenamefont {Lahanas}, \citenamefont {Nanopoulos},\ and\ \citenamefont
		{Tamvakis}}]{Ellis:1983sf}%
	\BibitemOpen
	\bibfield  {author} {\bibinfo {author} {\bibfnamefont {J.~R.}\ \bibnamefont
			{Ellis}}, \bibinfo {author} {\bibfnamefont {A.~B.}\ \bibnamefont {Lahanas}},
		\bibinfo {author} {\bibfnamefont {D.~V.}\ \bibnamefont {Nanopoulos}},\ and\
		\bibinfo {author} {\bibfnamefont {K.}~\bibnamefont {Tamvakis}},\ }\bibinfo
	{title} {{No-Scale Supersymmetric Standard Model}},\ \href
	{https://doi.org/10.1016/0370-2693(84)91378-9} {\bibfield  {journal}
		{\bibinfo  {journal} {Phys. Lett. B}\ }\textbf {\bibinfo {volume} {134}},\
		\bibinfo {pages} {429} (\bibinfo {year} {1984})}\BibitemShut {NoStop}%
	\bibitem [{\citenamefont {Lahanas}\ and\ \citenamefont
		{Nanopoulos}(1987)}]{Lahanas:1986uc}%
	\BibitemOpen
	\bibfield  {author} {\bibinfo {author} {\bibfnamefont {A.~B.}\ \bibnamefont
			{Lahanas}}\ and\ \bibinfo {author} {\bibfnamefont {D.~V.}\ \bibnamefont
			{Nanopoulos}},\ }\bibinfo {title} {{The Road to No Scale Supergravity}},\
	\href {https://doi.org/10.1016/0370-1573(87)90034-2} {\bibfield  {journal}
		{\bibinfo  {journal} {Phys. Rept.}\ }\textbf {\bibinfo {volume} {145}},\
		\bibinfo {pages} {1} (\bibinfo {year} {1987})}\BibitemShut {NoStop}%
	\bibitem [{\citenamefont {Witten}(1985)}]{Witten:1985xb}%
	\BibitemOpen
	\bibfield  {author} {\bibinfo {author} {\bibfnamefont {E.}~\bibnamefont
			{Witten}},\ }\bibinfo {title} {{Dimensional Reduction of Superstring
			Models}},\ \href {https://doi.org/10.1016/0370-2693(85)90976-1} {\bibfield
		{journal} {\bibinfo  {journal} {Phys. Lett. B}\ }\textbf {\bibinfo {volume}
			{155}},\ \bibinfo {pages} {151} (\bibinfo {year} {1985})}\BibitemShut
	{NoStop}%
	\bibitem [{\citenamefont {Li}\ {\it et~al.}(1997)\citenamefont {Li},
		\citenamefont {Lopez},\ and\ \citenamefont {Nanopoulos}}]{Li:1997sk}%
	\BibitemOpen
	\bibfield  {author} {\bibinfo {author} {\bibfnamefont {T.-j.}\ \bibnamefont
			{Li}}, \bibinfo {author} {\bibfnamefont {J.~L.}\ \bibnamefont {Lopez}},\ and\
		\bibinfo {author} {\bibfnamefont {D.~V.}\ \bibnamefont {Nanopoulos}},\
	}\bibinfo {title} {{Compactifications of M-theory and their phenomenological
			consequences}},\ \href {https://doi.org/10.1103/PhysRevD.56.2602} {\bibfield
		{journal} {\bibinfo  {journal} {Phys. Rev. D}\ }\textbf {\bibinfo {volume}
			{56}},\ \bibinfo {pages} {2602} (\bibinfo {year} {1997})},\ \Eprint
	{https://arxiv.org/abs/hep-ph/9704247} {arXiv:hep-ph/9704247} \BibitemShut
	{NoStop}%
	\bibitem [{\citenamefont {Ellis}\ {\it et~al.}(2013{\natexlab{a}})\citenamefont
		{Ellis}, \citenamefont {Nanopoulos},\ and\ \citenamefont
		{Olive}}]{Ellis:2013xoa}%
	\BibitemOpen
	\bibfield  {author} {\bibinfo {author} {\bibfnamefont {J.}~\bibnamefont
			{Ellis}}, \bibinfo {author} {\bibfnamefont {D.~V.}\ \bibnamefont
			{Nanopoulos}},\ and\ \bibinfo {author} {\bibfnamefont {K.~A.}\ \bibnamefont
			{Olive}},\ }\bibinfo {title} {{No-Scale Supergravity Realization of the
			Starobinsky Model of Inflation}},\ \href
	{https://doi.org/10.1103/PhysRevLett.111.111301} {\bibfield  {journal}
		{\bibinfo  {journal} {Phys. Rev. Lett.}\ }\textbf {\bibinfo {volume} {111}},\
		\bibinfo {pages} {111301} (\bibinfo {year} {2013}{\natexlab{a}})},\ \bibinfo
	{note} {[Erratum: Phys.Rev.Lett. 111, 129902 (2013)]},\ \Eprint
	{https://arxiv.org/abs/1305.1247} {arXiv:1305.1247 [hep-th]} \BibitemShut
	{NoStop}%
	\bibitem [{\citenamefont {Ellis}\ {\it et~al.}(2013{\natexlab{b}})\citenamefont
		{Ellis}, \citenamefont {Nanopoulos},\ and\ \citenamefont
		{Olive}}]{Ellis:2013nxa}%
	\BibitemOpen
	\bibfield  {author} {\bibinfo {author} {\bibfnamefont {J.}~\bibnamefont
			{Ellis}}, \bibinfo {author} {\bibfnamefont {D.~V.}\ \bibnamefont
			{Nanopoulos}},\ and\ \bibinfo {author} {\bibfnamefont {K.~A.}\ \bibnamefont
			{Olive}},\ }\bibinfo {title} {{Starobinsky-like Inflationary Models as
			Avatars of No-Scale Supergravity}},\ \href
	{https://doi.org/10.1088/1475-7516/2013/10/009} {J. Cosmol. Astropart. Phys.\
		\bibinfo {volume} {10}\bibfield  {year} {\bibinfo  {year} { (\textbf
				{2013})}\ }\bibfield  {pages} {\bibinfo  {pages} {009}},\ }\Eprint
	{https://arxiv.org/abs/1307.3537} {arXiv:1307.3537 [hep-th]} \BibitemShut
	{NoStop}%
	\bibitem [{\citenamefont {Ellis}\ {\it et~al.}(2016)\citenamefont {Ellis},
		\citenamefont {Garcia}, \citenamefont {Nanopoulos},\ and\ \citenamefont
		{Olive}}]{Ellis:2015xna}%
	\BibitemOpen
	\bibfield  {author} {\bibinfo {author} {\bibfnamefont {J.}~\bibnamefont
			{Ellis}}, \bibinfo {author} {\bibfnamefont {M.~A.~G.}\ \bibnamefont
			{Garcia}}, \bibinfo {author} {\bibfnamefont {D.~V.}\ \bibnamefont
			{Nanopoulos}},\ and\ \bibinfo {author} {\bibfnamefont {K.~A.}\ \bibnamefont
			{Olive}},\ }\bibinfo {title} {{No-Scale Inflation}},\ \href
	{https://doi.org/10.1088/0264-9381/33/9/094001} {\bibfield  {journal}
		{\bibinfo  {journal} {Class. Quant. Grav.}\ }\textbf {\bibinfo {volume}
			{33}},\ \bibinfo {pages} {094001} (\bibinfo {year} {2016})},\ \Eprint
	{https://arxiv.org/abs/1507.02308} {arXiv:1507.02308 [hep-ph]} \BibitemShut
	{NoStop}%
	\bibitem [{\citenamefont {Ellis}\ {\it et~al.}(2020)\citenamefont {Ellis},
		\citenamefont {Garcia}, \citenamefont {Nagata}, \citenamefont {V.},
		\citenamefont {Olive},\ and\ \citenamefont {Verner}}]{Ellis:2020lnc}%
	\BibitemOpen
	\bibfield  {author} {\bibinfo {author} {\bibfnamefont {J.}~\bibnamefont
			{Ellis}}, \bibinfo {author} {\bibfnamefont {M.~A.~G.}\ \bibnamefont
			{Garcia}}, \bibinfo {author} {\bibfnamefont {N.}~\bibnamefont {Nagata}},
		\bibinfo {author} {\bibfnamefont {D.~V.}\ \bibnamefont
			{Nanopoulos}}, \bibinfo
		{author} {\bibfnamefont {K.~A.}\ \bibnamefont {Olive}},\ and\ \bibinfo
		{author} {\bibfnamefont {S.}~\bibnamefont {Verner}},\ }\bibinfo {title}
	{{Building models of inflation in no-scale supergravity}},\ \href
	{https://doi.org/10.1142/S0218271820300116} {\bibfield  {journal} {\bibinfo
			{journal} {Int. J. Mod. Phys. D}\ }\textbf {\bibinfo {volume} {29}},\
		\bibinfo {pages} {2030011} (\bibinfo {year} {2020})},\ \Eprint
	{https://arxiv.org/abs/2009.01709} {arXiv:2009.01709 [hep-ph]} \BibitemShut
	{NoStop}%
	\bibitem [{\citenamefont {Kawasaki}\ {\it et~al.}(2000)\citenamefont
		{Kawasaki}, \citenamefont {Yamaguchi},\ and\ \citenamefont
		{Yanagida}}]{Kawasaki:2000yn}%
	\BibitemOpen
	\bibfield  {author} {\bibinfo {author} {\bibfnamefont {M.}~\bibnamefont
			{Kawasaki}}, \bibinfo {author} {\bibfnamefont {M.}~\bibnamefont
			{Yamaguchi}},\ and\ \bibinfo {author} {\bibfnamefont {T.}~\bibnamefont
			{Yanagida}},\ }\bibinfo {title} {{Natural chaotic inflation in
			supergravity}},\ \href {https://doi.org/10.1103/PhysRevLett.85.3572}
	{\bibfield  {journal} {\bibinfo  {journal} {Phys. Rev. Lett.}\ }\textbf
		{\bibinfo {volume} {85}},\ \bibinfo {pages} {3572} (\bibinfo {year}
		{2000})},\ \Eprint {https://arxiv.org/abs/hep-ph/0004243}
	{arXiv:hep-ph/0004243} \BibitemShut {NoStop}%
	\bibitem [{\citenamefont {Yamaguchi}\ and\ \citenamefont
		{Yokoyama}(2001)}]{Yamaguchi:2000vm}%
	\BibitemOpen
	\bibfield  {author} {\bibinfo {author} {\bibfnamefont {M.}~\bibnamefont
			{Yamaguchi}}\ and\ \bibinfo {author} {\bibfnamefont {J.}~\bibnamefont
			{Yokoyama}},\ }\bibinfo {title} {{New inflation in supergravity with a
			chaotic initial condition}},\ \href
	{https://doi.org/10.1103/PhysRevD.63.043506} {\bibfield  {journal} {\bibinfo
			{journal} {Phys. Rev. D}\ }\textbf {\bibinfo {volume} {63}},\ \bibinfo
		{pages} {043506} (\bibinfo {year} {2001})},\ \Eprint
	{https://arxiv.org/abs/hep-ph/0007021} {arXiv:hep-ph/0007021} \BibitemShut
	{NoStop}%
	\bibitem [{\citenamefont {Yamaguchi}(2001)}]{Yamaguchi:2001pw}%
	\BibitemOpen
	\bibfield  {author} {\bibinfo {author} {\bibfnamefont {M.}~\bibnamefont
			{Yamaguchi}},\ }\bibinfo {title} {{Natural double inflation in
			supergravity}},\ \href {https://doi.org/10.1103/PhysRevD.64.063502}
	{\bibfield  {journal} {\bibinfo  {journal} {Phys. Rev. D}\ }\textbf {\bibinfo
			{volume} {64}},\ \bibinfo {pages} {063502} (\bibinfo {year} {2001})},\
	\Eprint {https://arxiv.org/abs/hep-ph/0103045} {arXiv:hep-ph/0103045}
	\BibitemShut {NoStop}%
	\bibitem [{\citenamefont {Kawasaki}\ and\ \citenamefont
		{Yamaguchi}(2002)}]{Kawasaki:2001as}%
	\BibitemOpen
	\bibfield  {author} {\bibinfo {author} {\bibfnamefont {M.}~\bibnamefont
			{Kawasaki}}\ and\ \bibinfo {author} {\bibfnamefont {M.}~\bibnamefont
			{Yamaguchi}},\ }\bibinfo {title} {{A Supersymmetric topological inflation
			model}},\ \href {https://doi.org/10.1103/PhysRevD.65.103518} {\bibfield
		{journal} {\bibinfo  {journal} {Phys. Rev. D}\ }\textbf {\bibinfo {volume}
			{65}},\ \bibinfo {pages} {103518} (\bibinfo {year} {2002})},\ \Eprint
	{https://arxiv.org/abs/hep-ph/0112093} {arXiv:hep-ph/0112093} \BibitemShut
	{NoStop}%
	\bibitem [{\citenamefont {Kallosh}\ and\ \citenamefont
		{Linde}(2010)}]{Kallosh:2010ug}%
	\BibitemOpen
	\bibfield  {author} {\bibinfo {author} {\bibfnamefont {R.}~\bibnamefont
			{Kallosh}}\ and\ \bibinfo {author} {\bibfnamefont {A.}~\bibnamefont
			{Linde}},\ }\bibinfo {title} {{New models of chaotic inflation in
			supergravity}},\ \href {https://doi.org/10.1088/1475-7516/2010/11/011} {J.
		Cosmol. Astropart. Phys.\ \bibinfo {volume} {11}\bibfield  {year} {\bibinfo
			{year} { (\textbf {2010})}\ }\bibfield  {pages} {\bibinfo  {pages} {011}},\
	}\Eprint {https://arxiv.org/abs/1008.3375} {arXiv:1008.3375 [hep-th]}
	\BibitemShut {NoStop}%
	\bibitem [{\citenamefont {Kallosh}\ {\it et~al.}(2011)\citenamefont {Kallosh},
		\citenamefont {Linde},\ and\ \citenamefont {Rube}}]{Kallosh:2010xz}%
	\BibitemOpen
	\bibfield  {author} {\bibinfo {author} {\bibfnamefont {R.}~\bibnamefont
			{Kallosh}}, \bibinfo {author} {\bibfnamefont {A.}~\bibnamefont {Linde}},\
		and\ \bibinfo {author} {\bibfnamefont {T.}~\bibnamefont {Rube}},\ }\bibinfo
	{title} {{General inflaton potentials in supergravity}},\ \href
	{https://doi.org/10.1103/PhysRevD.83.043507} {\bibfield  {journal} {\bibinfo
			{journal} {Phys. Rev. D}\ }\textbf {\bibinfo {volume} {83}},\ \bibinfo
		{pages} {043507} (\bibinfo {year} {2011})},\ \Eprint
	{https://arxiv.org/abs/1011.5945} {arXiv:1011.5945 [hep-th]} \BibitemShut
	{NoStop}%
	\bibitem [{\citenamefont {Li}\ {\it et~al.}(2014{\natexlab{a}})\citenamefont
		{Li}, \citenamefont {Li},\ and\ \citenamefont {Nanopoulos}}]{Li:2013nfa}%
	\BibitemOpen
	\bibfield  {author} {\bibinfo {author} {\bibfnamefont {T.}~\bibnamefont
			{Li}}, \bibinfo {author} {\bibfnamefont {Z.}~\bibnamefont {Li}},\ and\
		\bibinfo {author} {\bibfnamefont {D.~V.}\ \bibnamefont {Nanopoulos}},\
	}\bibinfo {title} {{Supergravity Inflation with Broken Shift Symmetry and
			Large Tensor-to-Scalar Ratio}},\ \href
	{https://doi.org/10.1088/1475-7516/2014/02/028} {J. Cosmol. Astropart. Phys.\
		\bibinfo {volume} {02}\bibfield  {year} {\bibinfo  {year} { (\textbf
				{2014})}\ }\bibfield  {pages} {\bibinfo  {pages} {028}},\ }\Eprint
	{https://arxiv.org/abs/1311.6770} {arXiv:1311.6770 [hep-ph]} \BibitemShut
	{NoStop}%
	\bibitem [{\citenamefont {Nakayama}\ {\it
			et~al.}(2013{\natexlab{a}})\citenamefont {Nakayama}, \citenamefont
		{Takahashi},\ and\ \citenamefont {Yanagida}}]{Nakayama:2013jka}%
	\BibitemOpen
	\bibfield  {author} {\bibinfo {author} {\bibfnamefont {K.}~\bibnamefont
			{Nakayama}}, \bibinfo {author} {\bibfnamefont {F.}~\bibnamefont
			{Takahashi}},\ and\ \bibinfo {author} {\bibfnamefont {T.~T.}\ \bibnamefont
			{Yanagida}},\ }\bibinfo {title} {{Polynomial Chaotic Inflation in the Planck
			Era}},\ \href {https://doi.org/10.1016/j.physletb.2013.06.050} {\bibfield
		{journal} {\bibinfo  {journal} {Phys. Lett. B}\ }\textbf {\bibinfo {volume}
			{725}},\ \bibinfo {pages} {111} (\bibinfo {year} {2013}{\natexlab{a}})},\
	\Eprint {https://arxiv.org/abs/1303.7315} {arXiv:1303.7315 [hep-ph]}
	\BibitemShut {NoStop}%
	\bibitem [{\citenamefont {Nakayama}\ {\it
			et~al.}(2013{\natexlab{b}})\citenamefont {Nakayama}, \citenamefont
		{Takahashi},\ and\ \citenamefont {Yanagida}}]{Nakayama:2013txa}%
	\BibitemOpen
	\bibfield  {author} {\bibinfo {author} {\bibfnamefont {K.}~\bibnamefont
			{Nakayama}}, \bibinfo {author} {\bibfnamefont {F.}~\bibnamefont
			{Takahashi}},\ and\ \bibinfo {author} {\bibfnamefont {T.~T.}\ \bibnamefont
			{Yanagida}},\ }\bibinfo {title} {{Polynomial Chaotic Inflation in
			Supergravity}},\ \href {https://doi.org/10.1088/1475-7516/2013/08/038} {J.
		Cosmol. Astropart. Phys.\ \bibinfo {volume} {08}\bibfield  {year} {\bibinfo
			{year} { (\textbf {2013})}\ }\bibfield  {pages} {\bibinfo  {pages} {038}},\
	}\Eprint {https://arxiv.org/abs/1305.5099} {arXiv:1305.5099 [hep-ph]}
	\BibitemShut {NoStop}%
	\bibitem [{\citenamefont {Takahashi}(2013)}]{Takahashi:2013cxa}%
	\BibitemOpen
	\bibfield  {author} {\bibinfo {author} {\bibfnamefont {F.}~\bibnamefont
			{Takahashi}},\ }\bibinfo {title} {{New inflation in supergravity after Planck
			and LHC}},\ \href {https://doi.org/10.1016/j.physletb.2013.10.026} {\bibfield
		{journal} {\bibinfo  {journal} {Phys. Lett. B}\ }\textbf {\bibinfo {volume}
			{727}},\ \bibinfo {pages} {21} (\bibinfo {year} {2013})},\ \Eprint
	{https://arxiv.org/abs/1308.4212} {arXiv:1308.4212 [hep-ph]} \BibitemShut
	{NoStop}%
	\bibitem [{\citenamefont {Li}\ {\it et~al.}(2014{\natexlab{b}})\citenamefont
		{Li}, \citenamefont {Li},\ and\ \citenamefont {Nanopoulos}}]{Li:2014xna}%
	\BibitemOpen
	\bibfield  {author} {\bibinfo {author} {\bibfnamefont {T.}~\bibnamefont
			{Li}}, \bibinfo {author} {\bibfnamefont {Z.}~\bibnamefont {Li}},\ and\
		\bibinfo {author} {\bibfnamefont {D.~V.}\ \bibnamefont {Nanopoulos}},\
	}\bibinfo {title} {{Natural Inflation with Natural Trans-Planckian Axion
			Decay Constant from Anomalous $U(1)_X$}},\ \href
	{https://doi.org/10.1007/JHEP07(2014)052} {J. High Energ. Phys.\ \bibinfo
		{volume} {07}\bibfield  {year} {\bibinfo  {year} { (\textbf {2014})}\
		}\bibfield  {pages} {\bibinfo  {pages} {052}},\ }\Eprint
	{https://arxiv.org/abs/1405.1804} {arXiv:1405.1804 [hep-th]} \BibitemShut
	{NoStop}%
	\bibitem [{\citenamefont {Pallis}\ and\ \citenamefont
		{Shafi}(2014)}]{Pallis:2014xva}%
	\BibitemOpen
	\bibfield  {author} {\bibinfo {author} {\bibfnamefont {C.}~\bibnamefont
			{Pallis}}\ and\ \bibinfo {author} {\bibfnamefont {Q.}~\bibnamefont {Shafi}},\
	}\bibinfo {title} {{From Hybrid to Quadratic Inflation With High-Scale
			Supersymmetry Breaking}},\ \href
	{https://doi.org/10.1016/j.physletb.2014.07.031} {\bibfield  {journal}
		{\bibinfo  {journal} {Phys. Lett. B}\ }\textbf {\bibinfo {volume} {736}},\
		\bibinfo {pages} {261} (\bibinfo {year} {2014})},\ \Eprint
	{https://arxiv.org/abs/1405.7645} {arXiv:1405.7645 [hep-ph]} \BibitemShut
	{NoStop}%
	\bibitem [{\citenamefont {Li}\ {\it et~al.}(2015{\natexlab{a}})\citenamefont
		{Li}, \citenamefont {Li},\ and\ \citenamefont {Nanopoulos}}]{Li:2015mwa}%
	\BibitemOpen
	\bibfield  {author} {\bibinfo {author} {\bibfnamefont {T.}~\bibnamefont
			{Li}}, \bibinfo {author} {\bibfnamefont {Z.}~\bibnamefont {Li}},\ and\
		\bibinfo {author} {\bibfnamefont {D.~V.}\ \bibnamefont {Nanopoulos}},\
	}\bibinfo {title} {{Symmetry Breaking Indication for Supergravity Inflation
			in Light of the Planck 2015}},\ \href
	{https://doi.org/10.1088/1475-7516/2015/09/006} {J. Cosmol. Astropart. Phys.\
		\bibinfo {volume} {09}\bibfield  {year} {\bibinfo  {year} { (\textbf
				{2015})}\ }\bibfield  {pages} {\bibinfo  {pages} {006}},\ }\Eprint
	{https://arxiv.org/abs/1502.05005} {arXiv:1502.05005 [hep-ph]} \BibitemShut
	{NoStop}%
	\bibitem [{\citenamefont {Li}\ {\it et~al.}(2015{\natexlab{b}})\citenamefont
		{Li}, \citenamefont {Li},\ and\ \citenamefont {Nanopoulos}}]{Li:2014vpa}%
	\BibitemOpen
	\bibfield  {author} {\bibinfo {author} {\bibfnamefont {T.}~\bibnamefont
			{Li}}, \bibinfo {author} {\bibfnamefont {Z.}~\bibnamefont {Li}},\ and\
		\bibinfo {author} {\bibfnamefont {D.~V.}\ \bibnamefont {Nanopoulos}},\
	}\bibinfo {title} {{Helical Phase Inflation}},\ \href
	{https://doi.org/10.1103/PhysRevD.91.061303} {\bibfield  {journal} {\bibinfo
			{journal} {Phys. Rev. D}\ }\textbf {\bibinfo {volume} {91}},\ \bibinfo
		{pages} {061303} (\bibinfo {year} {2015}{\natexlab{b}})},\ \Eprint
	{https://arxiv.org/abs/1409.3267} {arXiv:1409.3267 [hep-th]} \BibitemShut
	{NoStop}%
	\bibitem [{\citenamefont {Li}\ {\it et~al.}(2015{\natexlab{c}})\citenamefont
		{Li}, \citenamefont {Li},\ and\ \citenamefont {Nanopoulos}}]{Li:2014unh}%
	\BibitemOpen
	\bibfield  {author} {\bibinfo {author} {\bibfnamefont {T.}~\bibnamefont
			{Li}}, \bibinfo {author} {\bibfnamefont {Z.}~\bibnamefont {Li}},\ and\
		\bibinfo {author} {\bibfnamefont {D.~V.}\ \bibnamefont {Nanopoulos}},\
	}\bibinfo {title} {{Helical Phase Inflation and Monodromy in Supergravity
			Theory}},\ \href {https://doi.org/10.1155/2015/397410} {\bibfield  {journal}
		{\bibinfo  {journal} {Adv. High Energy Phys.}\ }\textbf {\bibinfo {volume}
			{2015}},\ \bibinfo {pages} {397410} (\bibinfo {year} {2015}{\natexlab{c}})},\
	\Eprint {https://arxiv.org/abs/1412.5093} {arXiv:1412.5093 [hep-th]}
	\BibitemShut {NoStop}%
	\bibitem [{\citenamefont {Li}\ {\it et~al.}(2015{\natexlab{d}})\citenamefont
		{Li}, \citenamefont {Li},\ and\ \citenamefont {Nanopoulos}}]{Li:2015taa}%
	\BibitemOpen
	\bibfield  {author} {\bibinfo {author} {\bibfnamefont {T.}~\bibnamefont
			{Li}}, \bibinfo {author} {\bibfnamefont {Z.}~\bibnamefont {Li}},\ and\
		\bibinfo {author} {\bibfnamefont {D.~V.}\ \bibnamefont {Nanopoulos}},\
	}\bibinfo {title} {{Helical Phase Inflation via Non-Geometric Flux
			Compactifications: from Natural to Starobinsky-like Inflation}},\ \href
	{https://doi.org/10.1007/JHEP10(2015)138} {J. High Energ. Phys.\ \bibinfo
		{volume} {10}\bibfield  {year} {\bibinfo  {year} { (\textbf {2015})}\
		}\bibfield  {pages} {\bibinfo  {pages} {138}},\ }\Eprint
	{https://arxiv.org/abs/1507.04687} {arXiv:1507.04687 [hep-th]} \BibitemShut
	{NoStop}%
	\bibitem [{\citenamefont {Sabir}\ {\it et~al.}(2020)\citenamefont {Sabir},
		\citenamefont {Ahmed}, \citenamefont {Gong}, \citenamefont {Li},\ and\
		\citenamefont {Lin}}]{Sabir:2019xwk}%
	\BibitemOpen
	\bibfield  {author} {\bibinfo {author} {\bibfnamefont {M.}~\bibnamefont
			{Sabir}}, \bibinfo {author} {\bibfnamefont {W.}~\bibnamefont {Ahmed}},
		\bibinfo {author} {\bibfnamefont {Y.}~\bibnamefont {Gong}}, \bibinfo {author}
		{\bibfnamefont {T.}~\bibnamefont {Li}},\ and\ \bibinfo {author}
		{\bibfnamefont {J.}~\bibnamefont {Lin}},\ }\bibinfo {title} {{Helical phase
			inflation and its observational constraints}},\ \href
	{https://doi.org/10.1088/1475-7516/2020/09/038} {J. Cosmol. Astropart. Phys.\
		\bibinfo {volume} {09}\bibfield  {year} {\bibinfo  {year} { (\textbf
				{2020})}\ }\bibfield  {pages} {\bibinfo  {pages} {038}},\ }\Eprint
	{https://arxiv.org/abs/1908.05201} {arXiv:1908.05201 [hep-ph]} \BibitemShut
	{NoStop}%
	\bibitem [{\citenamefont {Broy}\ {\it et~al.}(2016)\citenamefont {Broy},
		\citenamefont {Ciupke}, \citenamefont {Pedro},\ and\ \citenamefont
		{Westphal}}]{Broy:2015zba}%
	\BibitemOpen
	\bibfield  {author} {\bibinfo {author} {\bibfnamefont {B.~J.}\ \bibnamefont
			{Broy}}, \bibinfo {author} {\bibfnamefont {D.}~\bibnamefont {Ciupke}},
		\bibinfo {author} {\bibfnamefont {F.~G.}\ \bibnamefont {Pedro}},\ and\
		\bibinfo {author} {\bibfnamefont {A.}~\bibnamefont {Westphal}},\ }\bibinfo
	{title} {{Starobinsky-Type Inflation from $\alpha'$-Corrections}},\ \href
	{https://doi.org/10.1088/1475-7516/2016/01/001} {J. Cosmol. Astropart. Phys.\
		\bibinfo {volume} {01}\bibfield  {year} {\bibinfo  {year} { (\textbf
				{2016})}\ }\bibfield  {pages} {\bibinfo  {pages} {001}},\ }\Eprint
	{https://arxiv.org/abs/1509.00024} {arXiv:1509.00024 [hep-th]} \BibitemShut
	{NoStop}%
	\bibitem [{\citenamefont {Ellis}\ {\it et~al.}(2014)\citenamefont {Ellis},
		\citenamefont {Mavromatos},\ and\ \citenamefont
		{Nanopoulos}}]{Ellis:2014cma}%
	\BibitemOpen
	\bibfield  {author} {\bibinfo {author} {\bibfnamefont {J.}~\bibnamefont
			{Ellis}}, \bibinfo {author} {\bibfnamefont {N.~E.}\ \bibnamefont
			{Mavromatos}},\ and\ \bibinfo {author} {\bibfnamefont {D.~V.}\ \bibnamefont
			{Nanopoulos}},\ }\bibinfo {title} {{Starobinsky-Like Inflation in
			Dilaton-Brane Cosmology}},\ \href
	{https://doi.org/10.1016/j.physletb.2014.04.014} {\bibfield  {journal}
		{\bibinfo  {journal} {Phys. Lett. B}\ }\textbf {\bibinfo {volume} {732}},\
		\bibinfo {pages} {380} (\bibinfo {year} {2014})},\ \Eprint
	{https://arxiv.org/abs/1402.5075} {arXiv:1402.5075 [hep-th]} \BibitemShut
	{NoStop}%
	\bibitem [{\citenamefont {Cicoli}\ {\it et~al.}(2011)\citenamefont {Cicoli},
		\citenamefont {Pedro},\ and\ \citenamefont {Tasinato}}]{Cicoli:2011ct}%
	\BibitemOpen
	\bibfield  {author} {\bibinfo {author} {\bibfnamefont {M.}~\bibnamefont
			{Cicoli}}, \bibinfo {author} {\bibfnamefont {F.~G.}\ \bibnamefont {Pedro}},\
		and\ \bibinfo {author} {\bibfnamefont {G.}~\bibnamefont {Tasinato}},\
	}\bibinfo {title} {{Poly-instanton Inflation}},\ \href
	{https://doi.org/10.1088/1475-7516/2011/12/022} {J. Cosmol. Astropart. Phys.\
		\bibinfo {volume} {12}\bibfield  {year} {\bibinfo  {year} { (\textbf
				{2011})}\ }\bibfield  {pages} {\bibinfo  {pages} {022}},\ }\Eprint
	{https://arxiv.org/abs/1110.6182} {arXiv:1110.6182 [hep-th]} \BibitemShut
	{NoStop}%
	\bibitem [{\citenamefont {Gao}\ and\ \citenamefont
		{Shukla}(2013)}]{Gao:2013hn}%
	\BibitemOpen
	\bibfield  {author} {\bibinfo {author} {\bibfnamefont {X.}~\bibnamefont
			{Gao}}\ and\ \bibinfo {author} {\bibfnamefont {P.}~\bibnamefont {Shukla}},\
	}\bibinfo {title} {{On Non-Gaussianities in Two-Field Poly-Instanton
			Inflation}},\ \href {https://doi.org/10.1007/JHEP03(2013)061} {J. High Energ.
		Phys.\ \bibinfo {volume} {03}\bibfield  {year} {\bibinfo  {year} { (\textbf
				{2013})}\ }\bibfield  {pages} {\bibinfo  {pages} {061}},\ }\Eprint
	{https://arxiv.org/abs/1301.6076} {arXiv:1301.6076 [hep-th]} \BibitemShut
	{NoStop}%
	\bibitem [{\citenamefont {Cicoli}\ {\it et~al.}(2013)\citenamefont {Cicoli},
		\citenamefont {Downes},\ and\ \citenamefont {Dutta}}]{Cicoli:2013oba}%
	\BibitemOpen
	\bibfield  {author} {\bibinfo {author} {\bibfnamefont {M.}~\bibnamefont
			{Cicoli}}, \bibinfo {author} {\bibfnamefont {S.}~\bibnamefont {Downes}},\
		and\ \bibinfo {author} {\bibfnamefont {B.}~\bibnamefont {Dutta}},\ }\bibinfo
	{title} {{Power Suppression at Large Scales in String Inflation}},\ \href
	{https://doi.org/10.1088/1475-7516/2013/12/007} {J. Cosmol. Astropart. Phys.\
		\bibinfo {volume} {12}\bibfield  {year} {\bibinfo  {year} { (\textbf
				{2013})}\ }\bibfield  {pages} {\bibinfo  {pages} {007}},\ }\Eprint
	{https://arxiv.org/abs/1309.3412} {arXiv:1309.3412 [hep-th]} \BibitemShut
	{NoStop}%
	\bibitem [{\citenamefont {Kobayashi}\ {\it et~al.}(2017)\citenamefont
		{Kobayashi}, \citenamefont {Uemura},\ and\ \citenamefont
		{Yamamoto}}]{Kobayashi:2017jeb}%
	\BibitemOpen
	\bibfield  {author} {\bibinfo {author} {\bibfnamefont {T.}~\bibnamefont
			{Kobayashi}}, \bibinfo {author} {\bibfnamefont {S.}~\bibnamefont {Uemura}},\
		and\ \bibinfo {author} {\bibfnamefont {J.}~\bibnamefont {Yamamoto}},\
	}\bibinfo {title} {{Polyinstanton axion inflation}},\ \href
	{https://doi.org/10.1103/PhysRevD.96.026007} {\bibfield  {journal} {\bibinfo
			{journal} {Phys. Rev. D}\ }\textbf {\bibinfo {volume} {96}},\ \bibinfo
		{pages} {026007} (\bibinfo {year} {2017})},\ \Eprint
	{https://arxiv.org/abs/1705.04088} {arXiv:1705.04088 [hep-ph]} \BibitemShut
	{NoStop}%
	\bibitem [{\citenamefont {Bhattacharya}\ {\it et~al.}(2020)\citenamefont
		{Bhattacharya}, \citenamefont {Dutta}, \citenamefont {Gangopadhyay},
		\citenamefont {Maharana},\ and\ \citenamefont
		{Singh}}]{Bhattacharya:2020gnk}%
	\BibitemOpen
	\bibfield  {author} {\bibinfo {author} {\bibfnamefont {S.}~\bibnamefont
			{Bhattacharya}}, \bibinfo {author} {\bibfnamefont {K.}~\bibnamefont {Dutta}},
		\bibinfo {author} {\bibfnamefont {M.~R.}\ \bibnamefont {Gangopadhyay}},
		\bibinfo {author} {\bibfnamefont {A.}~\bibnamefont {Maharana}},\ and\
		\bibinfo {author} {\bibfnamefont {K.}~\bibnamefont {Singh}},\ }\bibinfo
	{title} {{Fibre Inflation and Precision CMB Data}},\ \href
	{https://doi.org/10.1103/PhysRevD.102.123531} {\bibfield  {journal} {\bibinfo
			{journal} {Phys. Rev. D}\ }\textbf {\bibinfo {volume} {102}},\ \bibinfo
		{pages} {123531} (\bibinfo {year} {2020})},\ \Eprint
	{https://arxiv.org/abs/2003.05969} {arXiv:2003.05969 [astro-ph.CO]}
	\BibitemShut {NoStop}%
	\bibitem [{\citenamefont {Cicoli}\ and\ \citenamefont
		{Di~Valentino}(2020)}]{Cicoli:2020bao}%
	\BibitemOpen
	\bibfield  {author} {\bibinfo {author} {\bibfnamefont {M.}~\bibnamefont
			{Cicoli}}\ and\ \bibinfo {author} {\bibfnamefont {E.}~\bibnamefont
			{Di~Valentino}},\ }\bibinfo {title} {{Fitting string inflation to real
			cosmological data: The fiber inflation case}},\ \href
	{https://doi.org/10.1103/PhysRevD.102.043521} {\bibfield  {journal} {\bibinfo
			{journal} {Phys. Rev. D}\ }\textbf {\bibinfo {volume} {102}},\ \bibinfo
		{pages} {043521} (\bibinfo {year} {2020})},\ \Eprint
	{https://arxiv.org/abs/2004.01210} {arXiv:2004.01210 [astro-ph.CO]}
	\BibitemShut {NoStop}%
	\bibitem [{\citenamefont {Cicoli}\ {\it et~al.}(2022)\citenamefont {Cicoli},
		\citenamefont {Pedro},\ and\ \citenamefont {Pedron}}]{Cicoli:2022sih}%
	\BibitemOpen
	\bibfield  {author} {\bibinfo {author} {\bibfnamefont {M.}~\bibnamefont
			{Cicoli}}, \bibinfo {author} {\bibfnamefont {F.~G.}\ \bibnamefont {Pedro}},\
		and\ \bibinfo {author} {\bibfnamefont {N.}~\bibnamefont {Pedron}},\ }\bibinfo
	{title} {{Secondary GWs and PBHs in string inflation: formation and
			detectability}},\ \Eprint {https://arxiv.org/abs/2203.00021}
	{arXiv:2203.00021 [hep-th]} \BibitemShut {NoStop}%
	\bibitem [{\citenamefont {Kallosh}\ {\it et~al.}(2013)\citenamefont {Kallosh},
		\citenamefont {Linde},\ and\ \citenamefont {Roest}}]{Kallosh:2013yoa}%
	\BibitemOpen
	\bibfield  {author} {\bibinfo {author} {\bibfnamefont {R.}~\bibnamefont
			{Kallosh}}, \bibinfo {author} {\bibfnamefont {A.}~\bibnamefont {Linde}},\
		and\ \bibinfo {author} {\bibfnamefont {D.}~\bibnamefont {Roest}},\ }\bibinfo
	{title} {{Superconformal Inflationary $\alpha$-Attractors}},\ \href
	{https://doi.org/10.1007/JHEP11(2013)198} {J. High Energ. Phys.\ \bibinfo
		{volume} {11}\bibfield  {year} {\bibinfo  {year} { (\textbf {2013})}\
		}\bibfield  {pages} {\bibinfo  {pages} {198}},\ }\Eprint
	{https://arxiv.org/abs/1311.0472} {arXiv:1311.0472 [hep-th]} \BibitemShut
	{NoStop}%
	\bibitem [{\citenamefont {Kallosh}\ and\ \citenamefont
		{Linde}(2015)}]{Kallosh:2015lwa}%
	\BibitemOpen
	\bibfield  {author} {\bibinfo {author} {\bibfnamefont {R.}~\bibnamefont
			{Kallosh}}\ and\ \bibinfo {author} {\bibfnamefont {A.}~\bibnamefont
			{Linde}},\ }\bibinfo {title} {{Planck, LHC, and $\alpha$-attractors}},\ \href
	{https://doi.org/10.1103/PhysRevD.91.083528} {\bibfield  {journal} {\bibinfo
			{journal} {Phys. Rev. D}\ }\textbf {\bibinfo {volume} {91}},\ \bibinfo
		{pages} {083528} (\bibinfo {year} {2015})},\ \Eprint
	{https://arxiv.org/abs/1502.07733} {arXiv:1502.07733 [astro-ph.CO]}
	\BibitemShut {NoStop}%
	\bibitem [{\citenamefont {Linde}(2015)}]{Linde:2015uga}%
	\BibitemOpen
	\bibfield  {author} {\bibinfo {author} {\bibfnamefont {A.}~\bibnamefont
			{Linde}},\ }\bibinfo {title} {{Single-field $\alpha$-attractors}},\ \href
	{https://doi.org/10.1088/1475-7516/2015/05/003} {J. Cosmol. Astropart. Phys.\
		\bibinfo {volume} {05}\bibfield  {year} {\bibinfo  {year} { (\textbf
				{2015})}\ }\bibfield  {pages} {\bibinfo  {pages} {003}},\ }\Eprint
	{https://arxiv.org/abs/1504.00663} {arXiv:1504.00663 [hep-th]} \BibitemShut
	{NoStop}%
	\bibitem [{\citenamefont {Linde}(2017)}]{Linde:2016uec}%
	\BibitemOpen
	\bibfield  {author} {\bibinfo {author} {\bibfnamefont {A.}~\bibnamefont
			{Linde}},\ }\bibinfo {title} {{Random Potentials and Cosmological
			Attractors}},\ \href {https://doi.org/10.1088/1475-7516/2017/02/028} {J.
		Cosmol. Astropart. Phys.\ \bibinfo {volume} {02}\bibfield  {year} {\bibinfo
			{year} { (\textbf {2017})}\ }\bibfield  {pages} {\bibinfo  {pages} {028}},\
	}\Eprint {https://arxiv.org/abs/1612.04505} {arXiv:1612.04505 [hep-th]}
	\BibitemShut {NoStop}%
	\bibitem [{\citenamefont {Ellis}\ {\it et~al.}(2019)\citenamefont {Ellis},
		\citenamefont {Nanopoulos}, \citenamefont {Olive},\ and\ \citenamefont
		{Verner}}]{Ellis:2019bmm}%
	\BibitemOpen
	\bibfield  {author} {\bibinfo {author} {\bibfnamefont {J.}~\bibnamefont
			{Ellis}}, \bibinfo {author} {\bibfnamefont {D.~V.}\ \bibnamefont
			{Nanopoulos}}, \bibinfo {author} {\bibfnamefont {K.~A.}\ \bibnamefont
			{Olive}},\ and\ \bibinfo {author} {\bibfnamefont {S.}~\bibnamefont
			{Verner}},\ }\bibinfo {title} {{Unified No-Scale Attractors}},\ \href
	{https://doi.org/10.1088/1475-7516/2019/09/040} {J. Cosmol. Astropart. Phys.\
		\bibinfo {volume} {09}\bibfield  {year} {\bibinfo  {year} { (\textbf
				{2019})}\ }\bibfield  {pages} {\bibinfo  {pages} {040}},\ }\Eprint
	{https://arxiv.org/abs/1906.10176} {arXiv:1906.10176 [hep-th]} \BibitemShut
	{NoStop}%
	\bibitem [{\citenamefont {Tang}\ and\ \citenamefont {Wu}(2020)}]{Tang:2019olx}%
	\BibitemOpen
	\bibfield  {author} {\bibinfo {author} {\bibfnamefont {Y.}~\bibnamefont
			{Tang}}\ and\ \bibinfo {author} {\bibfnamefont {Y.-L.}\ \bibnamefont {Wu}},\
	}\bibinfo {title} {{Conformal $\alpha$-attractor inflation with Weyl gauge
			field}},\ \href {https://doi.org/10.1088/1475-7516/2020/03/067} {J. Cosmol.
		Astropart. Phys.\ \bibinfo {volume} {03}\bibfield  {year} {\bibinfo  {year} {
				(\textbf {2020})}\ }\bibfield  {pages} {\bibinfo  {pages} {067}},\ }\Eprint
	{https://arxiv.org/abs/1912.07610} {arXiv:1912.07610 [hep-ph]} \BibitemShut
	{NoStop}%
	\bibitem [{\citenamefont {Akrami}\ {\it et~al.}(2021)\citenamefont {Akrami},
		\citenamefont {Casas}, \citenamefont {Deng},\ and\ \citenamefont
		{Vardanyan}}]{Akrami:2020zxw}%
	\BibitemOpen
	\bibfield  {author} {\bibinfo {author} {\bibfnamefont {Y.}~\bibnamefont
			{Akrami}}, \bibinfo {author} {\bibfnamefont {S.}~\bibnamefont {Casas}},
		\bibinfo {author} {\bibfnamefont {S.}~\bibnamefont {Deng}},\ and\ \bibinfo
		{author} {\bibfnamefont {V.}~\bibnamefont {Vardanyan}},\ }\bibinfo {title}
	{{Quintessential $\alpha$-attractor inflation: forecasts for Stage IV galaxy
			surveys}},\ \href {https://doi.org/10.1088/1475-7516/2021/04/006} {J. Cosmol.
		Astropart. Phys.\ \bibinfo {volume} {04}\bibfield  {year} {\bibinfo  {year} {
				(\textbf {2021})}\ }\bibfield  {pages} {\bibinfo  {pages} {006}},\ }\Eprint
	{https://arxiv.org/abs/2010.15822} {arXiv:2010.15822 [astro-ph.CO]}
	\BibitemShut {NoStop}%
	\bibitem [{\citenamefont {Rodrigues}\ {\it et~al.}(2021)\citenamefont
		{Rodrigues}, \citenamefont {Santos~da Costa},\ and\ \citenamefont
		{Alcaniz}}]{Rodrigues:2020fle}%
	\BibitemOpen
	\bibfield  {author} {\bibinfo {author} {\bibfnamefont {J.~G.}\ \bibnamefont
			{Rodrigues}}, \bibinfo {author} {\bibfnamefont {S.}~\bibnamefont {Santos~da
				Costa}},\ and\ \bibinfo {author} {\bibfnamefont {J.~S.}\ \bibnamefont
			{Alcaniz}},\ }\bibinfo {title} {{Observational constraints on $\alpha$
			-attractor inflationary models with a Higgs-like potential}},\ \href
	{https://doi.org/10.1016/j.physletb.2021.136156} {\bibfield  {journal}
		{\bibinfo  {journal} {Phys. Lett. B}\ }\textbf {\bibinfo {volume} {815}},\
		\bibinfo {pages} {136156} (\bibinfo {year} {2021})},\ \Eprint
	{https://arxiv.org/abs/2007.10763} {arXiv:2007.10763 [astro-ph.CO]}
	\BibitemShut {NoStop}%
	\bibitem [{\citenamefont {Li}(1998)}]{Li:1998sq}%
	\BibitemOpen
	\bibfield  {author} {\bibinfo {author} {\bibfnamefont {T.-j.}\ \bibnamefont
			{Li}},\ }\bibinfo {title} {{Compactification and supersymmetry breaking in M
			theory}},\ \href {https://doi.org/10.1103/PhysRevD.57.7539} {\bibfield
		{journal} {\bibinfo  {journal} {Phys. Rev. D}\ }\textbf {\bibinfo {volume}
			{57}},\ \bibinfo {pages} {7539} (\bibinfo {year} {1998})},\ \Eprint
	{https://arxiv.org/abs/hep-th/9801123} {arXiv:hep-th/9801123} \BibitemShut
	{NoStop}%
	\bibitem [{\citenamefont {Li}\ {\it et~al.}(2015{\natexlab{e}})\citenamefont
		{Li}, \citenamefont {Li},\ and\ \citenamefont {Nanopoulos}}]{Li:2014owa}%
	\BibitemOpen
	\bibfield  {author} {\bibinfo {author} {\bibfnamefont {T.}~\bibnamefont
			{Li}}, \bibinfo {author} {\bibfnamefont {Z.}~\bibnamefont {Li}},\ and\
		\bibinfo {author} {\bibfnamefont {D.~V.}\ \bibnamefont {Nanopoulos}},\
	}\bibinfo {title} {{Chaotic Inflation in No-Scale Supergravity with String
			Inspired Moduli Stabilization}},\ \href
	{https://doi.org/10.1140/epjc/s10052-015-3291-2} {\bibfield  {journal}
		{\bibinfo  {journal} {Eur. Phys. J. C}\ }\textbf {\bibinfo {volume} {75}},\
		\bibinfo {pages} {55} (\bibinfo {year} {2015}{\natexlab{e}})},\ \Eprint
	{https://arxiv.org/abs/1405.0197} {arXiv:1405.0197 [hep-th]} \BibitemShut
	{NoStop}%
	\bibitem [{\citenamefont {Wu}\ {\it et~al.}(2021)\citenamefont {Wu},
		\citenamefont {Gong},\ and\ \citenamefont {Li}}]{Wu:2021zta}%
	\BibitemOpen
	\bibfield  {author} {\bibinfo {author} {\bibfnamefont {L.}~\bibnamefont
			{Wu}}, \bibinfo {author} {\bibfnamefont {Y.}~\bibnamefont {Gong}},\ and\
		\bibinfo {author} {\bibfnamefont {T.}~\bibnamefont {Li}},\ }\bibinfo {title}
	{{Primordial black holes and secondary gravitational waves from string
			inspired general no-scale supergravity}},\ \href
	{https://doi.org/10.1103/PhysRevD.104.123544} {\bibfield  {journal} {\bibinfo
			{journal} {Phys. Rev. D}\ }\textbf {\bibinfo {volume} {104}},\ \bibinfo
		{pages} {123544} (\bibinfo {year} {2021})},\ \Eprint
	{https://arxiv.org/abs/2105.07694} {arXiv:2105.07694 [gr-qc]} \BibitemShut
	{NoStop}%
	\bibitem [{\citenamefont {Gaillard}\ {\it et~al.}(1995)\citenamefont
		{Gaillard}, \citenamefont {Murayama},\ and\ \citenamefont
		{Olive}}]{Gaillard:1995az}%
	\BibitemOpen
	\bibfield  {author} {\bibinfo {author} {\bibfnamefont {M.~K.}\ \bibnamefont
			{Gaillard}}, \bibinfo {author} {\bibfnamefont {H.}~\bibnamefont {Murayama}},\
		and\ \bibinfo {author} {\bibfnamefont {K.~A.}\ \bibnamefont {Olive}},\
	}\bibinfo {title} {{Preserving flat directions during inflation}},\ \href
	{https://doi.org/10.1016/0370-2693(95)00773-E} {\bibfield  {journal}
		{\bibinfo  {journal} {Phys. Lett. B}\ }\textbf {\bibinfo {volume} {355}},\
		\bibinfo {pages} {71} (\bibinfo {year} {1995})},\ \Eprint
	{https://arxiv.org/abs/hep-ph/9504307} {arXiv:hep-ph/9504307} \BibitemShut
	{NoStop}%
	\bibitem [{\citenamefont {Diamandis}\ {\it et~al.}(1986)\citenamefont
		{Diamandis}, \citenamefont {Ellis}, \citenamefont {Lahanas},\ and\
		\citenamefont {Nanopoulos}}]{Diamandis:1986zg}%
	\BibitemOpen
	\bibfield  {author} {\bibinfo {author} {\bibfnamefont {G.~A.}\ \bibnamefont
			{Diamandis}}, \bibinfo {author} {\bibfnamefont {J.~R.}\ \bibnamefont
			{Ellis}}, \bibinfo {author} {\bibfnamefont {A.~B.}\ \bibnamefont {Lahanas}},\
		and\ \bibinfo {author} {\bibfnamefont {D.~V.}\ \bibnamefont {Nanopoulos}},\
	}\bibinfo {title} {{Vanishing Scalar Masses in No Scale Supergravity}},\
	\href {https://doi.org/10.1016/0370-2693(86)90521-6} {\bibfield  {journal}
		{\bibinfo  {journal} {Phys. Lett. B}\ }\textbf {\bibinfo {volume} {173}},\
		\bibinfo {pages} {303} (\bibinfo {year} {1986})}\BibitemShut {NoStop}%
	\bibitem [{\citenamefont {Garg}\ and\ \citenamefont
		{Mohanty}(2015)}]{Garg:2015mra}%
	\BibitemOpen
	\bibfield  {author} {\bibinfo {author} {\bibfnamefont {I.}~\bibnamefont
			{Garg}}\ and\ \bibinfo {author} {\bibfnamefont {S.}~\bibnamefont {Mohanty}},\
	}\bibinfo {title} {{No scale SUGRA SO(10) derived Starobinsky Model of
			Inflation}},\ \href {https://doi.org/10.1016/j.physletb.2015.10.011}
	{\bibfield  {journal} {\bibinfo  {journal} {Phys. Lett. B}\ }\textbf
		{\bibinfo {volume} {751}},\ \bibinfo {pages} {7} (\bibinfo {year} {2015})},\
	\Eprint {https://arxiv.org/abs/1504.07725} {arXiv:1504.07725 [hep-ph]}
	\BibitemShut {NoStop}%
	\bibitem [{\citenamefont {Garg}\ and\ \citenamefont
		{Mohanty}(2018)}]{Garg:2017tds}%
	\BibitemOpen
	\bibfield  {author} {\bibinfo {author} {\bibfnamefont {I.}~\bibnamefont
			{Garg}}\ and\ \bibinfo {author} {\bibfnamefont {S.}~\bibnamefont {Mohanty}},\
	}\bibinfo {title} {{No-scale SUGRA Inflation and Type-I seesaw}},\ \href
	{https://doi.org/10.1142/S0217751X18501270} {\bibfield  {journal} {\bibinfo
			{journal} {Int. J. Mod. Phys. A}\ }\textbf {\bibinfo {volume} {33}},\
		\bibinfo {pages} {1850127} (\bibinfo {year} {2018})},\ \Eprint
	{https://arxiv.org/abs/1711.01979} {arXiv:1711.01979 [hep-ph]} \BibitemShut
	{NoStop}%
	\bibitem [{\citenamefont {Khalil}\ {\it et~al.}(2019)\citenamefont {Khalil},
		\citenamefont {Moursy}, \citenamefont {Saha},\ and\ \citenamefont
		{Sil}}]{Khalil:2018iip}%
	\BibitemOpen
	\bibfield  {author} {\bibinfo {author} {\bibfnamefont {S.}~\bibnamefont
			{Khalil}}, \bibinfo {author} {\bibfnamefont {A.}~\bibnamefont {Moursy}},
		\bibinfo {author} {\bibfnamefont {A.~K.}\ \bibnamefont {Saha}},\ and\
		\bibinfo {author} {\bibfnamefont {A.}~\bibnamefont {Sil}},\ }\bibinfo {title}
	{{$ U(1)_R$ inspired inflation model in no-scale supergravity}},\ \href
	{https://doi.org/10.1103/PhysRevD.99.095022} {\bibfield  {journal} {\bibinfo
			{journal} {Phys. Rev. D}\ }\textbf {\bibinfo {volume} {99}},\ \bibinfo
		{pages} {095022} (\bibinfo {year} {2019})},\ \Eprint
	{https://arxiv.org/abs/1810.06408} {arXiv:1810.06408 [hep-ph]} \BibitemShut
	{NoStop}%
	\bibitem [{\citenamefont {Linde}(1983)}]{Linde:1983gd}%
	\BibitemOpen
	\bibfield  {author} {\bibinfo {author} {\bibfnamefont {A.~D.}\ \bibnamefont
			{Linde}},\ }\bibinfo {title} {{Chaotic Inflation}},\ \href
	{https://doi.org/10.1016/0370-2693(83)90837-7} {\bibfield  {journal}
		{\bibinfo  {journal} {Phys. Lett. B}\ }\textbf {\bibinfo {volume} {129}},\
		\bibinfo {pages} {177} (\bibinfo {year} {1983})}\BibitemShut {NoStop}%
	\bibitem [{\citenamefont {Creminelli}\ {\it et~al.}(2015)\citenamefont
		{Creminelli}, \citenamefont {Dubovsky}, \citenamefont {L\'opez~Nacir},
		\citenamefont {Simonovi\'c}, \citenamefont {Trevisan}, \citenamefont
		{Villadoro},\ and\ \citenamefont {Zaldarriaga}}]{Creminelli:2014nqa}%
	\BibitemOpen
	\bibfield  {author} {\bibinfo {author} {\bibfnamefont {P.}~\bibnamefont
			{Creminelli}}, \bibinfo {author} {\bibfnamefont {S.}~\bibnamefont
			{Dubovsky}}, \bibinfo {author} {\bibfnamefont {D.}~\bibnamefont
			{L\'opez~Nacir}}, \bibinfo {author} {\bibfnamefont {M.}~\bibnamefont
			{Simonovi\'c}}, \bibinfo {author} {\bibfnamefont {G.}~\bibnamefont
			{Trevisan}}, \bibinfo {author} {\bibfnamefont {G.}~\bibnamefont
			{Villadoro}},\ and\ \bibinfo {author} {\bibfnamefont {M.}~\bibnamefont
			{Zaldarriaga}},\ }\bibinfo {title} {{Implications of the scalar tilt for the
			tensor-to-scalar ratio}},\ \href {https://doi.org/10.1103/PhysRevD.92.123528}
	{\bibfield  {journal} {\bibinfo  {journal} {Phys. Rev. D}\ }\textbf {\bibinfo
			{volume} {92}},\ \bibinfo {pages} {123528} (\bibinfo {year} {2015})},\
	\Eprint {https://arxiv.org/abs/1412.0678} {arXiv:1412.0678 [astro-ph.CO]}
	\BibitemShut {NoStop}%
	\bibitem [{\citenamefont {Li}\ {\it et~al.}(2015{\natexlab{f}})\citenamefont
		{Li}, \citenamefont {Sun}, \citenamefont {Tian},\ and\ \citenamefont
		{Wu}}]{Li:2014zfa}%
	\BibitemOpen
	\bibfield  {author} {\bibinfo {author} {\bibfnamefont {T.}~\bibnamefont
			{Li}}, \bibinfo {author} {\bibfnamefont {Z.}~\bibnamefont {Sun}}, \bibinfo
		{author} {\bibfnamefont {C.}~\bibnamefont {Tian}},\ and\ \bibinfo {author}
		{\bibfnamefont {L.}~\bibnamefont {Wu}},\ }\bibinfo {title} {{The
			Renormalizable Three-Term Polynomial Inflation with Large Tensor-to-Scalar
			Ratio}},\ \href {https://doi.org/10.1140/epjc/s10052-015-3508-4} {\bibfield
		{journal} {\bibinfo  {journal} {Eur. Phys. J. C}\ }\textbf {\bibinfo {volume}
			{75}},\ \bibinfo {pages} {301} (\bibinfo {year} {2015}{\natexlab{f}})},\
	\Eprint {https://arxiv.org/abs/1407.8063} {arXiv:1407.8063 [hep-ph]}
	\BibitemShut {NoStop}%
\end{thebibliography}
%

\end{document}